\documentclass{JINST}

\RequirePackage[T1]{fontenc}
\RequirePackage{graphicx}
\RequirePackage{mathptmx}      
\RequirePackage{flushend}
\RequirePackage[numbers,sort&compress]{natbib}

\usepackage{atlasphysics} 
\usepackage{subfigure}
\usepackage{wrapfig}
\usepackage{rotating}
\usepackage{mathptmx}
\usepackage{helvet}
\usepackage{array}
\usepackage{tabularx}
\usepackage{lineno}
\usepackage{amssymb, amsmath}

\newcommand{\expfor}[2]{$#1\!\times\! 10^{#2}$}

\newcommand{\antikt}{anti-$k_{t}$}

\newcommand{\EM}    {{\rm EM}}

\newcommand{\Etmiss}   {\ensuremath{{E}_{\mathrm{T}}^{\mathrm{miss}}}}

\newcommand{\ptjet}  {\ensuremath{\pt^\mathrm{jet}}}

\newcommand{\Qmean}  {\ensuremath{ \langle Q \rangle }}
\newcommand{\HECQuality}  {\ensuremath{f_{{\rm Q}}^{{\rm HEC}} }}
\newcommand{\LArQuality}  {\ensuremath{f_{{\rm Q}}^{{\rm LAr}} }}
\newcommand{\chf}  {\ensuremath{f_{\rm ch} }}
\newcommand{\fmax}  {\ensuremath{f_{\rm max}}}
\newcommand{\emf}  {\ensuremath{f_{\EM}}}
\newcommand{\hecf}  {\ensuremath{f_{\rm HEC}}}
\newcommand{\negE}  {\ensuremath{E_{\rm neg}}}
\newcommand{\timing}  {\ensuremath{t_{\rm jet}}}

\date{\today}

\title{Characterisation and mitigation of beam-induced backgrounds 
observed in the ATLAS detector during the 2011 proton-proton run}
\author{ATLAS Collaboration \\
E-mail: Atlas.Publications@cern.ch}

\abstract{
This paper presents a summary of beam-induced backgrounds observed in the ATLAS detector
and discusses methods to tag and remove background contaminated events in data.
Trigger-rate based monitoring of beam-related backgrounds is presented. 
The correlations of backgrounds with machine conditions, such as residual pressure in the beam-pipe, are discussed. 
Results from dedicated beam-background simulations are shown, and
their qualitative agreement with data is evaluated. Data taken during the passage of
unpaired, i.e. non-colliding, proton bunches is used to obtain background-enriched data samples. 
These are used to identify characteristic features of beam-induced backgrounds, which then are exploited to
develop dedicated background tagging tools. These tools, based on observables in the Pixel detector, the
muon spectrometer and the calorimeters, are described in detail and their efficiencies are evaluated.
Finally an example of an application of these techniques to a monojet analysis is given, which demonstrates the importance of
such event cleaning techniques for some new physics searches.}

\keywords{Accelerator modeling and simulations, Analysis and statistical methods, Pattern recognition, cluster finding, calibration and fitting methods, Performance of High Energy Physics Detectors}

\begin{document}

\section{Introduction}

In this paper, analyses of beam induced backgrounds (BIB) seen in the ATLAS
detector during the 2011 proton-proton run are presented.
At every particle accelerator, including the LHC\,\cite{lhc}, particles are lost from the beam
by various processes. During LHC high-luminosity running, the loss of beam intensity to proton-proton 
collisions at  the experiments has a non-negligible impact on the beam lifetime. Beam cleaning, i.e.
removing off-momentum and off-orbit particles is another important factor that reduces the
beam intensity. Most of the cleaning losses are localised in special insertions
far from the experiments, but a small fraction of the proton halo ends up on collimators 
close to the high-luminosity experiments. This distributed cleaning on one hand 
mitigates
halo losses in the immediate vicinity of the experiments, but by 
intercepting some of the halo these collimators themselves constitute a source of background entering the 
detector areas. 

Another important source of BIB is beam-gas
scattering, which takes place all around the accelerator. Beam-gas events
in the vicinity of the experiments inevitably lead to background in the detectors.

In ATLAS most of these backgrounds are mitigated by heavy shielding hermetically plugging 
the entrances of the LHC tunnel. 
However, in two areas of the detector, BIB can be a concern for operation and physics analyses:
\begin{itemize}
\item Background close to the beam-line can pass through the aperture left for the
beam and cause large longitudinal clusters of energy deposition, especially in pixel detectors close to the interaction point (IP), increasing
the detector occupancy and in extreme cases affecting the  track reconstruction by introducing spurious clusters.
\item High-energy muons are rather unaffected by the shielding material, but have the potential
to leave large energy deposits via radiative energy losses in the calorimeters, where the energy gets reconstructed as a jet. 
These fake jets\footnote{In this paper jet candidates originating from proton-proton collision events are called ``collision jets'' 
while jet candidates caused by BIB or other sources of non-collision backgrounds are referred to as ``fake jets''.} need to be identified 
and removed in physics analyses which rely on the measurement of missing transverse energy (\MET) and on jet identification. 
This paper presents techniques capable of tagging events with fake jets due to
BIB.
\end{itemize}

An increase in occupancy due to BIB, especially when associated with large local charge deposition, can increase the
dead-time of front-end electronics and lead to a degradation of data-taking efficiency.
In addition the triggers, especially those depending on \MET, can suffer from rate increases due to BIB.

This paper first presents an overview of the LHC beam structure, beam cleaning and interaction region 
layout, to the extent that is necessary to understand the background formation. A concise description
of the ATLAS detector, with emphasis on the sub-detectors most relevant for background studies is
given. This is followed by an in-depth discussion of BIB characteristics, presenting
also some generic simulation results, which illustrate the main features expected in the data.
The next sections present background monitoring with trigger rates, which reveal interesting
correlations with beam structure and vacuum conditions. This is followed by background observations with
the Pixel detector, which are compared with dedicated simulation results. The rest of the paper
is devoted to fake-jet rates in the calorimeters and various jet cleaning techniques, which are
effective with respect to BIB, but also other non-collisions backgrounds, like instrumental noise 
and cosmic muon induced showers.


\section{LHC and the ATLAS interaction region}

During the proton-proton run in 2011, the LHC operated at the nominal energy of 3.5\,TeV for both beams.
The Radio-Frequency (RF) cavities, providing the acceleration at the LHC, operate 
at a frequency of 400 MHz. This corresponds to buckets every 2.5\,ns,
of which nominally every tenth can contain a proton bunch. 
To reflect this sparse filling, groups of ten buckets, of which one can contain a proton bunch, are 
assigned the same Bunch Crossing IDentifier (BCID), of which there are 3564 in total.
The nominal bunch spacing in the 2011 proton run was 50\,ns, i.e. every second BCID was filled.
Due to limitations of the injection chain the bunches are collected in trains, each containing up to
36 bunches.
Typically four trains form one injected batch. The normal gap between trains within a batch is 
about 200\,ns, while the gap between batches is around 900\,ns. These train lengths and
gaps are dictated by the injector chain and the injection process.
In addition a 3\,$\rm \mu$s long gap is left, corresponding to the rise-time of the kicker magnets of the
beam abort system. The first BCID after the abort gap is by
definition numbered as 1.

\begin{figure}
  \centering
  {\includegraphics[width=0.9\textwidth]{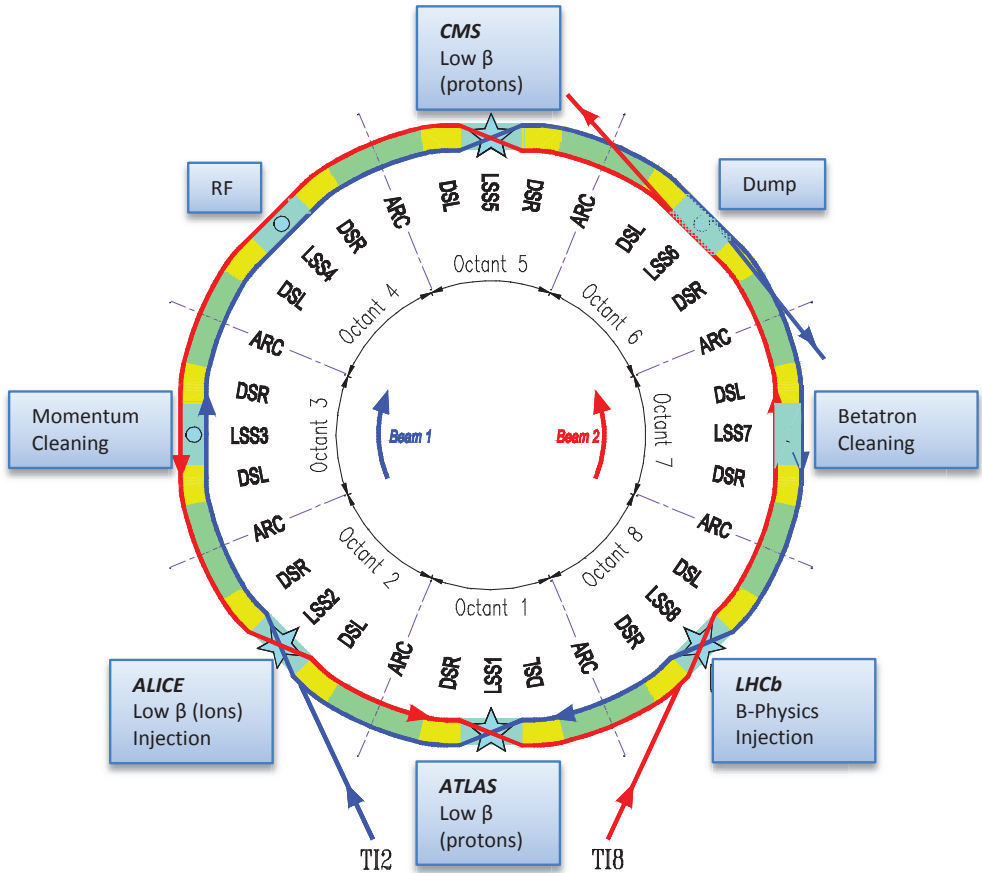}}            
  \caption{The general layout of the LHC\,\protect\cite{lhcfigref}. The dispersion suppressors (DSL and DSR) are 
sections between the straight section and the regular arc. In this paper they are considered to be
part of the arc, for simplicity. LSS denotes the Long Straight Section -- roughly 500\,m
long parts of the ring without net bending. All insertions (experiments, cleaning, dump, RF)
are located in the middle of these sections.
Beams are injected through transfer lines TI2 and TI8.}
  \label{lhc-layout}
\end{figure}

A general layout of the LHC, indicating the interaction regions with the experiments as well as the
beam cleaning insertions, is shown in Fig.\,\ref{lhc-layout}.

The beams are injected from the Super Proton Synchrotron (SPS) with an energy of 450\,GeV in several batches
and captured by the RF of the LHC. When the injection is complete the beams are
accelerated to full energy. When the maximum energy is reached the
next phase is the $\beta$-squeeze\footnote{The $\beta$-function determines the variation of the beam envelope around the ring and depends 
on the focusing properties of the magnet lattice -- for details see\,\protect\cite{wiedemann}}, 
during which the optics at the interaction points are changed from
an injection value of $\beta^*=11$\,m to a lower value, i.e. smaller beam size, at the IP.
Finally the beams are brought into collision, after which stable beams are declared and physics 
data-taking can commence. The phases prior to collisions, but at full energy, 
are relevant for background measurements because they allow the rates to be monitored
in the absence of the overwhelming signal rate from the proton-proton interactions.

The number of injected bunches varied from about 200 in early 2011 to
1380 during the final phases of the 2011 proton-proton run. Typically, 95\% of the bunches were colliding in ATLAS.
The pattern also included empty bunches and a small fraction of non-colliding, unpaired, bunches.
Nominally the empty bunches correspond to no protons passing through ATLAS, and are useful for monitoring of detector noise. 
The unpaired bunches are important for background monitoring in ATLAS.
It should be noted that these bunches were colliding in some other LHC experiments.
They were introduced by shifting some of the trains with respect to each other, such that unpaired bunches 
appeared in front of a train in one beam and at the end in the other. 
In some fill patterns 
some of these shifts overlapped such that interleaved bunches with only 25\,ns separation 
were introduced.

The average intensities of bunches in normal physics operation evolved over the year 
from \expfor{\sim 1.0}{11}\,{\it p}/bunch to \expfor{\sim 1.4}{11}\,{\it p}/bunch. The beam current at the end of the 
year was about 300\,mA and the peak luminosity in ATLAS was \expfor{3.5}{33}\,${\rm cm}^{-2}{\rm s}^{-1}$.

Due to the close bunch spacing, steering the beams head-on 
would create parasitic collisions outside of the IP. Therefore
a small crossing angle is used; in 2011 the full angle was 240\,$\rm \mu$rad in 
the vertical plane.
In the high-luminosity interaction regions the number of collisions is maximised by
the $\beta$-squeeze. In 2011 the value of $\beta^*$ was 1.5\,m initially and was
reduced to 1.0\,m in mid-September 2011. 

\begin{figure}
  \centering
  {\includegraphics[width=0.95\textwidth]{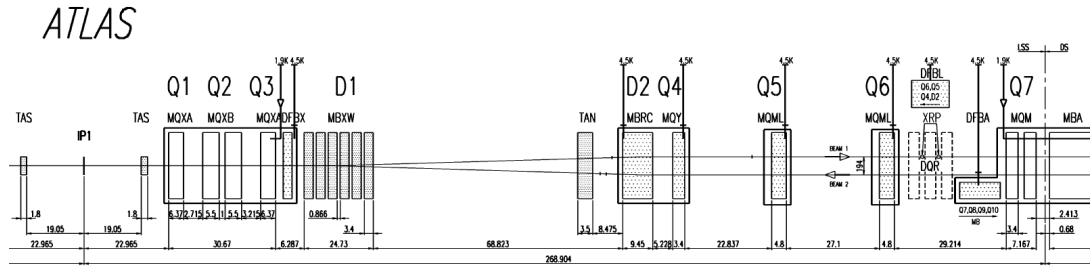}}            
  \caption{Detailed layout of the ATLAS interaction region\,\protect\cite{lhc}. The inner 
triplet consists of quadrupole magnets Q1, Q2 and Q3. The tertiary collimator (TCT) is not shown but is located 
between the neutral absorber (TAN) and the D2 magnet.}
  \label{ir1-layout}
\end{figure}

A detailed layout of the ATLAS interaction region (IR1) is shown in Fig.\,\ref{ir1-layout}.
Inside the inner triplet and up to the neutral absorber (TAN), both beams use the same 
beam pipe.
In the arc, beams travel in separate pipes with a horizontal separation of 194\,mm. 
The separation and recombination of the beams happens in dipole magnets D1 and D2 with 
distances to the IP of 59--83\,m and 153--162\,m, respectively. The D1 magnets are 
rather exceptional for the LHC, since they operate at room temperature in order to sustain 
the heat load due to debris from the interaction points. The TAS absorber, at 19\,m from
the IP, is a crucial element to protect the inner triplet against the heat load due to
collision products from the proton-proton interactions. It is a 1.8\,m long copper block with a 17\,mm radius
aperture for the beam. It is surrounded by massive steel shielding to reduce radiation 
levels in the experimental cavern\,\cite{ATLAS-Rad-Report}. The outer radius of this shielding
extends far enough to cover the tunnel mouth entirely, thereby shielding ATLAS from
low-energy components of BIB.

\begin{figure}
  \centering
  {\includegraphics[width=0.95\textwidth]{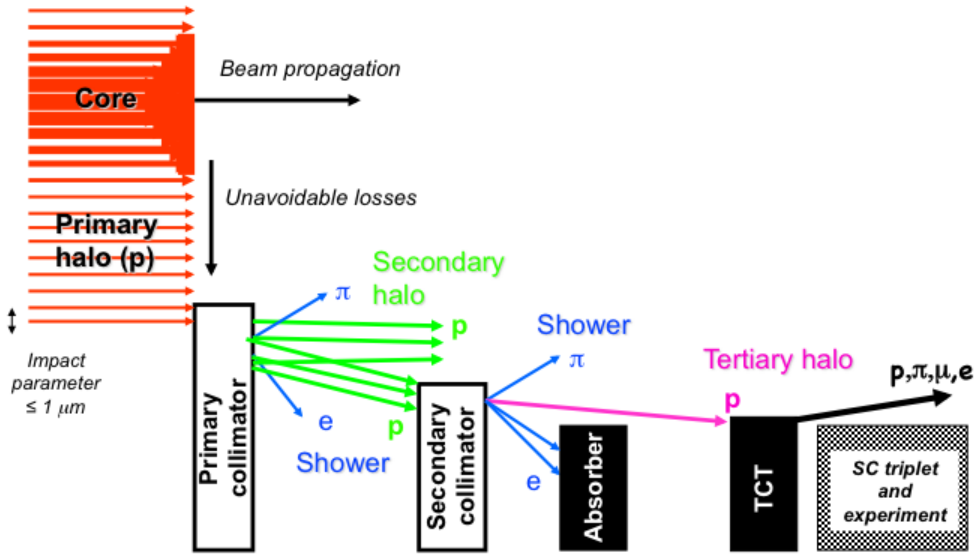}}            
  \caption{Schematic illustration of the LHC cleaning system. Primary and secondary collimators
and absorbers in the cleaning insertions remove most of the halo. Some tertiary halo escapes and
is intercepted close to the experiments by the TCT\protect\,\cite{assmannfig}.}
  \label{beamcleaning}
\end{figure}

The large stored beam energy of the LHC, in combination with the heat sensitivity of the superconducting magnets, 
requires highly efficient beam cleaning. This is achieved by two separate cleaning
insertions\,\cite{ralph1, ralph2,cleaning-ref}: betatron cleaning at LHC point 7 and momentum cleaning at point 3. In these
insertions a two-stage collimation takes place, as illustrated in Fig.\,\ref{beamcleaning}. Primary 
collimators (TCP) intercept particles that have left the beam core. Some of these particles are scattered and 
remain in the LHC acceptance, constituting the secondary halo, which hits the secondary collimators. Tungsten
absorbers are used to intercept any leakage from the collimators.
Although the combined local efficiency\footnote{Here the local efficiency ($\varepsilon_{\rm loc}$) is defined such that
on no element of the machine is the loss a fraction larger than $1-\varepsilon_{\rm loc}$ of the total.} 
of the the system is better than 99.9\,\%\,\cite{cleaning-ref}, 
some halo -- called tertiary halo -- escapes and is lost elsewhere in the machine. The inner 
triplets of the high-luminosity experiments represent limiting apertures where losses of tertiary halo 
would be most likely. In order to protect the quadrupoles, dedicated tertiary collimators (TCT) were
introduced at 145--148\,m from the high-luminosity IP's on the incoming beam side. The tungsten jaws of the TCT
were set in 2011 to 11.8\,$\sigma$, while the primary and secondary collimators at point 7 
intercepted the halo at 5.7\,$\sigma$ and 8.5\,$\sigma$, respectively.\footnote{Here $\sigma$ is the transverse betatronic beam standard 
deviation, assuming a normalised emittance of 3.5\,$\mu$m. 
In 2011 the LHC operated at smaller than nominal emittance, thus the actual physical apertures were larger in terms of $\sigma$.} 
Typical loss rates at the primary collimators were between $10^8$--$10^9$\,{\it p}/s during the 2011 high luminosity operation. These rates are 
comparable to about $10^8$ proton-proton events/s in both ATLAS and CMS, which indicates that the beam lifetime was influenced about 
equally by halo losses and proton-proton collisions.
The leakage fraction reaching the TCT was measured to be in the range $10^{-4}$--$10^{-3}$\,\cite{cleaning-ref,roderick-in-prep}, 
resulting in a loss rate on the order of $10^5$\,{\it p}/s on the TCT.

The dynamic residual pressure, i.e. in the presence of a nominal beam, in the LHC beam pipe is typically 
of the order of $10^{-9}$ mbar N$_2$-equivalent\footnote{The most abundant gases are H$_2$, CO, CO$_2$ and CH$_4$. 
For simplicity a common practice is to describe these with an N$_2$-equivalent, where the equivalence is calculated 
on the basis of the inelastic cross section at beam energy.} in the cold regions. 
In warm sections cryo-pumping, i.e. condensation on the cold pipe walls, is not available and pressures 
would be higher. Therefore most room-temperature sections of the vacuum chambers are coated with a special Non-Evaporative 
Getter (NEG) layer\,\cite{neg-benvenuti}, which maintains a good vacuum and significantly reduces 
secondary  electron yield. There are, however, some uncoated warm sections in the vicinity of the experiments. 
In 2010 and 2011 electron-cloud formation\,\cite{ecloud-prl,ecloud-evian} in these regions led to an increase of the 
residual pressure when the bunch spacing was decreased. As an emergency measure, in late 2010, small solenoids were 
placed around sections where electron-cloud formation was observed (58\,m from the IP). These solenoids curled up the 
low-energy electrons within the vacuum, suppressing the multiplication and thereby preventing electron-cloud build-up. 
During a campaign of dedicated "scrubbing" runs with high-intensity injection-energy beams, the surfaces
were conditioned and the vacuum improved. After this scrubbing, typical residual pressures in the warm sections 
remained below $10^{-8}$\,mbar N$_2$-equivalent in IR1 and were practically negligible in NEG coated 
sections -- as predicted by early simulations\,\cite{neg-rossi}.


\section{The ATLAS detector}
\label{atlas-sect}

The ATLAS detector\,\cite{atlasdet} at the LHC covers nearly the entire
solid angle around the interaction point with calorimeters extending up to a 
pseudorapidity $|\eta|=4.9$. Here $\eta=-\ln(\tan(\theta/2))$, with $\theta$ being 
the polar angle with respect to the nominal LHC beam-line. 

In the right-handed ATLAS coordinate system, with its origin at the nominal IP, the azimuthal angle $\phi$ is measured with respect to the  $x$-axis, which points towards 
the centre of the LHC ring.  
Side A of ATLAS is defined as the side of the incoming clockwise LHC beam-1, while the side of the incoming beam-2 is labelled C.
The $z$-axis in the ATLAS coordinate system points from C to A, i.e. along the beam-2 direction.

ATLAS consists of an inner tracking detector (ID) in the  $|\eta|<2.5$ region inside a 2\,T
superconducting solenoid, which is surrounded by electromagnetic and hadronic 
calorimeters, and an external muon spectrometer with three large superconducting toroid magnets. Each of these magnets consists of eight coils arranged radially and symmetrically around the beam axis.
The high-$\eta$ edge of the endcap toroids is at a radius of 0.83\,m and they extend to a radius of 5.4\,m. The barrel toroid is
at a radial distance beyond 4.3\,m and is thus not relevant for studies in this paper.

The ID is responsible for the high-resolution measurement of vertex positions and momenta of charged particles. 
It comprises a Pixel detector, a silicon tracker (SCT) and a Transition Radiation
Tracker (TRT). The Pixel detector consists of three barrel layers at mean radii of 50.5\,mm, 88.5\,mm and 122.5\,mm each with a half-length 
of 400.5\,mm.
The coverage in the forward region is provided by three Pixel disks per side at $z$-distances of 495\,mm, 580\,mm and 650\,mm from the IP 
and covering a radial range between 88.8--149.6\,mm. The Pixel sensors are 250\,$\mu$m thick and have a nominal pixel size 
of $r\phi\times z = 50\times 400\,\rm{\mu m}^2$. At the edge of the front-end chip there are linked pairs of ``ganged'' pixels which share 
a read-out channel. These ganged pixels are typically excluded in the analyses presented in this paper.

The ATLAS solenoid is surrounded by a high-granularity liquid-argon (LAr) electromagnetic calorimeter with lead 
as absorber material. The LAr barrel covers the radial range between 1.5\,m and 2\,m and has a half-length of 3.2\,m.
The hadronic calorimetry in the region $|\eta|<1.7$ is provided by 
a scintillator-tile calorimeter (TileCal), while 
hadronic endcap calorimeters (HEC) based on LAr technology are used in the region $1.5<|\eta|<3.2$. The absorber materials are iron and copper, respectively.
The barrel TileCal extends from $r=2.3$\,m to $r=4.3$\,m and has a total length of 8.4\,m. 
The endcap calorimeters cover up to $|\eta|=3.2$, beyond which the coverage is extended by the 
Forward Calorimeter (FCAL) up to $|\eta|=4.9$. The high-$\eta$ edge of the FCAL is at a radius of $\sim 70$\,mm
and the absorber materials are copper (electromagnetic part) and tungsten (hadronic part).
Thus the FCAL is likely to provide some shielding from BIB for the ID.
All calorimeters provide nanosecond timing resolution.

The muon spectrometer surrounds the calorimeters and is composed of a Monitored Drift Tube (MDT) system, covering the region 
of $|\eta|<2.7$ except for the innermost endcap layer where the coverage is limited to $|\eta|<2$.  In the
$|\eta|>2$ region of the innermost layer, Cathode-Strip Chambers (CSC) are used. The
CSCs cover the radial range 1--2\,m and are located at $|z|=8$\,m from the IP.
The timing resolution of the muon system is $2.5\,\textrm{ns}$ for the MDT and $7\,\textrm{ns}$ for the CSC.
The first-level muon trigger is provided
by Resistive Plate Chambers (RPC) up to  $|\eta|=1.05$ and Thin Gap Chambers (TGC) for $1.05<|\eta|<2.4$.

Another ATLAS sub-detector extensively used in beam-related studies is the Beam Conditions Monitor (BCM)\,\cite{bcmpap}. 
Its primary purpose is to monitor beam conditions and detect anomalous beam-losses which could result in detector damage. 
Aside from this protective function it is also used to monitor luminosity and BIB levels.
It consists of two detector stations (forward and backward) with four modules each.
A module consists of two polycrystalline chemical-vapour-deposition (pCVD) diamond sensors, glued together back-to-back
and read out in parallel.
The modules are positioned at $z=\pm 184$\,cm, corresponding to $z/c=6.13$\,ns distance to the interaction point.
The modules are at a radius of 55\,mm, i.e. at an $|\eta|$ of about 4.2 and arranged as a cross -- two modules on the vertical axis
and two on the horizontal. The active area of each sensor is $8\times8\,{\rm mm}^2$.
They provide a time resolution in the sub-ns range, and are thus well suited to identify BIB
by timing measurements.

In addition to these main detectors, ATLAS has dedicated detectors for forward physics and luminosity measurement (ALFA, LUCID, ZDC),
of which only LUCID was operated throughout the 2011 proton run. Despite the fact that LUCID is very close to the beam-line, it is
not particularly useful for background studies, mainly because collision activity entirely masks the small background signals.

An ATLAS data-taking session (run) ideally covers an entire stable beam period, which can last several hours. During this
time beam intensities and luminosity, and thereby the event rate, change significantly. To optimise the data-taking 
efficiency, the trigger rates are adjusted several times during a run by changing the trigger prescales. To cope with these changes and
those in detector conditions, a run is subdivided into luminosity blocks (LB). The typical length of a LB in the 2011 proton-proton run 
was 60\,seconds. The definition contains the intrinsic assumption that during a LB the luminosity changes by a negligible amount. 
Changes to trigger prescales and any other settings affecting the data-taking are always aligned with LB boundaries. 

In order to assure good quality of the analysed data, lists of runs and LBs with good
beam conditions and detector performance are used.
Furthermore, there are quality criteria for various reconstructed physics objects in the events that
help to distinguish between particle response and noise.
In the context of this paper, it is important to mention the quality criteria related to jets reconstructed
in the calorimeters.
The jet candidates used here are reconstructed using the \antikt{} jet clustering 
algorithm\,\cite{antikt} with a radius parameter $R=0.4$,
and topologically connected clusters of calorimeter cells\,\cite{topo} are used as input objects.
Energy deposits arising from particles showering in the calorimeters produce a characteristic pulse in the read-out of the calorimeter 
cells that can be used to distinguish ionisation signals from noise.
The measured pulse is compared with the expectation from simulation of the electronics response,
and the quadratic difference $Q_{\textrm{\scriptsize cell}}$ between the actual and expected pulse shape
is used to discriminate noise from real energy deposits.\footnote{
$Q_{\textrm{\scriptsize cell}}$ is computed online using the measured samples of the pulse shape in time as
\begin{equation}
Q_{\textrm{\scriptsize cell}} = \sum^{N}_{j=1} (s_{j}-Ag_{j}^{\textrm{\scriptsize phys}})^2
\end{equation}
where $A$ is the measured amplitude of the signal\,\cite{lar}, $s_{j}$ is the amplitude of each sample $j$,
and $g_{j}^{\textrm{\scriptsize phys}}$ is the normalised predicted ionisation shape.}
Several jet-level quantities can be derived from the following cell-level variables:
\begin {itemize}
\item $\hecf$:  Fraction of the jet energy in the HEC calorimeter.
\item $\Qmean$: The average jet quality is defined as the energy-squared weighted average of the pulse quality of the calorimeter cells ($Q_{\textrm{\scriptsize cell}}$) in the jet. This quantity is normalised such that $0<\Qmean<1$.
\item $\LArQuality$: Fraction of the energy in LAr calorimeter cells with poor signal shape quality ($Q_{\textrm{\scriptsize cell}}>4000$).
\item $\HECQuality$:  Fraction of the energy in the HEC calorimeter cells with poor signal shape quality ($Q_{\textrm{\scriptsize cell}}>4000$).
\item $\negE$: Energy of the jet originating from cells with negative energy
that can arise from electronic noise or early out-of-time pile-up\footnote{Out-of-time pile-up refers to proton-proton collisions occurring in BCIDs before or after the triggered collision event. }.
\end{itemize}

\section{Characteristics of BIB}
\label{sect-simulation}

At the LHC, BIB in the experimental regions are due mainly to three different processes\,\cite{drozhdin,mokhov-bg-elastic, Bruce-IPAC}:

\begin{itemize}
\item {\bf Tertiary halo}: protons that escape the cleaning insertions and are lost on limiting apertures,
typically the TCT situated at $|z|\approx150\,$m from the IP.
\item {\bf Elastic beam-gas}: elastic beam-gas scattering, as well as single diffractive scattering, can 
result in small-angle deflections of the protons. These can be lost on the next limiting aperture before reaching the 
cleaning insertions. These add to the loss rate on the TCTs.
\item {\bf Inelastic beam-gas}: inelastic beam-gas scattering results in showers of secondary particles. Most of these
have only fairly local effects, but high-energy muons produced in such events can travel large distances and
reach the detectors even from the LHC arcs.
\end{itemize}

By design, the TCT is the main source of BIB resulting from tertiary halo losses. Since it is in the straight section
with only the D1 dipole and inner triplet separating it from the IP, it is expected that the secondary particles produced in the TCT
arrive at rather small radii at the experiment. The losses on the TCT depend on the leakage from the primary collimators, but
also on other bottlenecks in the LHC ring. Since the betatron cleaning is at LHC point 7, halo of the clockwise beam-1 has to pass
two LHC octants to reach ATLAS, while beam-2 halo has six octants to cover, with the other low-$\beta$ experiment, CMS, 
on the way. Due to this asymmetry, BIB due to losses on the TCT cannot be assumed to be symmetric for both beams.

There is no well-defined distinction between halo and elastic beam-gas scattering because scattering at very small angles
feeds the halo, the formation of which is a multi-turn process as protons slowly drift out of the beam core until they hit the
primary collimators in the cleaning insertions at IP3 and IP7. Some scattering events, however, lead to enough deflection that 
the protons are lost on other limiting apertures before they reach the cleaning insertions. 
The most likely elements at which those protons can be lost close to the experiments are the TCTs. The rate of such losses is in 
addition to the regular tertiary halo. This component is not yet included in the simulations, but earlier studies based on
7\,TeV beam energy suggest that it is of similar magnitude as the tertiary halo\,\cite{mokhov-bg-elastic}. The same 7\,TeV 
simulations also indicate that the particle distributions at the experiment are very similar to those due to tertiary halo losses.

\begin{figure}
  \centering
  {\includegraphics[width=0.8\textwidth]{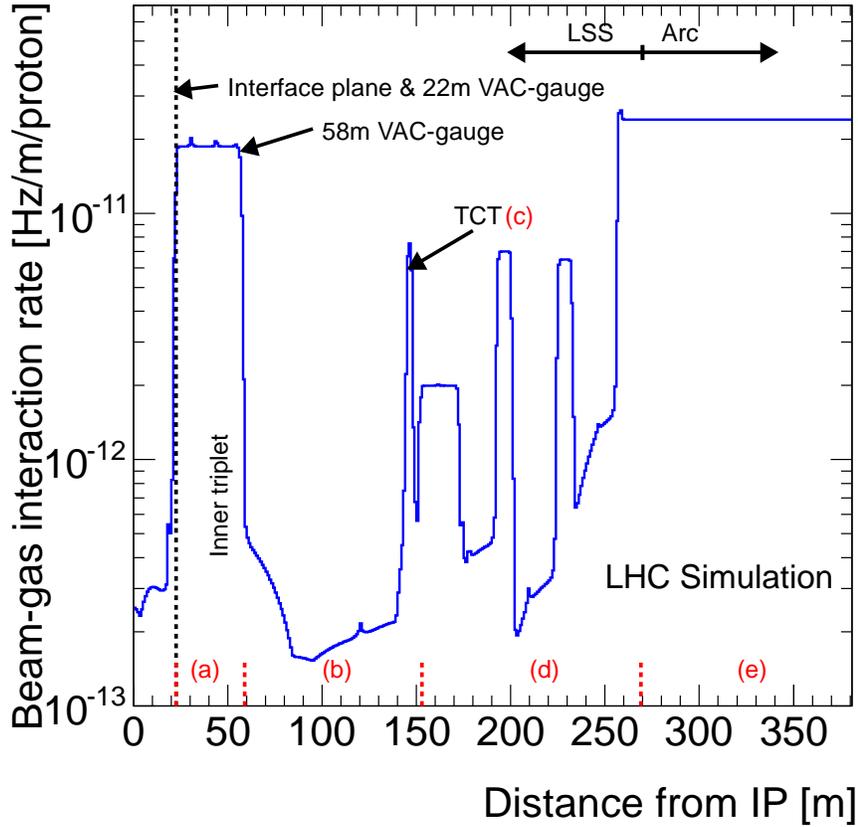}}            
  \caption{Inelastic beam-gas interaction rates per proton, calculated for beam-1 in LHC fill 2028.
The machine elements of main interest are indicated.  The pressure in the arc is assumed to be constant from 270\,m onwards.
The letters, a-e, identify the different sections, for which rates are given separately in other plots.}
  \label{pintFig}
\end{figure}

The inelastic beam-gas rate is a linear function of the beam intensity and of the residual pressure in the vacuum chamber.
The composition of the residual gas depends on the surface characteristics of the vacuum chamber and is different in warm and cryogenic sections and
in those with NEG coating.  
Although several pressure gauges are present around the LHC, detailed pressure maps can be obtained only from
simulation similar to those described in\,\cite{giuseppe}. The gauges can then be used to cross-check the simulation results at selected points. 
The maps allow the expected rate of beam-gas events to be determined. Such an interaction distribution, calculated for the conditions of LHC fill 2028, is
shown in Fig.\,\ref{pintFig}. The cryogenic regions, e.g. inner triplet (23--59m), the magnets D2 \& Q4, Q5 and Q6 at $\sim 170$\,m, 
$\sim 200$\,m and $\sim 220$\,m, respectively, and the arc ($>$269\,m), are clearly visible as regions with a higher rate, while the NEG coating of warm 
sections efficiently suppresses beam-gas interactions. The TCT, being a warm element without NEG coating, produces a prominent spike 
at $\sim 150$\,m. In the simulations it is assumed that the rate and distribution of beam-gas events are the same for both beams.

\subsection{BIB simulation methods}

The simulation of BIB follows the methods first outlined in\,\cite{drozhdin}, in particular the concept of a two-phase
approach with the machine and experiment simulations being separate steps. In the first phase the various sources of BIB are
simulated for the LHC geometry\,\cite{Bruce-IPAC, roderick-in-prep}. These simulations produce a file of particles crossing an interface plane at $z=22.6$\,m from
the IP. From this plane onwards, dedicated detector simulations are used to propagate the particles through the experimental area and the detector. 
Contrary to earlier studies\,\cite{mokhov-bg-elastic,drozhdin,mokhov1}, more powerful CPUs available today allow the machine simulations to be
performed without biasing.\footnote{There are several biasing techniques 
available in Monte Carlo simulations. All of these aim at increasing statistics in some regions of phase space at the
cost of others by modifying the physical probabilities and compensating this by assigning non-unity statistical weights to the
particles. As an example the life-time of charged pions can be decreased in order to increase muon statistics. 
The statistical weight of each produced muon is then smaller than one so that on average the sum of muon weights corresponds to the true
physical production rate.}
This has the advantage of preserving all
correlations within a single event and thus allows event-by-event studies of detector response. The beam halo formation and cleaning
are simulated with SixTrack\,\cite{sixtrack}, which combines optical tracking and Monte Carlo simulation of particle interactions in
the collimators. The inelastic interactions, either in the TCT based on the impact coordinates from SixTrack, or with residual
gas, are simulated with {\sc Fluka}\,\cite{fluka}. The further transport of secondary particles up to the interface plane is also done with {\sc Fluka}.

High-energy muons are the most likely particles to cause fake jet signals in the calorimeters. At sufficiently
large muon energies, typically above 100\,GeV, radiative energy losses start to dominate and these can result in
local depositions of a significant fraction of the muon energy via electromagnetic and, rarely, hadronic cascades\,\cite{pdg-mokhov}.

The TCTs are designed to intercept the tertiary halo. Thus they represent intense -- viewed from the IP, almost point-like --
sources of high-energy secondary particles. The TCTs are in the straight section and the high-energy particles have a 
strong Lorentz boost along $z$. Although they have to traverse the D1 magnet and the focusing quadrupoles before reaching the 
interface plane, most of the muons above 100\,GeV remain at radii below 2\,m.

The muons from inelastic beam-gas events, however, can originate either from the straight section or from the arc.
In the latter case they emerge tangentially to the ring or pass through several bending 
dipoles, depending on energy and charge. Both effects cause these muons to be spread out in the horizontal plane so that their radial distribution at 
the experiment shows long tails, especially towards the outside of the ring.

In the following, some simulation results are shown, based on the distribution of muons with momentum greater than 100\,GeV at 
the interface plane. The reason to restrict the discussion to muons is twofold:

\begin{enumerate}
\item The region between the interface plane and the IP is covered by heavy shielding and detector material.
All hadrons and EM-particles, except those within the 17\,mm TAS aperture or at radii outside the shielding, 
undergo scattering and result in a widely spread shower of secondary particles. Therefore the distributions of these particles 
at the interface plane do not directly reflect what can be seen in the detector data.
\item High-energy muons are very penetrating and rather unaffected by material, but they 
are also the cause of beam-related calorimeter background. Therefore the distribution
of high-energy muons is expected to reflect the fake jet distribution seen in data. 
The muon component is less significant for the ID, but its distribution can still reveal interesting effects. 
\end{enumerate}

\begin{figure}
  \centering
  {\includegraphics[width=0.8\textwidth]{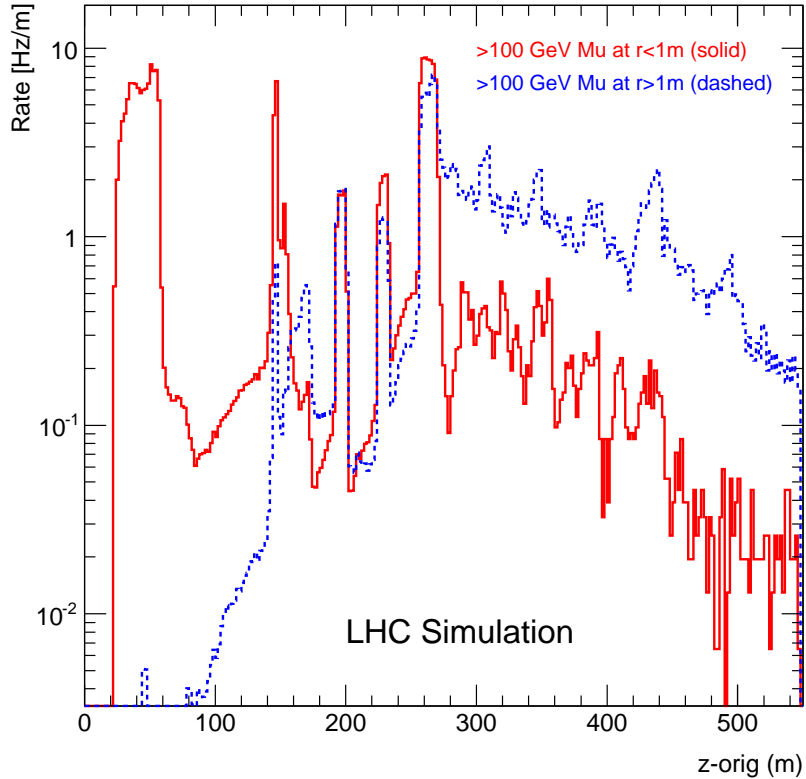}}            
  \caption{Simulated distribution of the $z$-coordinates of inelastic beam-gas events from which a muon with more than 100 GeV
has reached the interface plane at 22.6\,m. The two curves correspond to muons at radii below and above 1\,m at the interface plane.}
  \label{muons-from-z}
\end{figure}

Figure\,\ref{muons-from-z} shows the simulated $z$-distribution of inelastic beam-gas events resulting in a high-energy muon 
at the interface plane. In order to reach larger radii the muons have to originate from more distant events. Since the barrel 
calorimeters\footnote{Fake jets can be produced also in the endcap and forward calorimeters, but due to higher rapidity are
less likely to fake a high-$\pt$ jet.}, which detect the possible fake jets, cover radii above 1\,m, the fake jet rate is not expected to be sensitive 
to close-by beam-gas interactions and therefore not to the pressure in the inner triplet. This is discussed later in the context of correlations 
between background rates and pressures seen by the vacuum gauges at $|z|=22$\,m and $|z|=58$\,m.

\begin{figure}
  \centering
  {\includegraphics[width=0.8\textwidth]{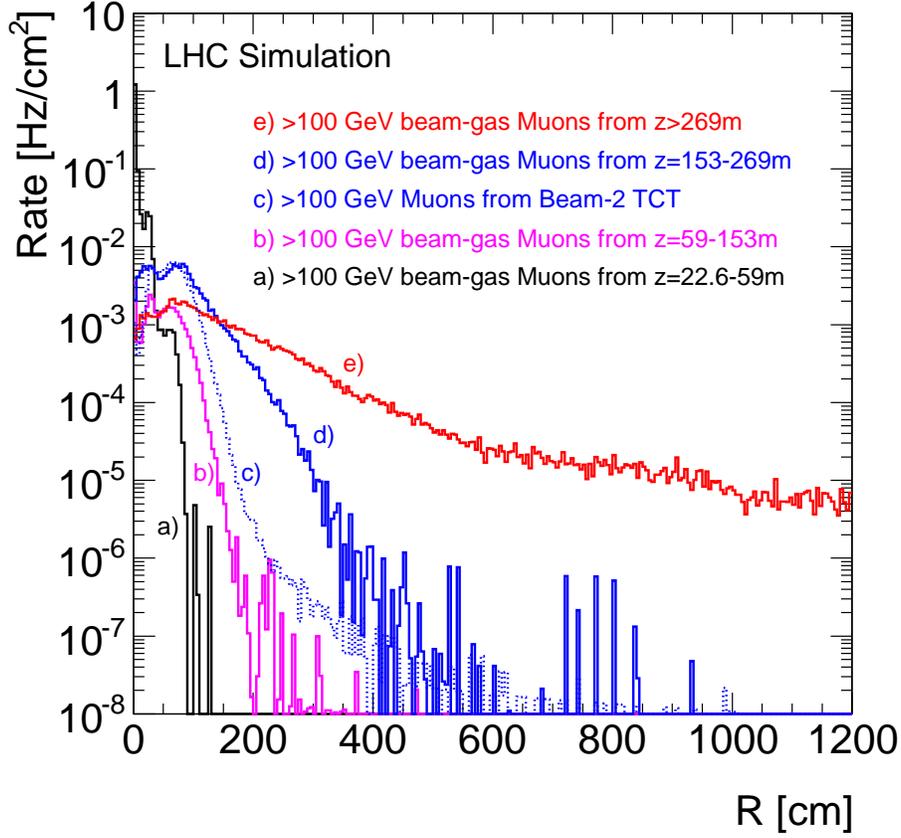}}            
  \caption{Simulated radial distribution at the interface plane of muons from inelastic beam-gas collisions and from the beam-2 TCT.
  The four solid curves correspond to muons originating from beam-gas events in different regions of the LSS and the adjacent LHC arc. The 
dashed curve shows the distribution of muons from the beam-2 TCT, normalised to $10^5$ {\it p}/s lost on the TCT.
The letters refer to the regions indicated in Fig.\,\protect\ref{pintFig}.}
  \label{muon-r-z}
\end{figure}

Figure\,\ref{muon-r-z} shows the simulated radial distributions of high-energy muons from inelastic beam-gas events taking place at
various distances from the IP. Figure\,\ref{pintFig} suggests that the regions with highest interaction rate are the inner 
triplet, the TCT region, the cold sections in the LSS beyond the TCT, and the arc. 
In NEG-coated warm regions the expected beam-gas rate is negligible, which allows the interesting sections to be
grouped into four wide
regions, as indicated at the bottom of Fig.\,\ref{pintFig}.
It is evident from Fig.\,\ref{muon-r-z} that at very small radii beam-gas interactions in the inner triplet dominate, but these do not give any contributions at 
radii beyond 1\,m. The radial range between $1$--$4$\,m, covered by the calorimeters, gets contributions from all three distant
regions, but the correlation between distance and radius is very strong and in the TileCal ($r=2$--$4$\,m) muons from the arc dominate 
by a large factor. Beyond a radius of 4\,m only the arc contributes to the high-energy muon rate.

The dashed curve in Fig.\,\ref{muon-r-z} shows the radial distribution of high-energy muons from interactions in the TCT, which
represents a practically point-like source situated at slightly less than 150\,m from the IP. It can be seen that the
radial distribution is quite consistent with that of beam-gas collisions in the $z=59$--$153$\,m region. The TCT losses lead to a fairly broad maximum
below $r=1$\,m, followed by a rapid drop, such that there are very few high-energy muons from the TCT at $r>3$\,m. The absolute level, normalised to
the average loss rate of $10^5$\,{\it p}/s on the TCT, is comparable to that expected from beam-gas collisions.

\begin{figure}
  \centering
  {\includegraphics[width=0.9\textwidth]{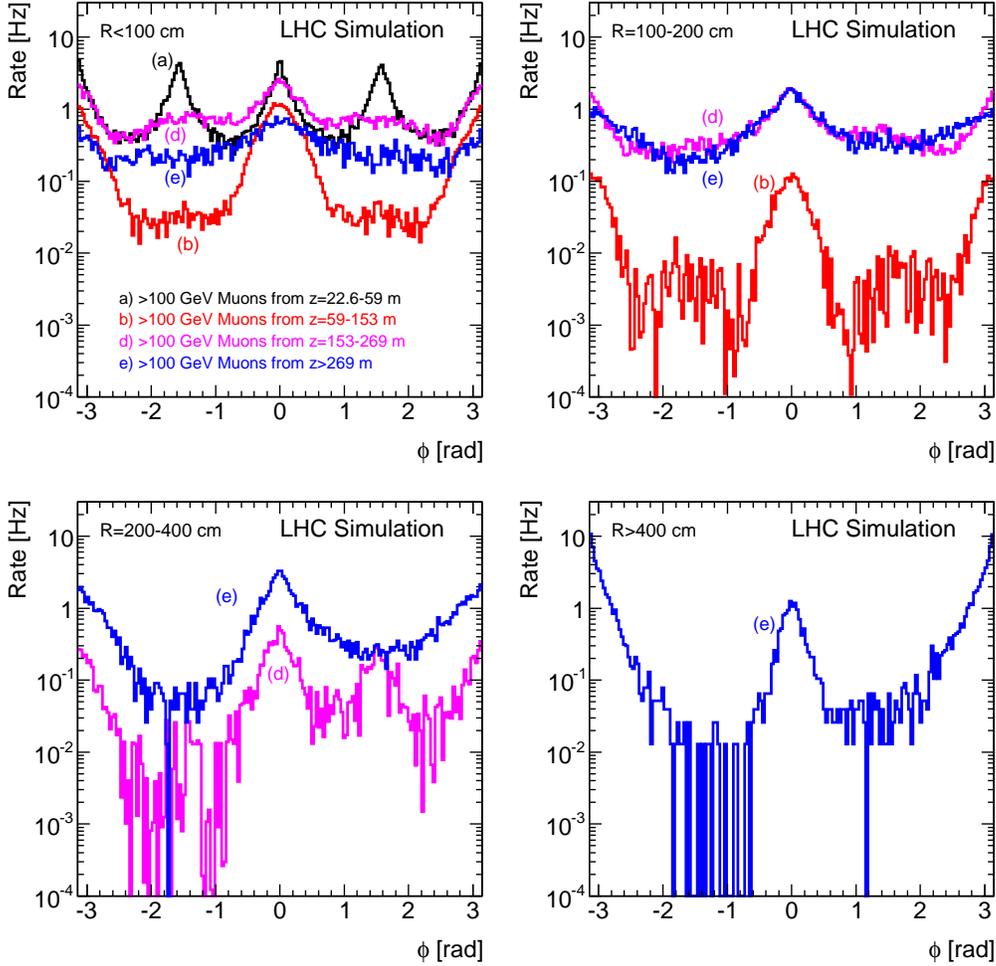}}            
  \caption{Simulated azimuthal distribution of beam-gas muons at the interface plane in four different radial ranges. 
  The regions of origin considered are the same as in Fig.\,\protect\ref{muon-r-z}. The values
  have not been normalised to unit area, but represent the rate over the entire surface - which is different in each plot.
  The letters refer to the regions indicated in Fig.\,\protect\ref{pintFig}. The 
  contribution from nearby regions drops quickly with radius. Histograms with negligible contribution have been suppressed.
}
  \label{phidist4}
\end{figure}

Figure\,\ref{phidist4} shows the simulated $\phi$-distribution of the high-energy muons for different radial ranges and regions of
origin of the muons. At radii below 1\,m the muons from the inner triplet show a structure with four spikes, created by 
the quadrupole fields of the focusing magnets. Muons from more distant locations are deflected 
in the horizontal plane by the separation and recombination dipoles creating a structure with two prominent spikes. The figure shows both charges together, but 
actually D1 separates, according to charge, the muons originating from within 59-153\,m.
Since D2 has the same bending power but in the opposite direction, muons from farther away are again mixed.
The same two-spiked structure is also seen at larger radii. Beyond $r=2$\,m a slight up-down
asymmetry is observed, which can be attributed to a non-symmetric position of the beam-line with respect to the tunnel floor and ceiling -- depending on the
region, the beam-line is about 1\,m above the floor and about 2\,m below the ceiling.
This causes a different free drift for upward- and downward-going pions and kaons to decay into muons before interacting in material. Since the floor 
is closer than the roof, fewer high-energy muons are expected in the lower hemisphere. 
A similar up-down asymmetry was already observed in calorimetric energy deposition when 450\,GeV low-intensity proton bunches were dumped
on the TCT during LHC beam commissioning\,\cite{splashes}, although in this case high-energy muons probably were a small contribution
to the total calorimeter energy.
Finally, at radii beyond 4\,m, only  muons from the arc contribute. The peak at $|\phi|=\pi$ is
clearly dominant, and is due to the muons being emitted tangentially to the outside of the ring.


\section{BIB monitoring with Level-1 trigger rates}
\label{sec:monitoring}

The system that provides the Level-1 (L1) trigger decision, the ATLAS Central Trigger Processor (CTP)\,\cite{ctp}, 
organises the BCIDs into Bunch Groups (BG) to account for the very different characteristics, trigger rates, and 
use-cases of colliding, unpaired, and empty bunches.
The BGs are adapted to the pattern of each LHC fill and their purpose
is to group together BCIDs with similar characteristics as far as trigger rates are concerned. In particular, the same
trigger item can have different prescales in different BGs.

\noindent
The BGs of interest for background studies are:

\begin{itemize}
\item {\bf BGRP0}, all BCIDs, except a few at the end of the abort gap
\item {\bf Paired}, a bunch in both LHC beams in the same BCID
\item {\bf Unpaired isolated} (UnpairedIso), a bunch in only one LHC beam with no bunch in the other beam within $\pm$\,3\,BCIDs.
\item {\bf Unpaired non-isolated} (UnpairedNonIso), a bunch in only one LHC beam with a nearby bunch (within three BCIDs) in the other beam.
\item {\bf Empty}, a BCID containing no bunch and separated from any bunch by at least five BCIDs. 
\end{itemize}

The L1 trigger items which were primarily used for background monitoring in the 2011 proton run are summarised in
Table\,\ref{L1triggers} and explained in the following.

\begin{table}
{\small
\begin{tabular}{l|l|l} \hline
Trigger item                  & Description                     & Usage in background studies  \\ \hline
L1\_BCM\_AC\_CA\_BGRP0        & BCM background-like coincidence & BIB level monitoring         \\
L1\_BCM\_AC\_CA\_UnpairedIso  & BCM background-like coincidence & BIB level monitoring         \\
L1\_BCM\_Wide\_UnpairedIso    & BCM collision-like coincidence  & Ghost collisions             \\
L1\_BCM\_Wide\_UnpairedNonIso & BCM collision-like coincidence  & Ghost collisions             \\
L1\_J10\_UnpairedIso          & Jet with $\pt>10$\,GeV at L1    & Fake jets \& ghost collisions \\
L1\_J10\_UnpairedNonIso       & Jet with $\pt>10$\,GeV at L1    & Fake jets \& ghost collisions \\ 
\hline
\end{tabular}}
\caption{ATLAS trigger items used during the 2011 proton runs for background studies and monitoring.}
\label{L1triggers}
\end{table}

The L1\_BCM\_AC\_CA trigger is defined to select particles travelling parallel to the beam, from side A to side C or vice-versa.
It requires a background-like coincidence of two hits, defined as
one (early) hit in a time window $-6.25\pm~2.73$\,ns before
the nominal collision time and the other (in-time) hit in a time window $+6.25\pm~2.73$\,ns after the nominal collision time.

Table\,\ref{L1triggers} lists two types of BCM background-like triggers -- one in BGRP0, 
and the other in the Unpair\-edIso BG. The motivation to move from L1\_BCM\_AC\_CA\_BGRP0, used in 2010\,\cite{conf-note-2010}, 
to unpaired bunches was that a study of 2010 data revealed a significant luminosity-related contamination due to
accidental background-like coincidences in the trigger on all bunches (BGRP0).
Although the time window of the trigger is narrow enough to discriminate collision products from the actually passing
bunch, each proton-proton event is followed by afterglow\,\cite{ATLASLUMI}, i.e. delayed tails of the particle cascades produced in the detector 
material. The afterglow in the BCM  is exponentially falling and the tail extends to 
$\sim 10$\,$\mu$s after the collision. With 50\,ns bunch spacing this afterglow piles up and becomes intense enough to have a non-negligible 
probability for causing an upstream hit in a later BCID that is in background-like coincidence with a true background hit in the downstream 
detector arm. In the rest of this paper, unless otherwise stated, the L1\_BCM\_AC\_CA\_UnpairedIso rate before prescaling is referred to as
BCM background rate.

A small fraction of the protons injected into the LHC escape their nominal bunches. If this happens in the 
injectors, the bunches usually end up in neighbouring RF buckets. If the bunches are within the same 25\,ns BCID as the
main bunch, they are referred to as satellites. If de- and re-bunching happens during RF capture in the LHC, the protons 
spread over a wide range of buckets and if they fall outside filled BCIDs, they are referred to as ghost charge.

The L1\_BCM\_Wide triggers require a collision-like coincidence, i.e. in-time hits on both sides of the IP. The time window
to accept hits extends from 0.39\,ns to 8.19\,ns after the nominal collision time. 

The L1\_J10 triggers fire on an energy deposition above $10\,\textrm{GeV}$, at approximately electromagnetic scale,
in the transverse plane in an $\eta$--$\phi$ region with a width of about $0.8\times0.8$ anywhere within $|\eta|<3.0$ and, with
reduced efficiency, up to $\eta|=3.2$.
Like the L1\_BCM\_Wide triggers, the two L1\_J10 triggers given in Table\,\ref{L1triggers} are active in UnpairedIso or UnpairedNonIso 
bunches, which makes them suitable for studies of ghost collisions rates in these two categories of unpaired bunches.

The original motivation for introducing the UnpairedIso BG was to stay clear of this ghost charge, while the UnpairedNonIso 
BG was intended to be used to estimate the amount of this component. However, as will be shown, an isolation by $\pm 3$ BCID is 
not always sufficient, and some of the UnpairedIso bunches still have signs of collision activity. Therefore 
Table\,\ref{L1triggers} lists the UnpairedIso BG as suitable for ghost charge studies.

\subsection{BCM background rates vs residual pressure}
  \label{sec:bcmp22}
\begin{figure}
  \centering
  {\includegraphics[width=0.8\textwidth]{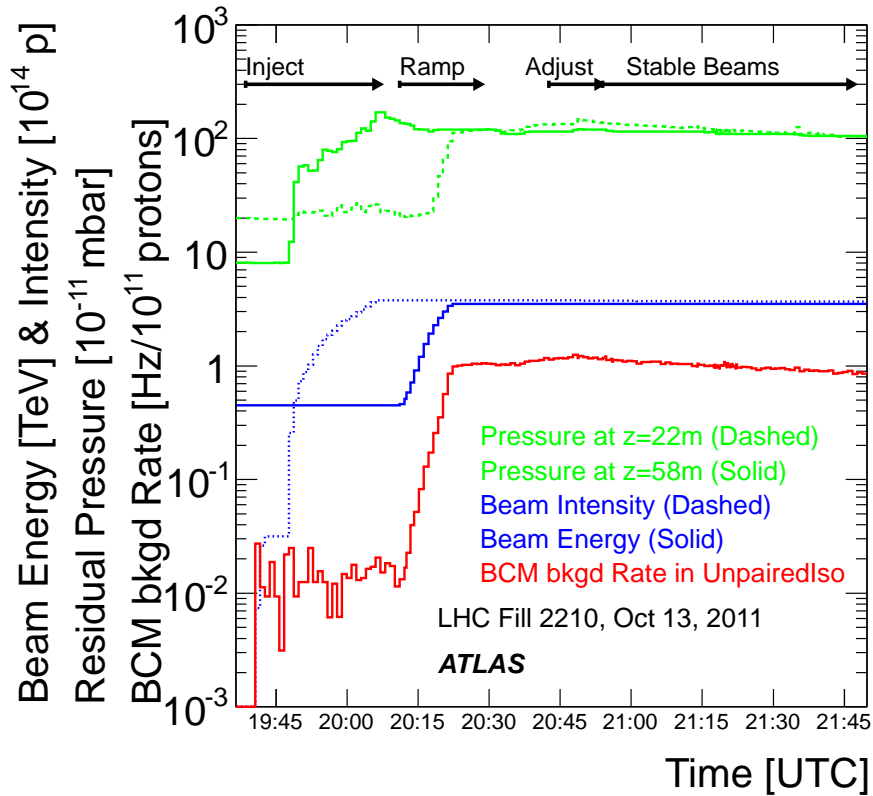}}            
  \caption{Pressures, beam parameters and BCM background rates during the start of
a typical LHC fill.}
  \label{fillstart}
\end{figure}

In order to understand the origin of the background seen by the BCM,
the evolution of the rates and residual pressure in various parts
of the beam pipe at the beginning of an LHC fill are studied. The vacuum gauges providing data
for this study are located at 58\,m, 22\,m and 18\,m from the IP. The pressures from these are
referred to as P58, P22 and P18, respectively.
Figure\,\ref{fillstart} shows a characteristic evolution of pressures and BCM background rate when 
the beams are injected, ramped and brought into collision.
P58 starts to increase as soon as beam is injected into the LHC.
The pressure, however, does not reflect itself in the background seen by the
BCM. Only when the beams are ramped from 450\,GeV to 3.5\,TeV, does P22
increase, presumably due to increased synchrotron radiation from the inner triplet. 
The observed BCM background increase is disproportionate to the pressure increase. This is explained
by the increasing beam energy, which causes the produced secondary particles, besides 
being more numerous, to have higher probability for inducing penetrating showers in the 
TAS, which is between the 22\,m point and the BCM. The pressure of the third gauge, located at 18\,m in a NEG-coated
section of the vacuum pipe, is not shown in Fig.\,\ref{fillstart}. The NEG-coating reduces the pressure by almost two orders of
magnitude, such that the residual gas within $\pm$\,19\,m does not contribute significantly to the
background rate. 
According to Fig.\,\ref{pintFig}, the pressure measured by the 22\,m gauge is constant through the entire 
inner triplet\footnote{The pressure simulation is based, among other aspects, on the distribution and intensity
of synchrotron radiation, which is assumed to be constant within the triplet.}. This and the correlation with P22 suggest 
that the background seen by the BCM is due mostly to beam-gas events in the inner triplet region. 

\begin{figure}
  \centering
  {\includegraphics[width=0.8\textwidth]{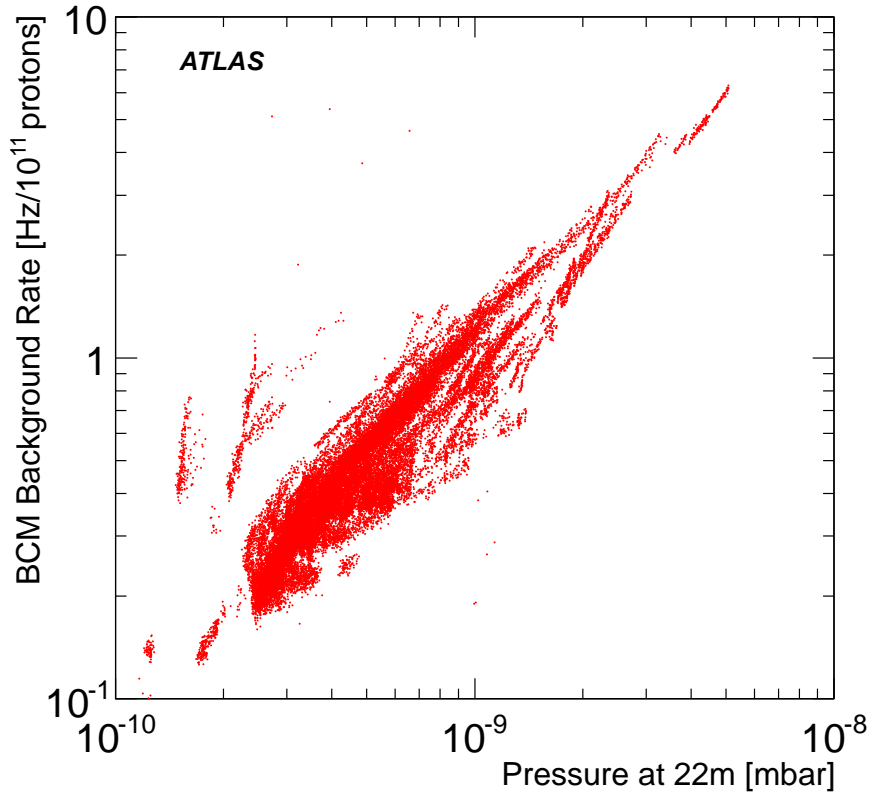}}            
  \caption{Correlation between P22 and BCM background rate. Each dot represents one LB.}
  \label{bcmhalop22}
\end{figure}

This conclusion is further supported by Fig.\,\ref{bcmhalop22} where the
BCM background rate versus P22 is shown. In the plot
each point represents one LB, i.e. about 60\,seconds of data-taking. Since
beam intensities decay during a fill, the pressures and background rate also decrease so that
individual LHC fills are seen in the plot as continuous lines of dots.
A clear, although not perfect, correlation can be observed. There are a few outliers
with low pressure and relatively high rate. All of these are associated with 
fills where P58 was abnormally high.

\begin{figure}
  \centering
  {\includegraphics[width=0.8\textwidth]{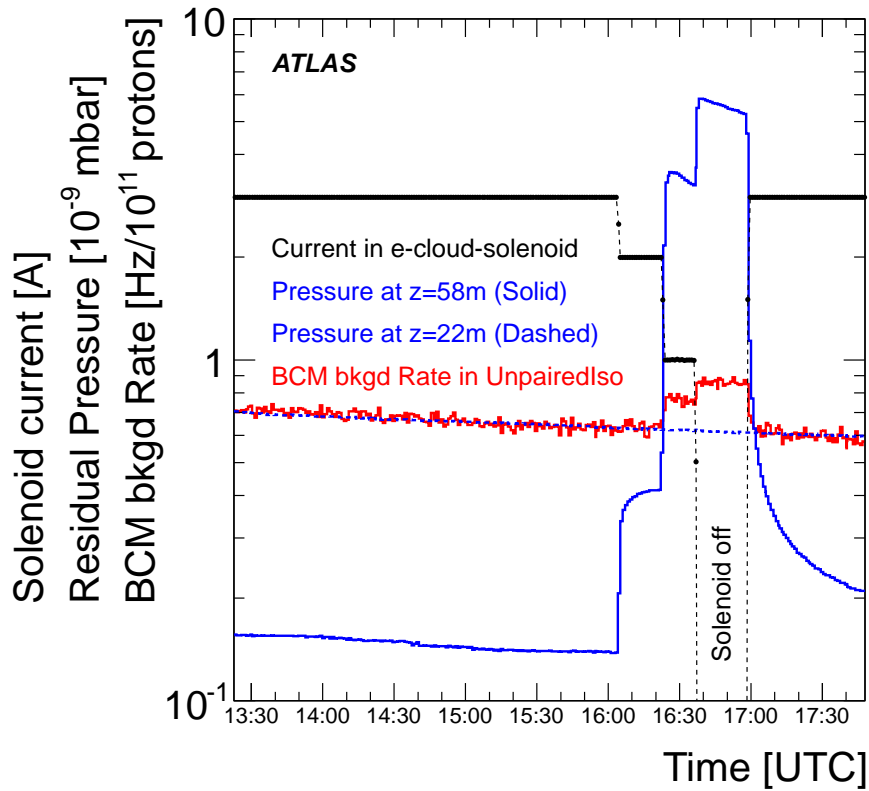}}            
  \caption{P58, P22 and background seen at the BCM during 
the ``solenoid test'' in LHC fill 1803. The dashed line is 
not a fit, but the actual P22 value, which just happens to
agree numerically with the BCM rate on this scale.}
  \label{solenoid}
\end{figure}

The relative influence of P22 and P58 on the BCM background
was studied in a special test, where the small solenoids around the beam pipe
at 58\,m, intended to suppress electron-cloud formation, were gradually turned off
and back on again. Figure\,\ref{solenoid} shows the results of this study. The solenoids
were turned off in three steps and due to the onset of electron-cloud formation the
pressure at 58\,m increased by a factor of about 50. At the same time the pressure at
22\,m showed only the gradual decrease due to intensity lifetime. With the solenoids turned off,
P58 was about nine times larger than P22. At the same time the BCM background rate increased
by only 30\%, while it showed perfect proportionality to P22 when the solenoids were on
and P58 suppressed. This allows quantifying the relative effect of P58 on the BCM background
to be about 3-4\% of that of P22. 
If these 3-4\% were taken into account in Fig.\,\ref{bcmhalop22}, the outliers described 
above would be almost entirely brought into the main distribution. 

In summary, the BCM background trigger can be considered to be a very good measurement of
beam-gas rate produced close to the experiment, while it has low efficiency
to monitor beam losses far away from the detector.

\subsection{BCM background rates during 2011}

\begin{figure}
  \centering
  {\includegraphics[width=0.85\textwidth]{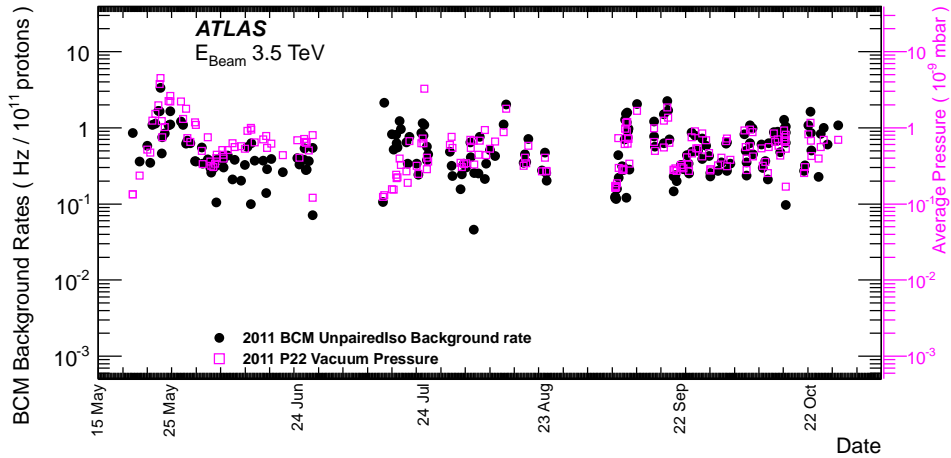}}            
  \caption{BCM background rate normalised to $10^{11}$ protons for the
    2011 proton-proton running period starting from mid-May. 
    The rate is shown together with the P22 average residual pressure.}
  \label{bcm-all2011}
\end{figure}

Figure\,\ref{bcm-all2011} shows the BCM background rate for the 2011 proton runs together with the P22 average residual pressure.
These rates are based on the L1\_BCM\_AC\_CA\_UnpairedIso trigger rates, which became available
after the May technical stop of the LHC. During the period covered by the plot, the number of unpaired bunches
and their location in the fill pattern changed considerably. No obvious correlation between the scatter of the data and 
these changes could be identified. No particular time structure or long-term trend can be observed in the 2011 data.
The average value of the intensity-normalised rate remains just below 1\,Hz throughout the year.

Except for a few outliers, due to abnormally high P58, the BCM background rate correlates well with the average 
P22 residual pressure, in agreement with Fig.\,\ref{bcmhalop22} and the discussion in Sect.\,\ref{sec:bcmp22}.


\subsection{Observation of ghost charge}
\label{sec:ghostcharge}

The BCM allows studies of the amount of ghost charge in nominally empty BCIDs. The
background-like trigger can be used to select beam-gas events created by ghost charge.
Since, for a given pressure, the beam-gas event rate is a function of bunch intensity 
only, this trigger yields directly the relative intensity of the ghost charge with respect
to a nominal bunch, in principle. The rate, however, is small and almost entirely absorbed in 
backgrounds, mainly the accidental afterglow coincidences discussed at the beginning of this section.
Another problem is that due to the width of the background trigger time window, 
only the charge in two or three RF buckets is seen, depending on how accurately the
window is centred around the nominal collision time.

A more sensitive method is to look at the collisions of a ghost bunch with
nominal bunches. Provided the emittance of the ghost bunches is the same as that of
nominal ones, the luminosity of these collisions, relative to normal
per-bunch luminosity gives directly the fraction of ghost charge in the bucket
with respect to a nominal bunch. 
The collisions probe the ghost charge only in the nominal RF bucket,
which is the only one colliding with the unpaired bunch. The charge in the
other nine RF buckets of the BCID is not seen. Data from the Longitudinal Density Monitors
of the LHC indicate that the ghost charge is quite uniformly distributed in 
all RF buckets of a non-colliding BCID\,\cite{BCNWG-note, adam}. 

\begin{figure}
  \centering
  \setlength{\unitlength}{1cm}
  \begin{picture}(10,18.5)
  \put(-4.5,0)
  {\includegraphics*[width=1.25\textwidth]{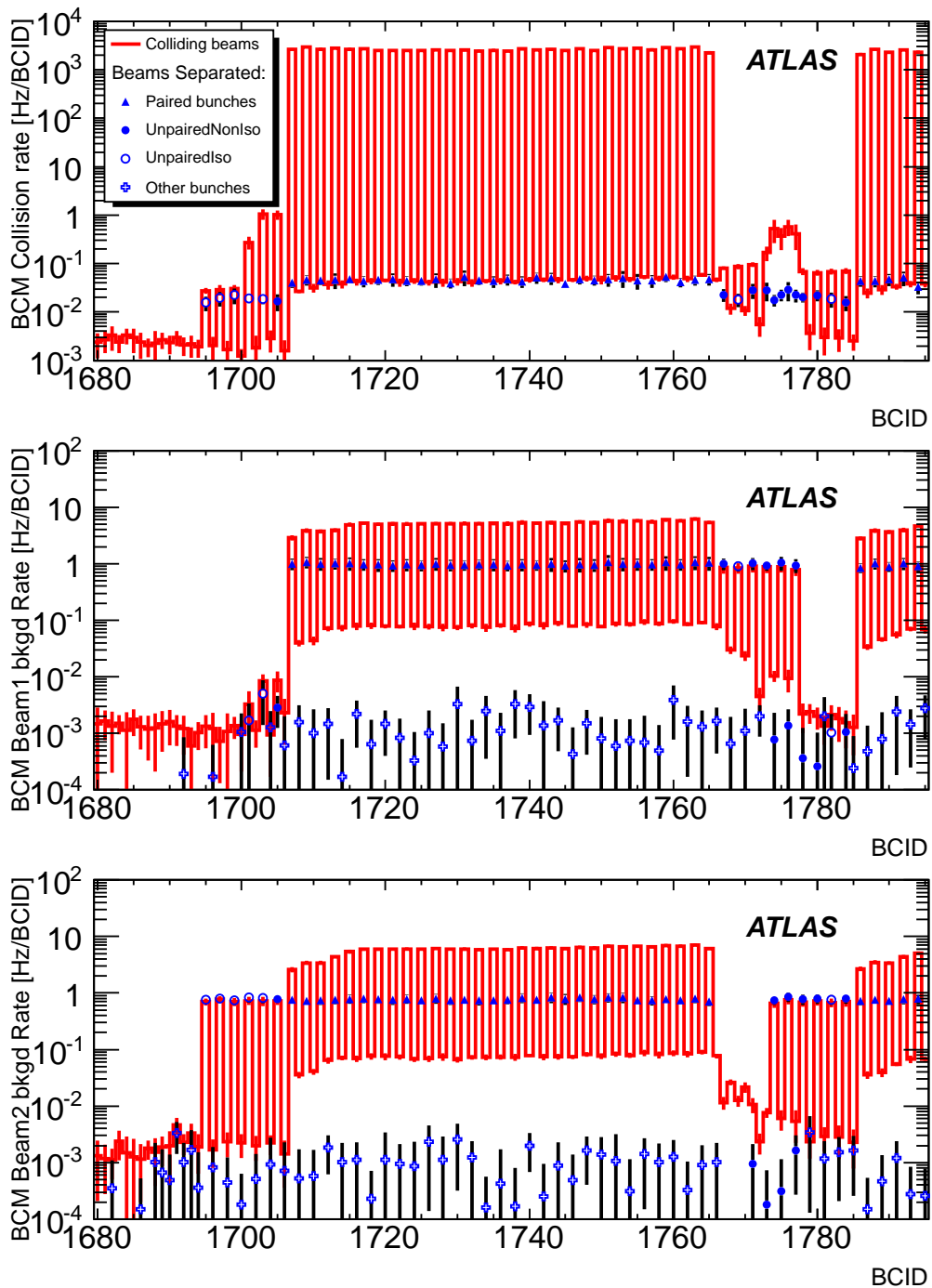}}
  \end{picture}
  \caption{BCM collision rate (top) and background rates for beam-1 (middle) and beam-2 (bottom)
per BCID before and during collisions. The data are averaged for several LHC fills over 
roughly 15\,minute periods: at full energy but before bringing beams into collision (symbols) and after 
declaring stable beams (histogram).
Thus the error bars reflect both 
the fill-to-fill variation and differences of the intensity decay during the averaging time. The data are not normalised by intensity, but
only fills with comparable luminosities at the start of the fill are used in the average.}
  \label{haloplot}
\end{figure}

Figure\,\ref{haloplot} shows a summary of BCM collision-like and background-like trigger rates for
a particularly interesting BCID range of a bunch pattern with 1317 colliding bunches. 
For this plot, several ATLAS runs with the same bunch-pattern
and comparable initial beam intensities have been averaged. 
The first train of a batch is shown with part of the second train.
The symbols show the trigger rates
with both beams at 3.5\,TeV but before they are brought into collision, while the 
histograms show the rates for the first $\sim$15\,minutes of stable beam collisions. This restriction 
to the start of collisions is necessary since the rates are not normalised by intensity, and 
a longer period would
have biased the histograms due to intensity decay. The groups of six unpaired bunches each in
front of the beam-2 trains (around BCID 1700 and 1780, respectively) and after the beam-1 train (around BCID 1770) 
can be clearly seen. These
show the same background trigger rate before and during collisions. As soon as the beams collide, the
collision rate in paired BCIDs rises, but the background rate also increases by about
an order of magnitude. As explained before, this increase is due to accidental background-like 
coincidences from afterglow. The gradual build-up of this excess is typical of
afterglow build-up within the train\,\cite{ATLASLUMI}.

The uppermost plot in Fig.\,\ref{haloplot}, showing the collision rate, reveals two 
interesting features:
\begin{itemize} 
\item Collision activity can be clearly seen in front of the train, in BCIDs 1701, 1703 and 1705.
This correlates with slightly increased background seen in the middle plot for
the same BCIDs. This slight excess seen both before and during collisions is indicative 
of ghost charge and since there are nominal unpaired bunches in beam-2 in the matching BCIDs, this 
results in genuine collisions. It is worth noting that a similar excess does not appear in 
front of the second train of the batch, seen on the very right in the plots. This is consistent
with no beam-1 ghost charge being visible in the middle plot around BCID 1780.
\item Another interesting feature is seen around BCID 1775, where a small peak is seen
in the collision rate. This peak correlates with a BCID range where beam-1 bunches are in
odd BCIDs and beam-2 in even BCIDs. Thus the bunches are interleaved with only 25\,ns spacing.
Therefore this peak is almost certainly due to ghost charge in the neighbouring BCID,
colliding with the nominal bunch in the other beam.
\end{itemize}

The two features described above are not restricted to single LHC fills, but
appear rather consistently in all fills with the same bunch pattern. Thus it seems reasonable
to assume that this ghost charge distribution is systematically produced in the injectors
or RF capture in the LHC.

Figure\,\ref{haloplot} suggests that the definition of an isolated bunch, used by ATLAS in 2011, is
not sufficient to suppress all collision activity. Instead of requiring nothing in the other beam
within $\pm$\,3\,BCIDs, a better definition would be to require an isolation by $\pm$\,7\,BCIDs.
In the rest of this paper, bunches with such stronger isolation are called {\em super-isolated}
(SuperIso).\footnote{For the start of 2012 data-taking the UnpairedIso BG was redefined to match this definition
of SuperIso.}


\subsection{Jet trigger rates in unpaired bunches}
\label{sec:jetrates}

The L1\_J10\_UnpairedIso trigger listed in Table\,\ref{L1triggers} is in principle a suitable trigger to monitor fake-jet
rates due to BIB muons. Unfortunately the L1\_J10 trigger rate has a large noise component 
due to a limited number of calorimeter channels which may be affected by a large source of instrumental
noise for a short period of time, on the order of seconds or minutes. While these noisy channels 
are relatively easy to deal with offline by considering the pulse shape of the signal, this
is not possible at trigger level.
In this study, done on the trigger rates alone, the fluctuations caused by these noise bursts 
are reduced by rejecting LBs where the intensity-normalised rate is more than 50\% higher than the 
5-minute average. 

Another feature of 
the J10 trigger is that the rates show a 
dependence on the total luminosity even in the empty bunches, i.e. there is a luminosity-dependent constant  pedestal in all BCIDs. 
While this level is insignificant with respect to the rate in colliding
BCIDs, it is a non-negligible fraction of the rates in the unpaired bunches. To remove this effect 
the rate in the empty BCIDs is averaged in each LB separately and this pedestal is
subtracted from the rates in the unpaired bunches.

\begin{figure}
  \centering
  {\includegraphics[width=0.7\textwidth]{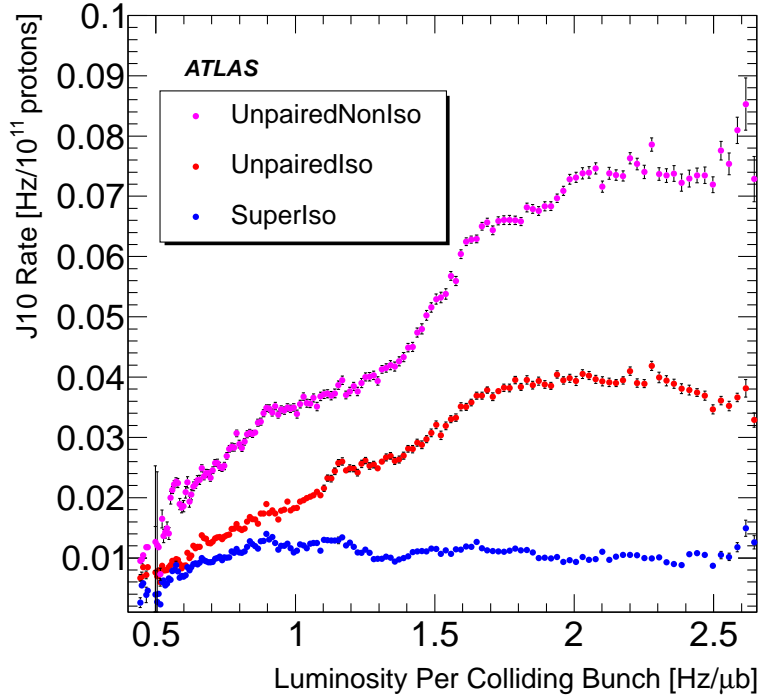}}            
  \caption{L1\_J10 trigger rate in different classes of unpaired bunch isolation as a function of
the luminosity of colliding bunches after subtraction of the luminosity-dependent pedestal, 
determined from the empty bunches. }
  \label{j10vslumi}
\end{figure}

Figure\,\ref{j10vslumi} shows these pedestal-subtracted L1\_J10 trigger rates in unpaired bunches, plotted 
against the luminosity of colliding bunches. Provided the intensity of ghost bunches is proportional to the
nominal ones, their emittance is the same as that of normal bunches and if all the rate is due to proton-proton 
collisions, a good correlation is expected.
Indeed, the UnpairedNonIso rates correlate rather well with the luminosity, indicating that a large 
fraction of the rate is due to bunch-ghost encounters. Even the
UnpairedIso rates show some correlation, especially at low luminosity. 
This suggests that even these isolated bunches are paired with some charge in the other beam which is consistent 
with Fig.\,\ref{haloplot}. 
In  superIso bunches, i.e. applying an even tighter isolation, the correlation mostly disappears and the rate 
is largely independent of luminosity. 

\begin{figure}
  \centering
  {\includegraphics[width=1.0\textwidth]{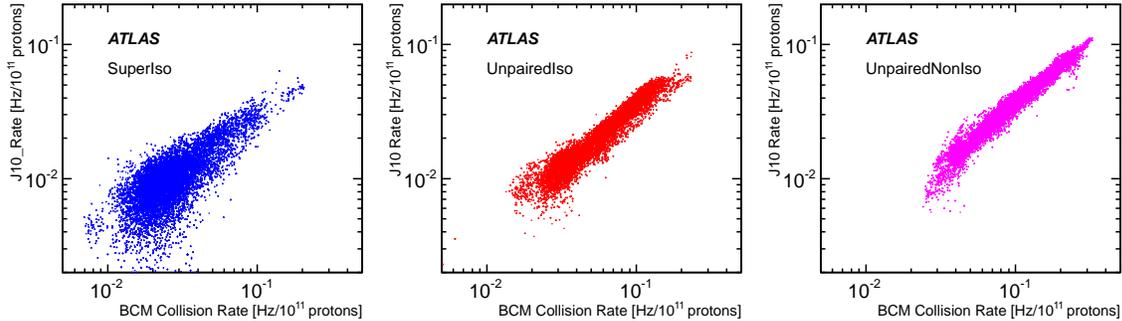}}            
  \caption{Correlation of L1\_J10 and BCM collision trigger rates in different classes of unpaired bunch isolation.}
  \label{j10vsbcmw}
\end{figure}

If the rates shown in Fig.\,\ref{j10vslumi} are dominated by collisions, then this should be
reflected as a good correlation between the J10 and BCM collision-like trigger rates. Figure\,\ref{j10vsbcmw}
shows that this is, indeed, the case. While the correlation is rather weak for the superIso bunches, it
becomes increasingly stronger with reduced isolation criteria.

\section{Studies of BIB with the ATLAS Pixel detector}

\subsection{Introduction}
Like the BCM, the ATLAS Pixel detector  is very close to the beam-line, so it is sensitive 
to similar background events. 
However, while the BCM consists of only eight active elements, the Pixel detector has
over 80 million read-out channels, each corresponding to at least one pixel. This fine
granularity enables a much more detailed study of the characteristics of the BIB events.

As shown in Sect.\,\ref{sec:monitoring}, the BCM background rate is dominated by beam-gas events in rather
close proximity to ATLAS. 
Energetic secondary particles from beam-gas events are likely to impinge on the
TAS and initiate showers. 
The particles emerging from the TAS towards the Pixel detector are essentially
parallel to the beam-line and therefore typically hit only individual
pixels in each endcap layer, but potentially leave long continuous tracks in
Pixel barrel sensors. If a beam-gas event takes place very close to the TAS,
it is geometrically possible for secondary particles to pass through the aperture and still hit
the inner Pixel layer.

In studies using 2010 data\,\cite{conf-note-2010} the characteristic features of high cluster 
multiplicity and the presence of long clusters in the $z$-direction in the barrel, were found to be a good indicator 
of background contamination in collision events.

The study in Ref.\,\cite{conf-note-2010} was done by considering paired and unpaired 
BCIDs separately. Comparing the hit multiplicity distributions for these two samples
allows the differences between BIB and collision events to be characterised. 
An independent method to identify BIB events is to use the early arrival time on the
upstream side of the detector.
While the time difference expected from the half length of the Pixel detector is too short to apply 
this method with the pixel timing alone, correlations with events selected by other, larger, ATLAS 
sub-detectors with nanosecond-level time resolution are observed. For example, BIB events 
identified by a significant time difference between the BCM stations on either side of ATLAS, are also 
found to exhibit large cluster multiplicity in the Pixel detector\,\cite{conf-note-2010}.

The characterisation of BIB-like events by comparing distributions for paired and unpaired bunches, coupled 
with the event timing in other sub-detectors, allows parameters to be determined for the efficient 
identification of BIB in the Pixel detector. 
The most striking feature in the Pixel barrel of BIB-like events, compared to collision products, is the 
shallow angle of incidence, which causes Pixel clusters to be elongated along $z$,
 where a cluster is defined as a group of neighbouring pixels in which charge is deposited. 
Since the pixels have a length of 400\,$\mu$m, or larger, in the $z$-direction, the charge per pixel tends to be larger
than for a particle with normal incidence on the 250\,$\mu$m thick sensor. More significantly however, 
a horizontal track is likely to hit many pixels causing the total cluster charge to be much larger than for 
typical ``collision'' clusters.

In the following, the different properties of pixel clusters generated by collisions and BIB events are examined to help 
develop a background identification algorithm, which relies only on the cluster properties. The BIB 
tagging efficiency is quantified and the tools are applied to study 2011 data.

\subsection{Pixel cluster properties}
\begin{figure}[htb!]
\centering
    \includegraphics*[width=0.85\textwidth]{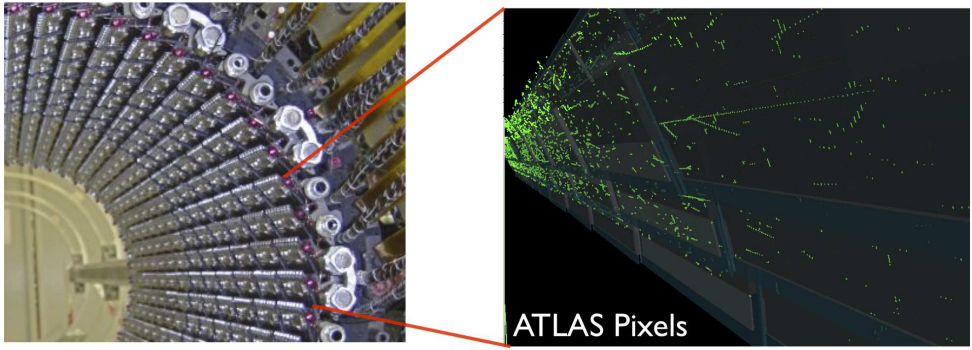}
\caption{A high-multiplicity BIB event in the Pixel detector, showing the typically long pixel clusters deposited in the barrel region.
On the left is the layout of the Pixel detector barrel viewed along the beam-line
and the right shows the event display in a zoomed region.}
\label{fig_pixel_eventdisplay}
\end{figure}

An example of a high-multiplicity BIB event is shown in 
Fig.~\ref{fig_pixel_eventdisplay}, in which the elongated clusters in the barrel region can be observed. 

\begin{figure}[htb!]
\centering
\mbox{
  \subfigure[]{
    \includegraphics*[width=0.49\textwidth]{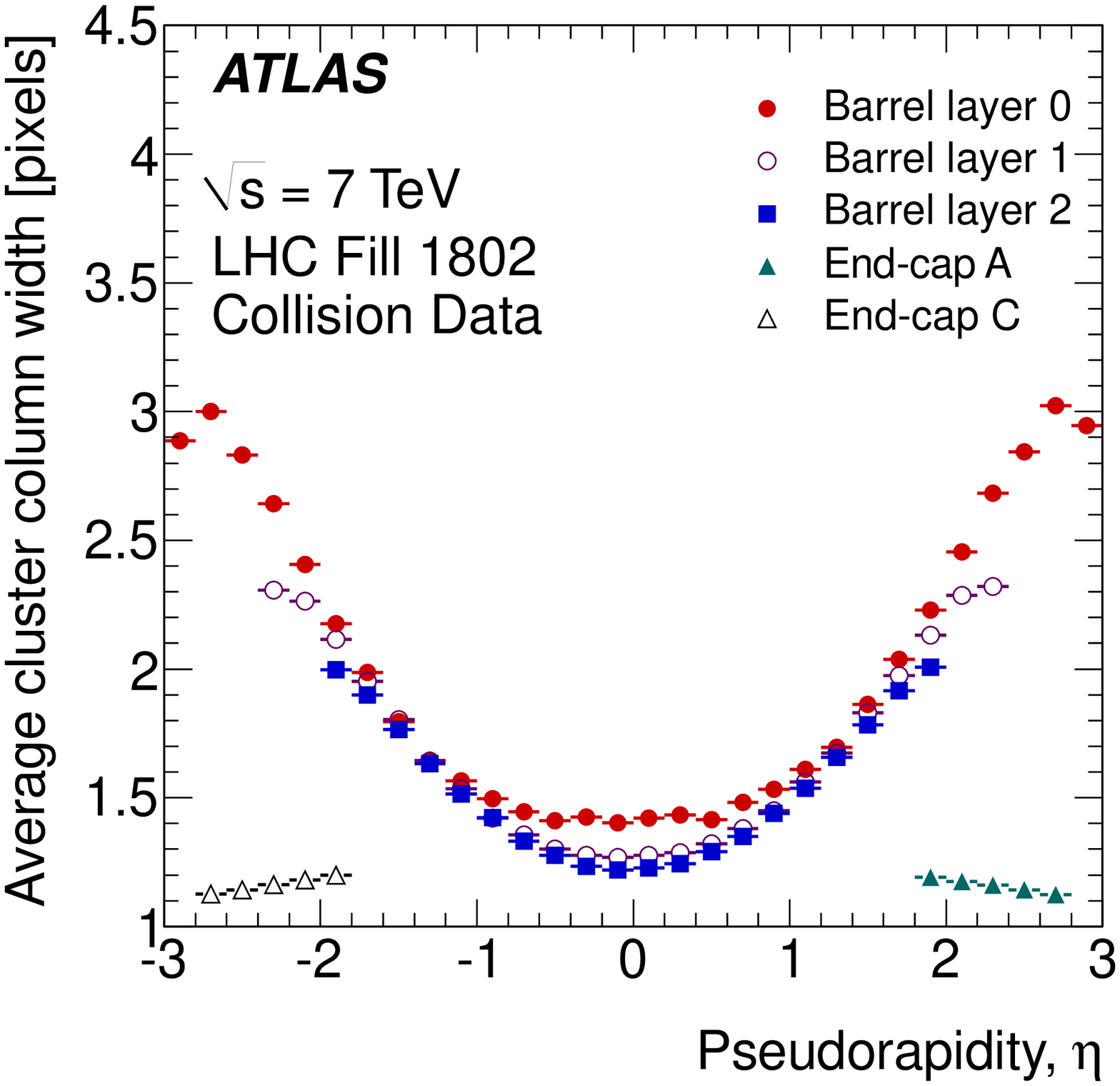}
    \label{fig_pix_cluslength_data_minbias}
  }
  \subfigure[]{
    \includegraphics*[width=0.49\textwidth]{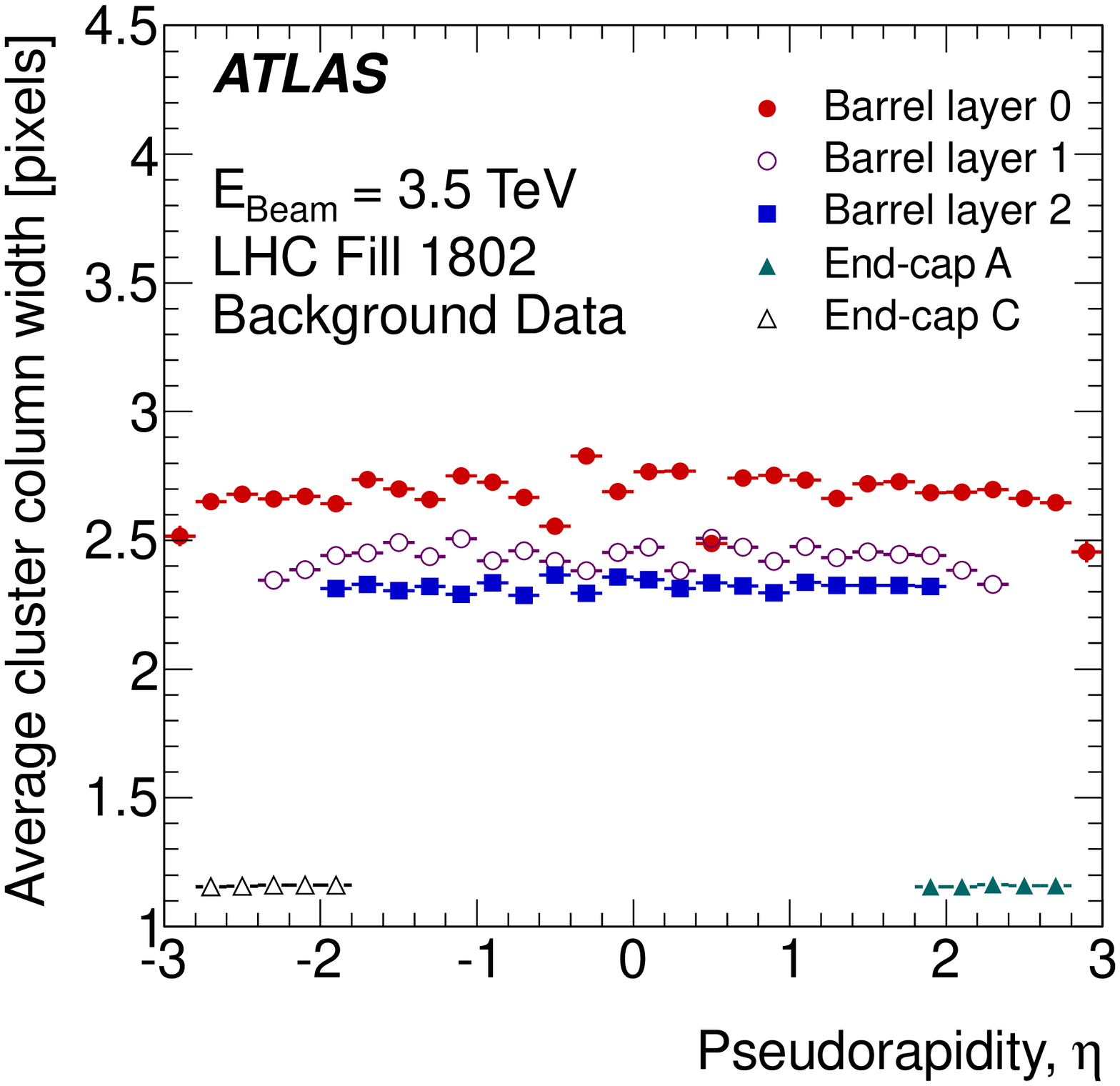}
    \label{fig_pix_cluslength_data_background}
  }
}
\mbox{
  \subfigure[]{
    \includegraphics*[width=0.49\textwidth]{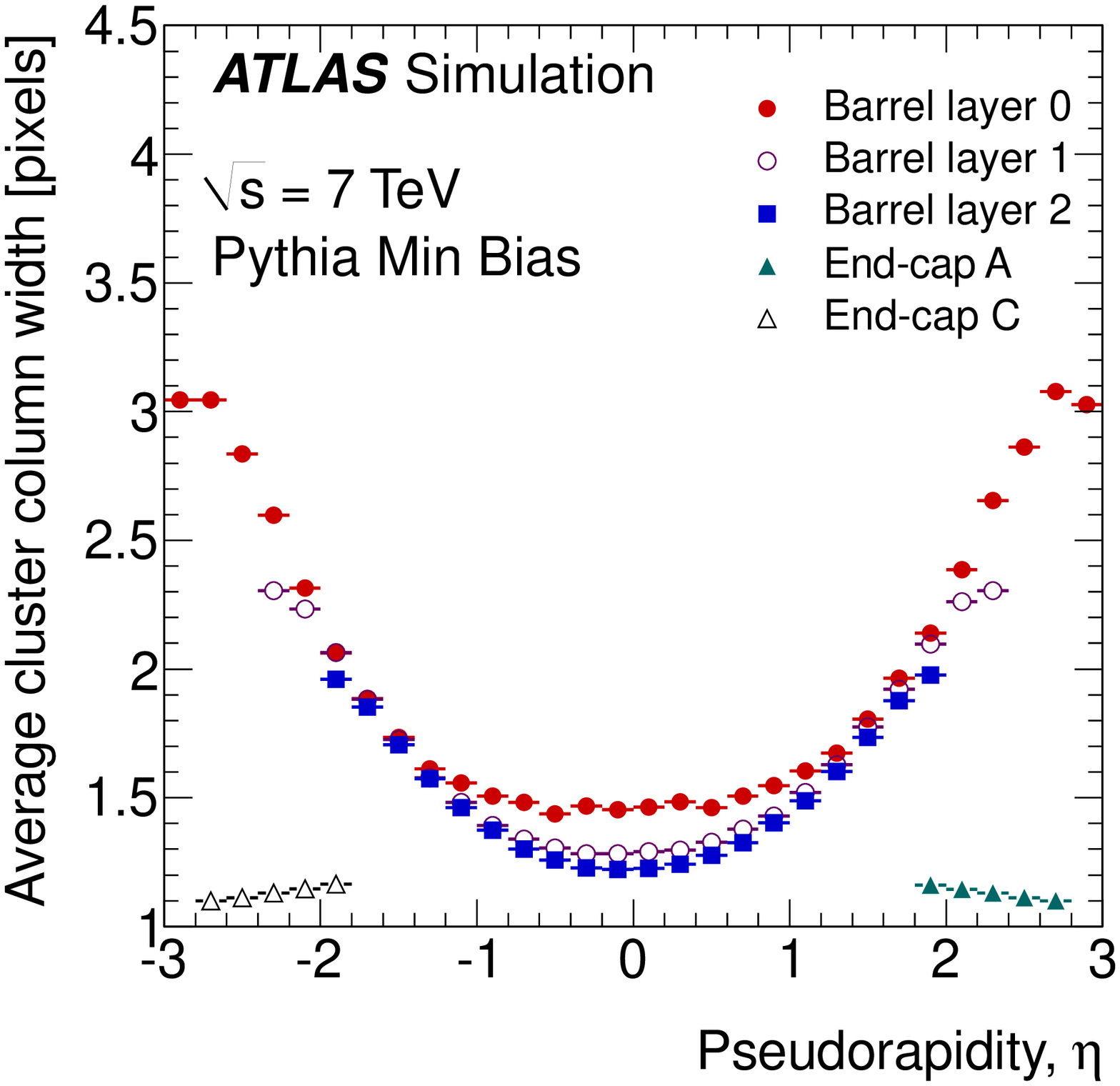}
    \label{fig_pix_cluslength_mc_minbias}
  }
  \subfigure[]{
    \includegraphics*[width=0.49\textwidth]{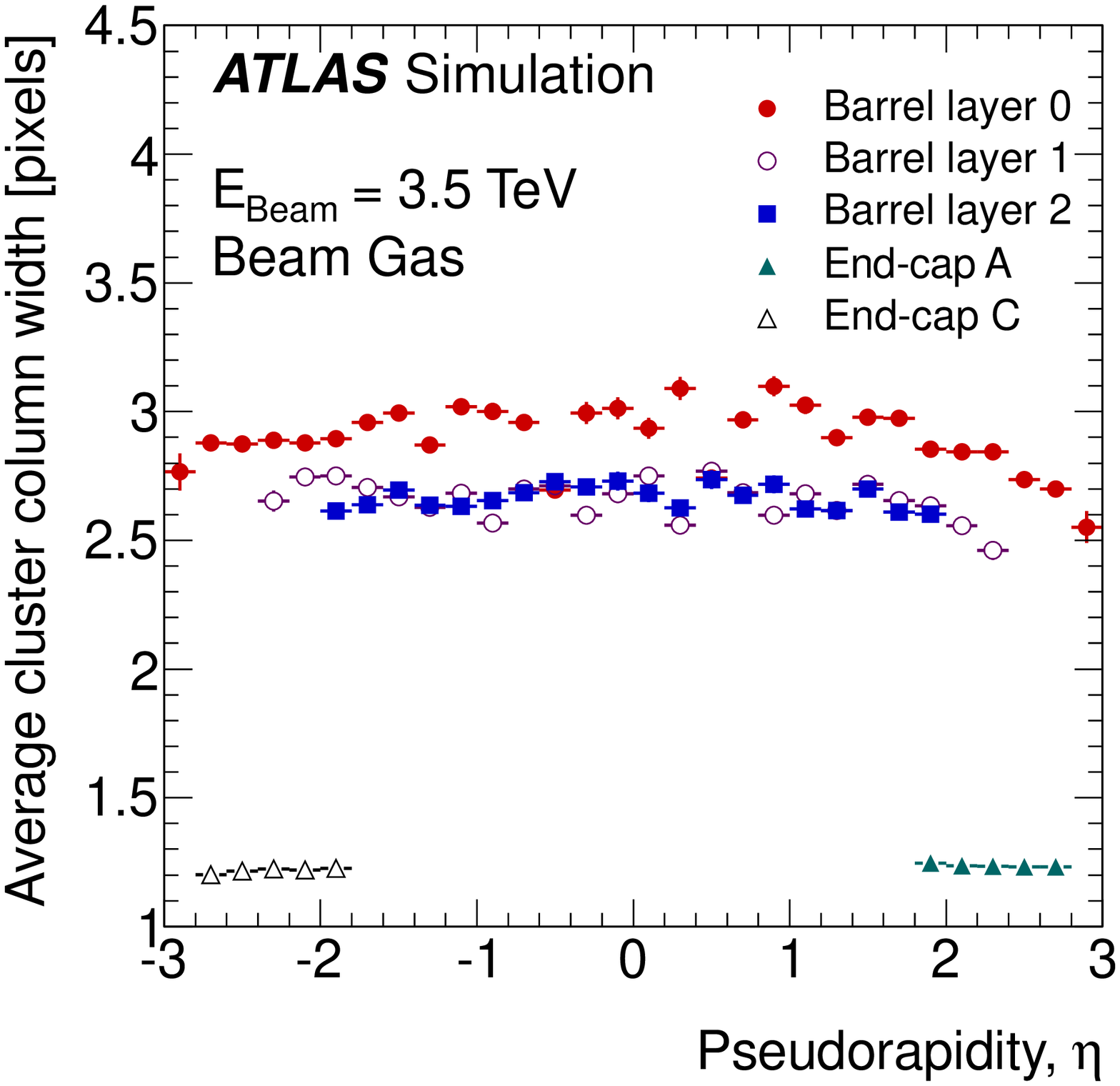}
    \label{fig_pix_cluslength_mc_beamgas}
  }
}
\caption{Pixel cluster width (in $\eta$ direction) versus pseudorapidity
for (a, c) collision and (b, d) background data and Monte Carlo simulation.}
\label{fig_pix_cluslength_eta}
\end{figure}

The differences in average cluster properties for collision-like and BIB-like events are shown in 
Fig.~\ref{fig_pix_cluslength_eta}. For each barrel layer and endcap, the pixel cluster column width in 
the $\eta$ direction is averaged over all clusters and plotted against the pseudorapidity of the cluster position. 
Ganged pixels are excluded and no requirement for the clusters to be associated with a track is applied.

For collisions, shown on the left of Fig.\,\ref{fig_pix_cluslength_eta}, the cluster width is
a function of $\eta$ simply for geometrical reasons
and the agreement between data and Monte Carlo simulation~\cite{Sjostrand:2006za}
 is good.

The distribution for BIB-like events is shown on the right side of Fig.\,\ref{fig_pix_cluslength_eta}.
The upper plot shows data in super-isolated unpaired bunches for events that are selected using 
the background identification tool, which is described in Sect.~\ref{sec:pixelclus}. 
The distribution is independent of $\eta$ as expected for BIB tracks. 
A detailed simulation\,\cite{roderick-in-prep}, described in Sect.\,\ref{sect-simulation},
was interfaced to the ATLAS detector simulation to check the cluster properties in beam-gas events.
Based on the assumption that BIB in the detector is dominated by showering in the TAS, a 20\,GeV energy
transport cut was used in the beam-gas simulations. This high cut allowed maximisation of the statistics by
discarding particles that would not have enough energy to penetrate the 1.8\,m of copper of the
TAS. Here it is assumed that particles passing through the TAS aperture, which might have low energy, 
do not change the average cluster properties significantly -- an assumption that remains to be verified by
further, more detailed, simulations.
The distributions are found to match very well the distributions observed in data.
It can also be seen from Fig.\,\ref{fig_pix_cluslength_eta} that the clusters in the endcaps are small and of
comparable size for both collision events and BIB. This is expected from the geometry, because at the $\eta$-values
covered by the endcap disks, the collision products have a very small angle with respect to the beam-line.
In the Barrel, layer 0 clusters are systematically larger than layer 1 and layer 2 for small $\eta$, due to the beam spot 
spread along the beam-line.

\begin{figure}[htb!]
\centering
\mbox{
  \subfigure[]{
    \includegraphics*[width=0.49\textwidth]{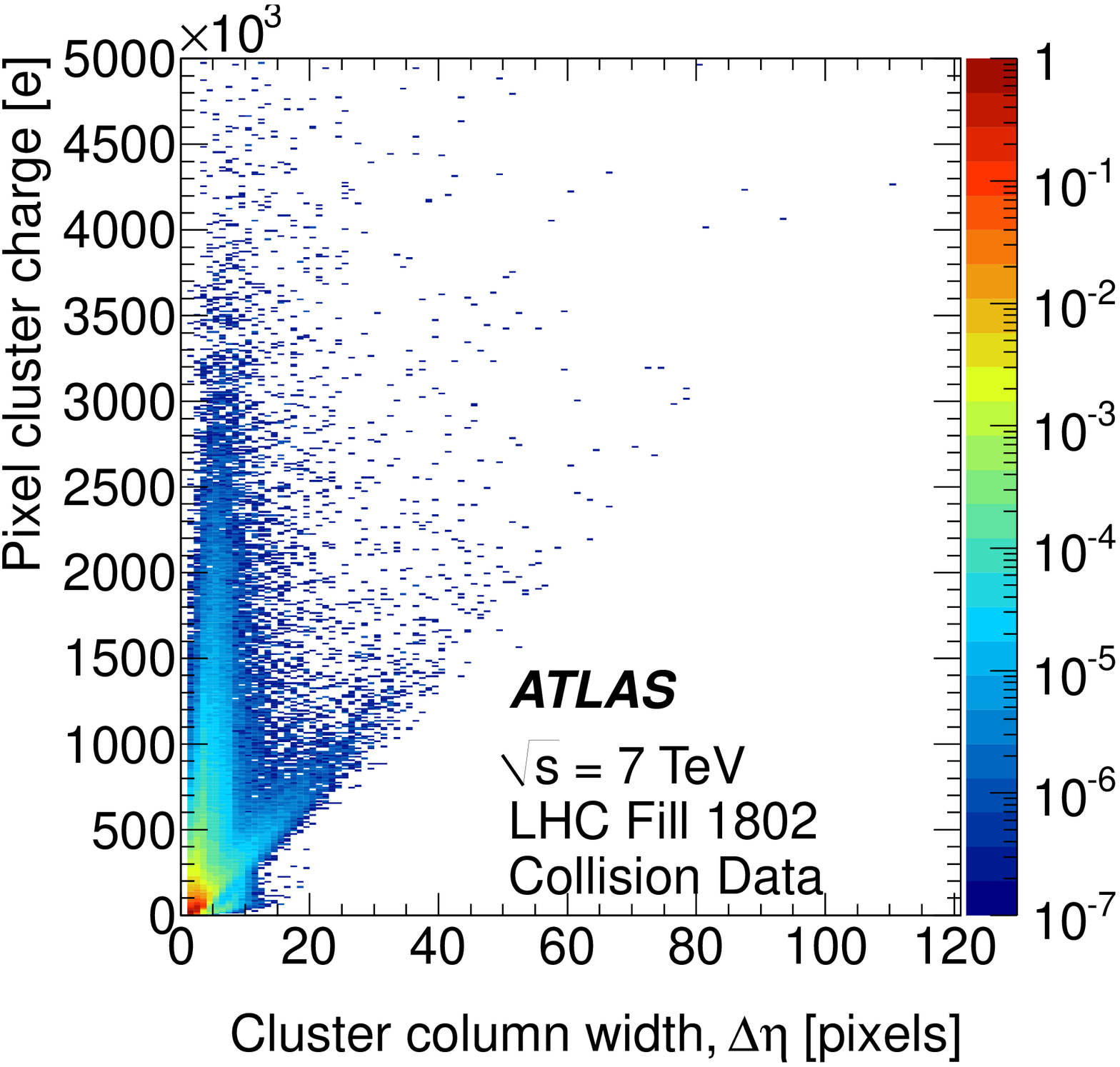}
    \label{fig_pix_cluscharge_data_minbias}
  }
  \subfigure[]{
    \includegraphics*[width=0.49\textwidth]{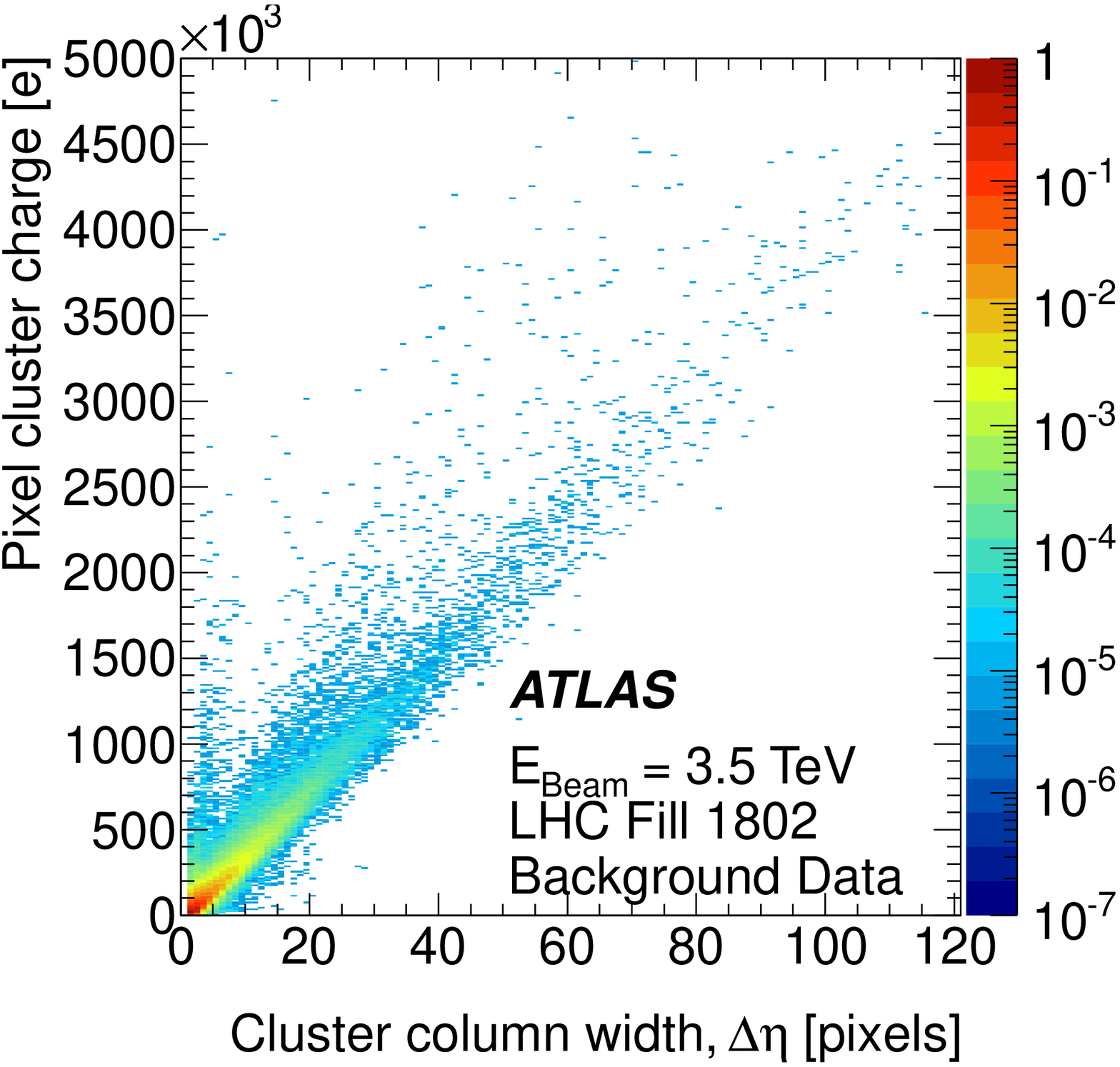}
    \label{fig_pix_cluscharge_data_background}
  }
}
\mbox{
  \subfigure[]{
    \includegraphics*[width=0.49\textwidth]{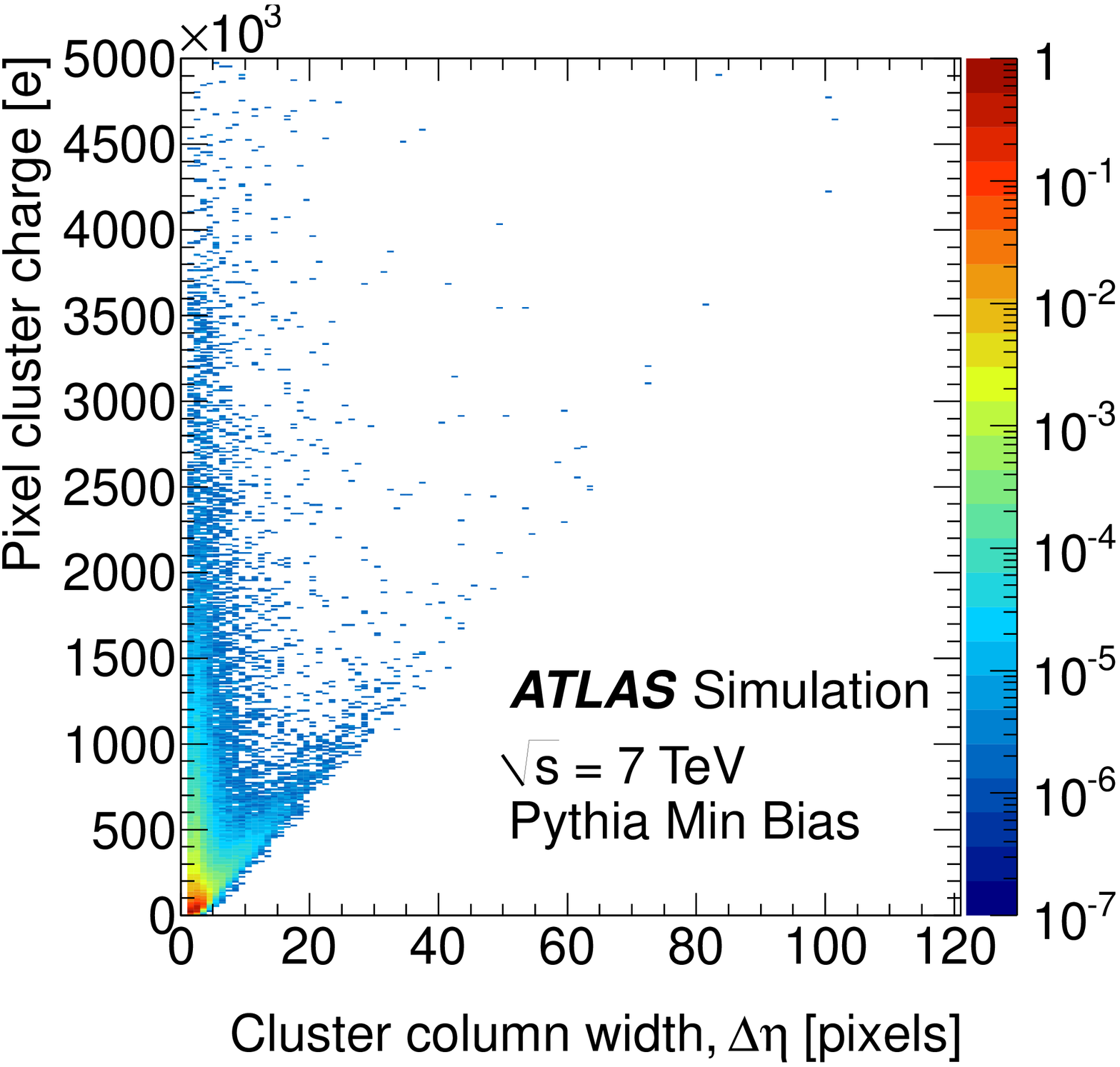}
    \label{fig_pix_cluscharge_mc_minbias}
  }
  \subfigure[]{
    \includegraphics*[width=0.49\textwidth]{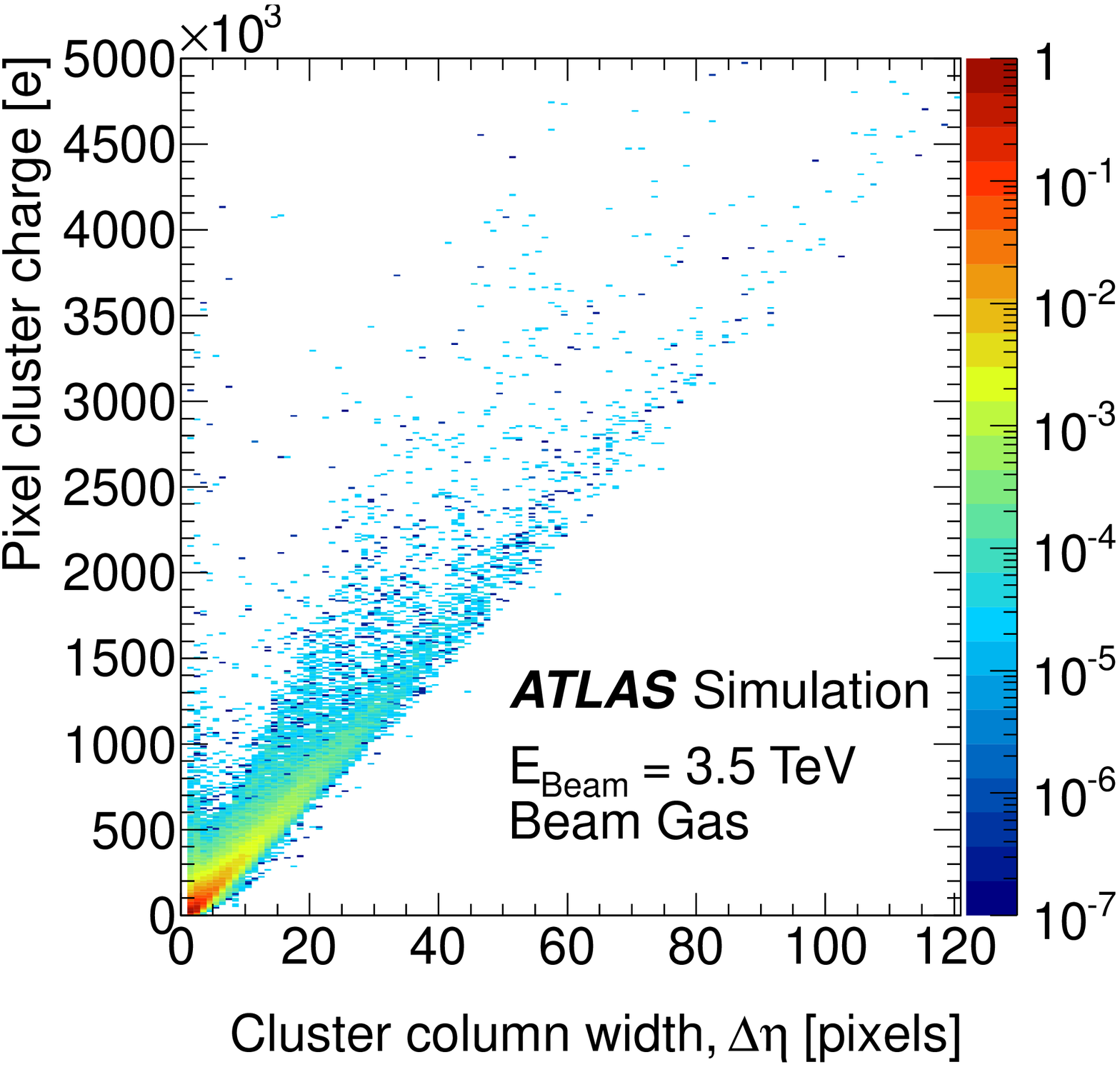}
    \label{fig_pix_cluscharge_mc_beamgas}
  }
}
\caption{Pixel cluster deposited charge versus cluster width (in $\eta$ direction), for Pixel barrel clusters.}
\label{fig_pix_cluscharge_cluslength}
\end{figure}

In the Pixel detector, the charge deposited in each pixel is measured from the time that the signal is above the 
discriminator threshold. After appropriate calibration, the charge is determined and summed over all pixels in 
the cluster. Figure\,\ref{fig_pix_cluscharge_cluslength} shows the charge versus the cluster column width for the outer barrel 
layer for the same data and Monte Carlo samples that are used for Fig.\,\ref{fig_pix_cluslength_eta}.
As expected, the majority of clusters are small both in terms of spatial extent and amount of charge.

However, differences between BIB and collision samples become apparent when clusters of larger size
or charge are considered. 
In the BIB events, a strong correlation is observed between cluster width and deposited 
charge, because the elongated clusters tend to align along the beam direction. Large clusters in 
collision events, however, may arise either from secondary particles such as $\delta$-rays or low-momentum loopers, or from
particles stopping in the sensor (Bragg-peak). Thus the clusters with large charges are not necessarily aligned 
with the beam direction. These features, seen in data, are qualitatively well reproduced by the Monte Carlo simulations.

\subsection{Pixel cluster compatibility method}
\label{sec:pixelclus}
The cluster characteristics of BIB particles have been exploited to develop a BIB identification algorithm, based 
on a check of the compatibility of the pixel cluster shape with BIB.

\begin{figure}[htb!]
\centering
\mbox{
  \subfigure[]{
    \includegraphics*[width=0.49\textwidth]{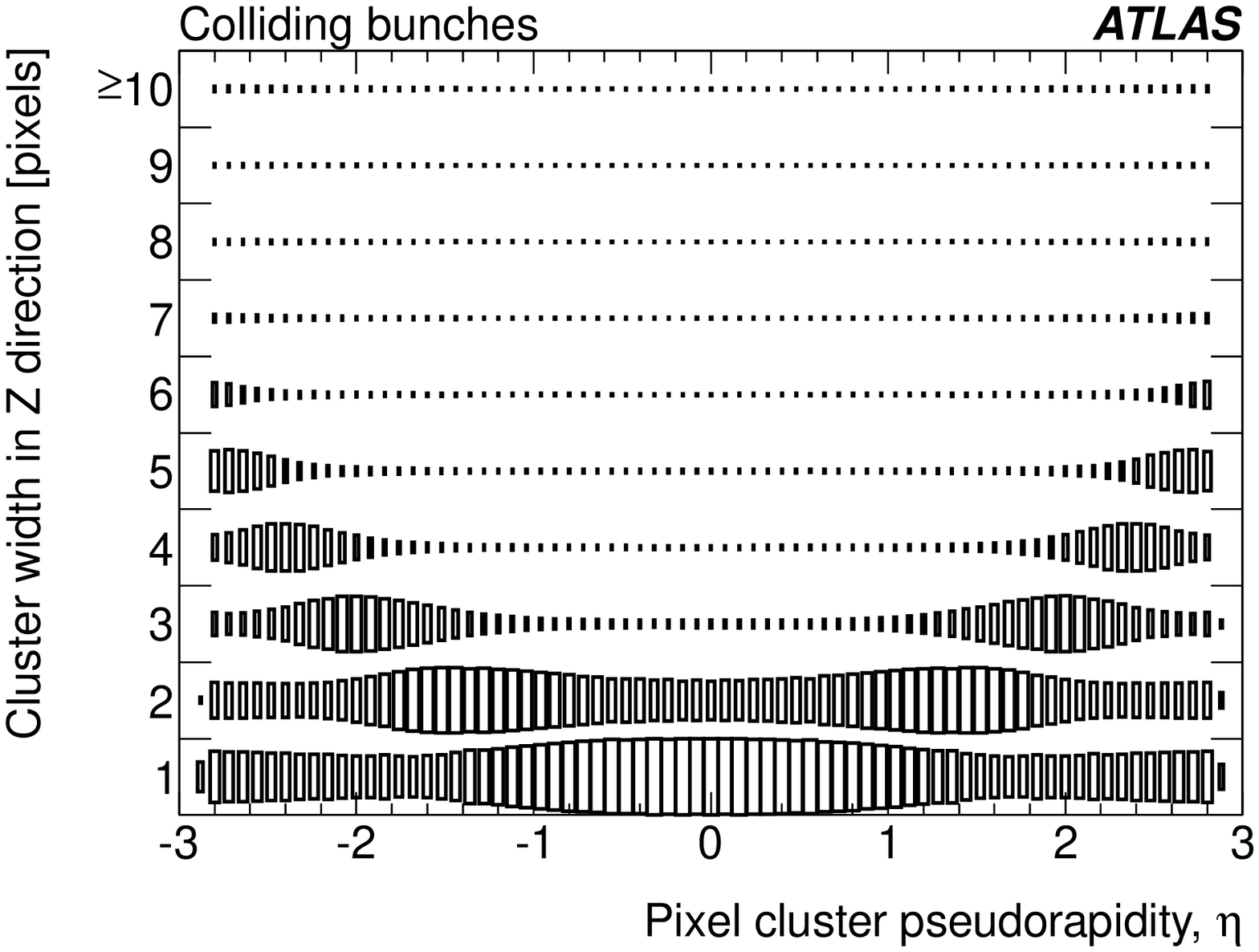}
    \label{fig_pix_cdgclusetacol2d}
  }
  \subfigure[]{
    \includegraphics*[width=0.49\textwidth]{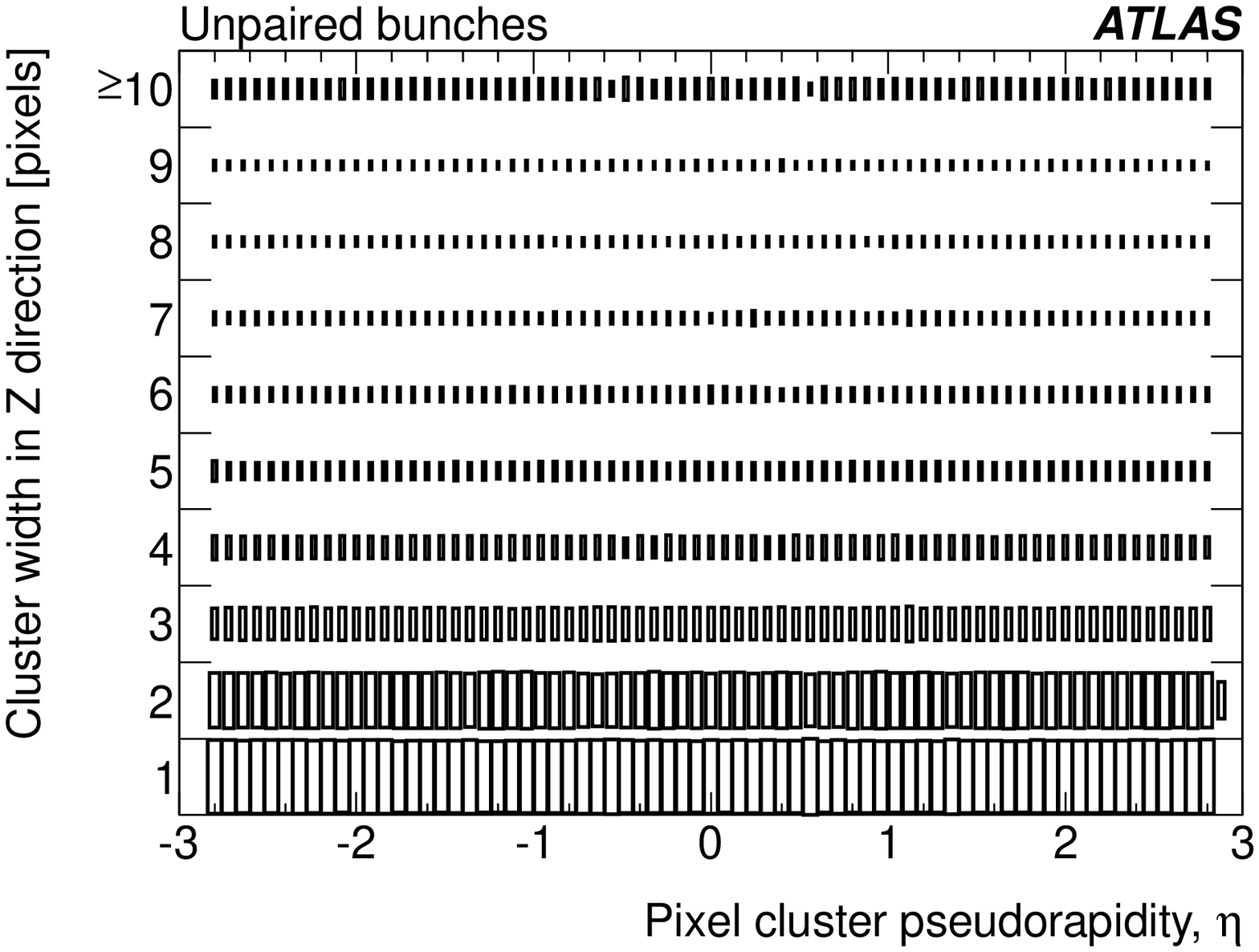}
    \label{fig_pix_bbgclusetacol2d}
  }
}
\mbox{
  \subfigure[]{
    \includegraphics*[width=0.59\textwidth]{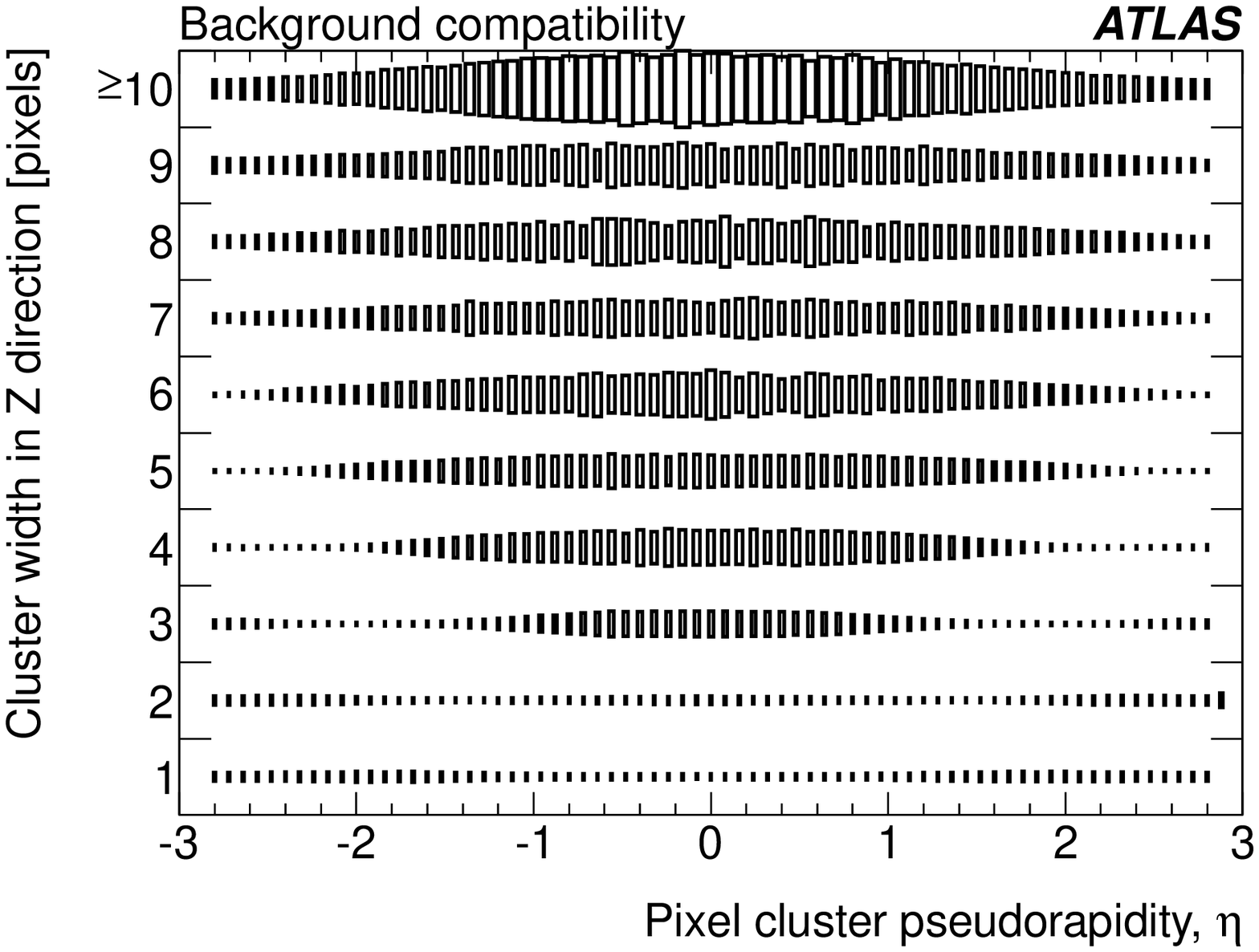}
    \label{fig_pix_bbgclusetacol2dcompat}
  }
}
\caption{Conditional probability distributions, $P^{c}$ and $P^{b}$, for clusters in the innermost pixel barrel for (a) colliding and (b) unpaired bunches respectively. The lower plot (c) shows the calculated BIB compatibility.}
\label{fig_pix_clus_compatibility}
\end{figure}

Only the cluster widths, $\Delta\eta$ and $\Delta\phi$, are necessary for an efficient 
selection of BIB. The algorithm processes all clusters in the event, independent of whether the 
cluster is associated with a track after reconstruction. Therefore, in addition to offline analysis, the 
algorithm is also suited for rapid online monitoring of the background. 

For each pixel cluster in the event, the algorithm computes the conditional probability to obtain the measured cluster width, $w = \Delta\eta$ (or $\Delta\phi$), (in units of pixels), 
given the cluster position in pseudorapidity, $\eta$, and the barrel layer. Only pixel clusters in the barrel layers are considered, as these provide the 
best discriminating power. The conditional probability associated with each possible source of the cluster, $P^{c}$ for collisions or $P^{b}$ for BIB, is retrieved from look-up tables, $T^{c}_{w}$ for collisions or $T^{b}_{w}$ for BIB:
\begin{eqnarray}
\label{eq:pixwidth}
P^{c,b}(w|\eta,{\rm layer}) = \frac{T^{c,b}_{w}(\eta,{\rm layer}) }{\sum_{w=1}^{N}T^{c,b}_{w}(\eta,{\rm layer})} 
\end{eqnarray}
where $T^{c,b}_{w}(\eta,{\rm layer})$ is the number of clusters with width $w$ for a given $\eta$ bin and 
barrel layer.

The values for $T^{c,b}_{w}$ were obtained using a data-driven method based on studies of colliding 
and unpaired bunches. The study was performed using LHC fill 1022 from 2010, in which the bunch configuration had 
only one pair of colliding bunches in $\textrm{BCID}=1$ and one unpaired bunch per beam, in BCID 892 and 1786. In this sparse 
pattern, the unpaired bunches satisfied the definition of being super-isolated.

The conditional probability distribution for pixel clusters in the innermost barrel layer is plotted in Fig.\,\ref{fig_pix_clus_compatibility}, 
for different cluster widths and for colliding, $P^{c}$ (left), and unpaired, $P^{b}$ (right), bunches. The probability distributions are shown for the 
cluster width in the $\eta$ direction only. The other barrel layers have similar distributions, 
with reduced pseudorapidity coverage.
As described by Eq.\,\eqref{eq:pixwidth}, the pixel cluster width distributions are normalised to the total number of pixel clusters in each pseudorapidity bin, 
so that the relative multiplicity of all cluster widths can be compared. It is seen that the fraction of clusters with a certain 
width depends strongly on $\eta$ for colliding bunches, whereas the probability to generate a certain width of 
cluster is independent of $\eta$ for clusters from BIB.

The conditional probability distributions are used to construct the compatibility of the cluster with BIB rather than with collisions. 
The BIB compatibility $C^{b}$, is defined as the ratio of conditional probabilities, and is calculated independently for 
the cluster $w=\Delta\eta$ and $w=\Delta\phi$ dimensions:
\begin{eqnarray}
C^{b}(w|\eta,{\rm layer}) = \frac{P^{b}(w|\eta,{\rm layer})}{P^{c}(w|\eta,{\rm layer})}
\end{eqnarray}
The resulting BIB compatibility is plotted in Fig.\,\ref{fig_pix_bbgclusetacol2dcompat} and has the expected distribution; 
the longest pixel clusters in the central  barrel region are the best indicators of BIB. Similar plots are obtained 
for the cluster widths in the orthogonal, $\phi$, direction, and both directions are exploited to calculate the background 
compatibility of the cluster.

After the compatibility is computed for each cluster in the event, the algorithm uses two methods to identify events containing BIB: 

\begin{itemize}
\item {\bf Simple counting method:} In the first method, each pixel cluster is taken to be compatible with BIB if the cluster compatibility in 
both dimensions exceeds the quality cuts $C^{b}(\Delta\eta|\eta,\allowbreak{\rm layer})>20$ and $C^{b}(\Delta\phi|\eta,{\rm layer})>4$. The entire event is tagged as a 
BIB candidate if it contains more than five BIB compatible clusters. 
The quality cuts are tuned in Monte Carlo simulation to efficiently select BIB events, while rejecting collisions.

\begin{figure}[htb!]
\centering
\mbox{
  \subfigure[]{
    \includegraphics*[width=0.49\textwidth]{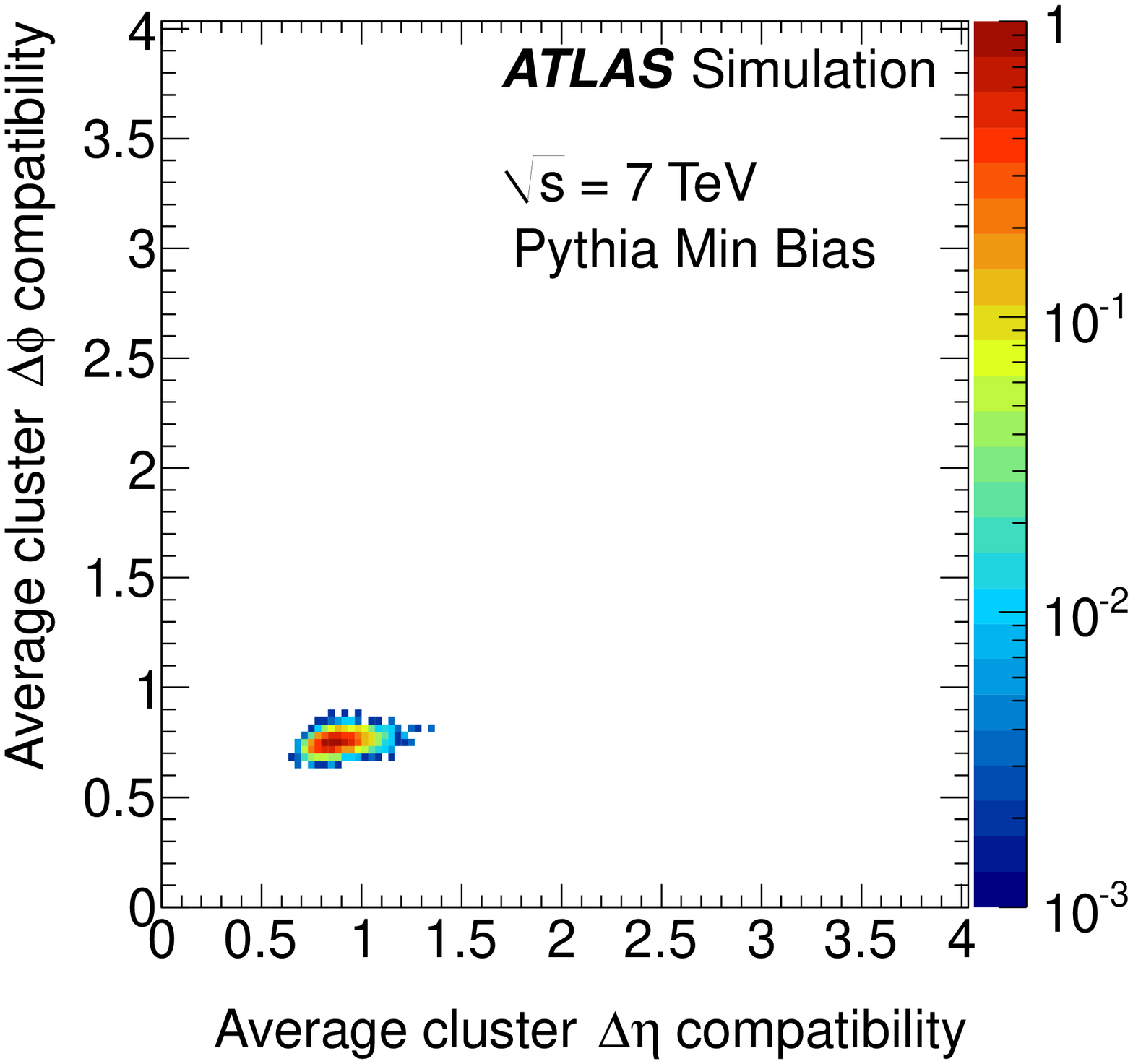}
    \label{fig_pix_cluscompat2d_mcminbias}
  }
  \subfigure[]{
    \includegraphics*[width=0.49\textwidth]{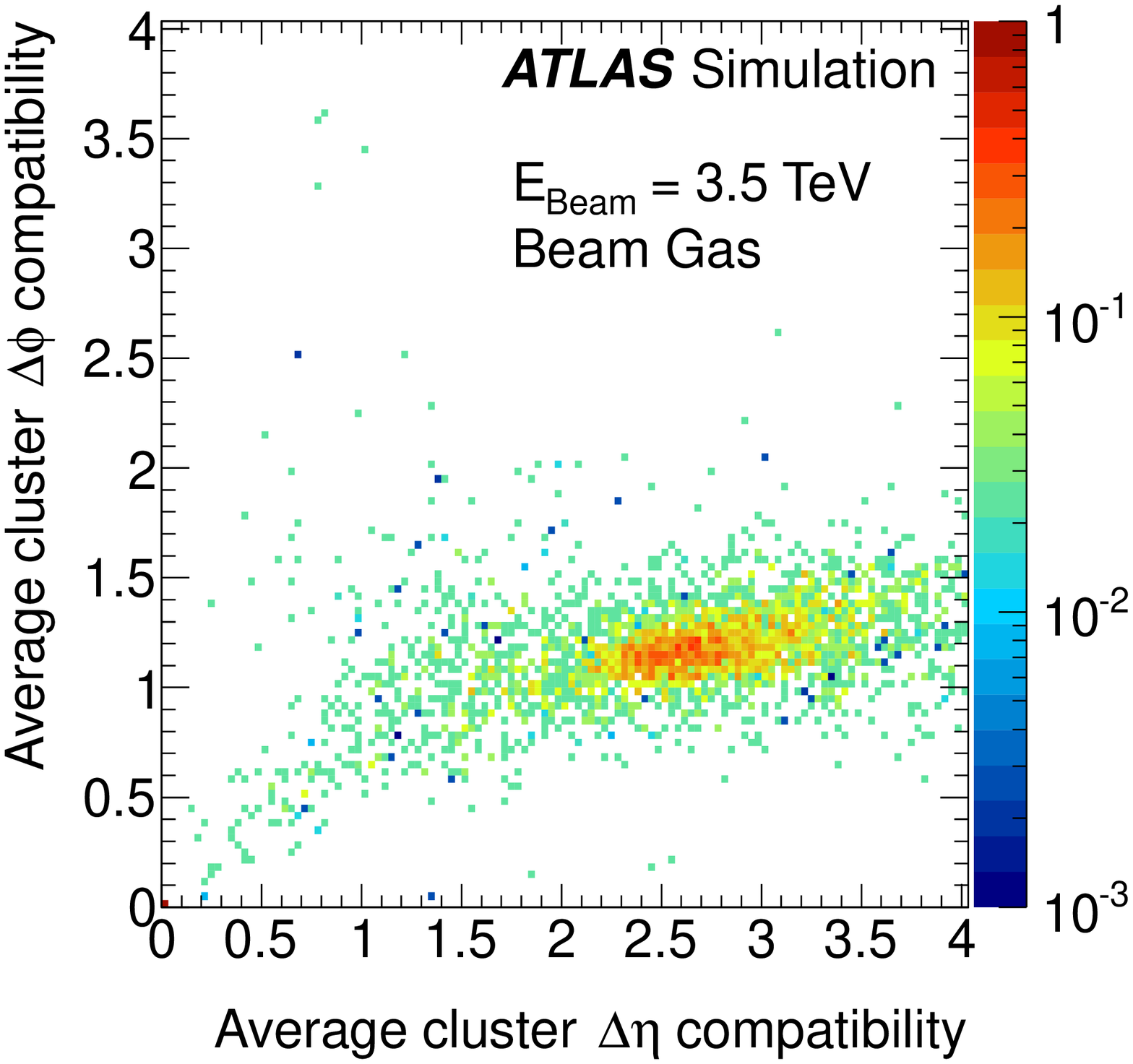}
    \label{fig_pix_cluscompat2d_mcbeamgas}
  }
}
\mbox{
  \subfigure[]{
    \includegraphics*[width=0.49\textwidth]{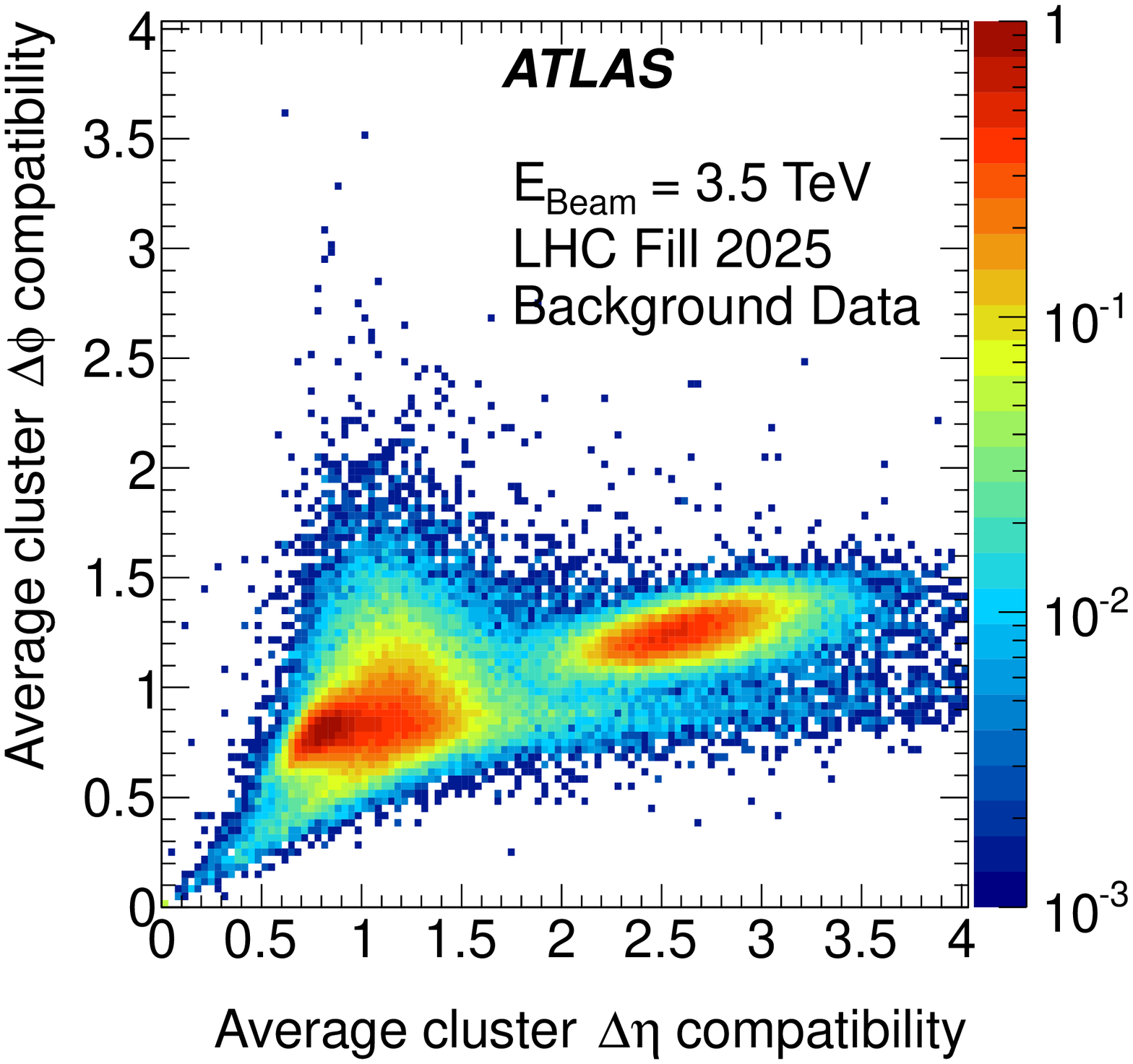}
    \label{fig_pix_cluscompat2d_data}
  \hfill
  }
}
\caption{The average pixel cluster compatibility distributions for (a) simulated collisions, (b) simulated beam-gas, and (c) background data. The intensity scale represents the number of events normalised by the maximum bin.}
\label{fig_pix_cluscompat}
\end{figure}

\item {\bf Cluster compatibility averaging:} In the second method, the cluster compatibilities in $\Delta\eta$ and $\Delta\phi$ are independently averaged 
over all clusters in the event. A two-dimensional 
compatibility distribution is obtained, shown in Figs.~\ref{fig_pix_cluscompat2d_mcminbias}, \ref{fig_pix_cluscompat2d_mcbeamgas} and \ref{fig_pix_cluscompat2d_data} 
respectively, for simulated collision events, simulated beam-gas events and a 2011 data run. It is seen from the Monte Carlo samples that the collision and BIB distributions are centred 
in different regions of the compatibility parameter space. The two regions remain distinct in the background data sample, one corresponding to the unpaired bunch colliding with 
ghost charge, as discussed in Sect.~\ref{sec:ghostcharge}, and possibly afterglow, while the other region is dominated by beam-background events. 
A two-dimensional cut is applied to select BIB candidates. 
\end{itemize}

The simple counting method essentially relies on a sufficient number ($>5$) of large BIB clusters in the central barrel regions to identify BIB events. The cluster
compatibility averaging method takes into account all clusters in the event, so it is suitable for identifying events containing fewer large BIB clusters together with many smaller 
BIB clusters, which may not be tagged by the simple counting method.
If a BIB event is overlaid with multiple collisions, the additional collision-like clusters pull the average compatibility for a BIB event toward the centre of the collision distribution. 
An increase in pile-up therefore reduces the efficiency for tagging a BIB event using only the cluster compatibility averaging method.
However, the tagging efficiency of the simple counting method is robust against pile-up, since an event containing a sufficient number of 
large BIB compatible clusters is always tagged.
At high pile-up the merging of collision-like clusters into bigger ones reduces the rejection power for collision events, because merged collision clusters 
are more likely to be mistaken as originating from BIB -- and the merging probability is a function of cluster density, which increases with pile-up.
Therefore, the combination of both methods is used in the final algorithm to ensure the best possible efficiency and rejection power 
over a wide range of conditions, including the number of BIB pixel clusters in the event.

\begin{figure}[htb!]
\centering
\mbox{
  \subfigure[]{
    \includegraphics*[width=0.49\textwidth]{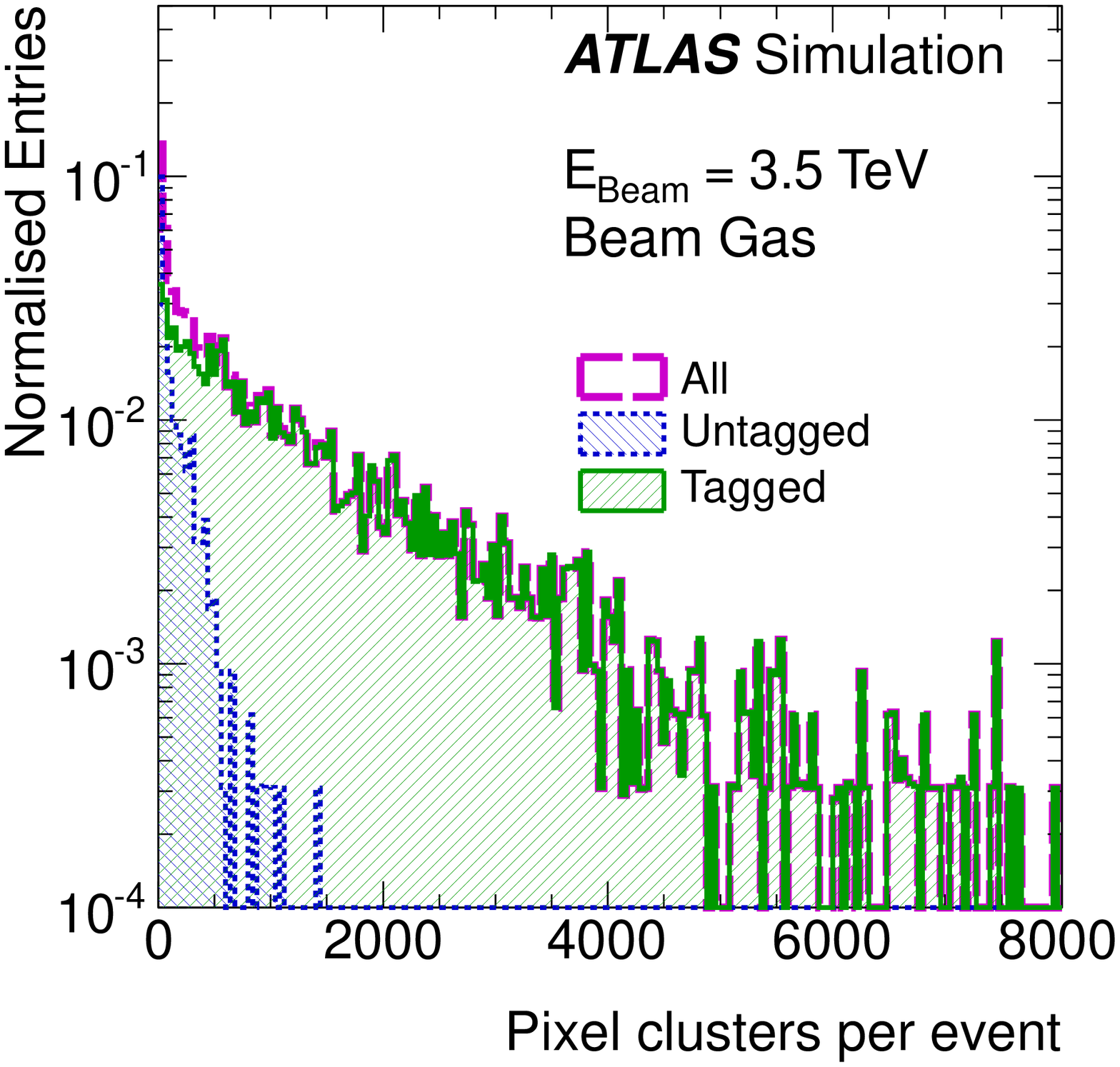}
    \label{fig_pix_pixclusnum_tag_mcbeamgas}
  }
  \subfigure[]{
    \includegraphics*[width=0.49\textwidth]{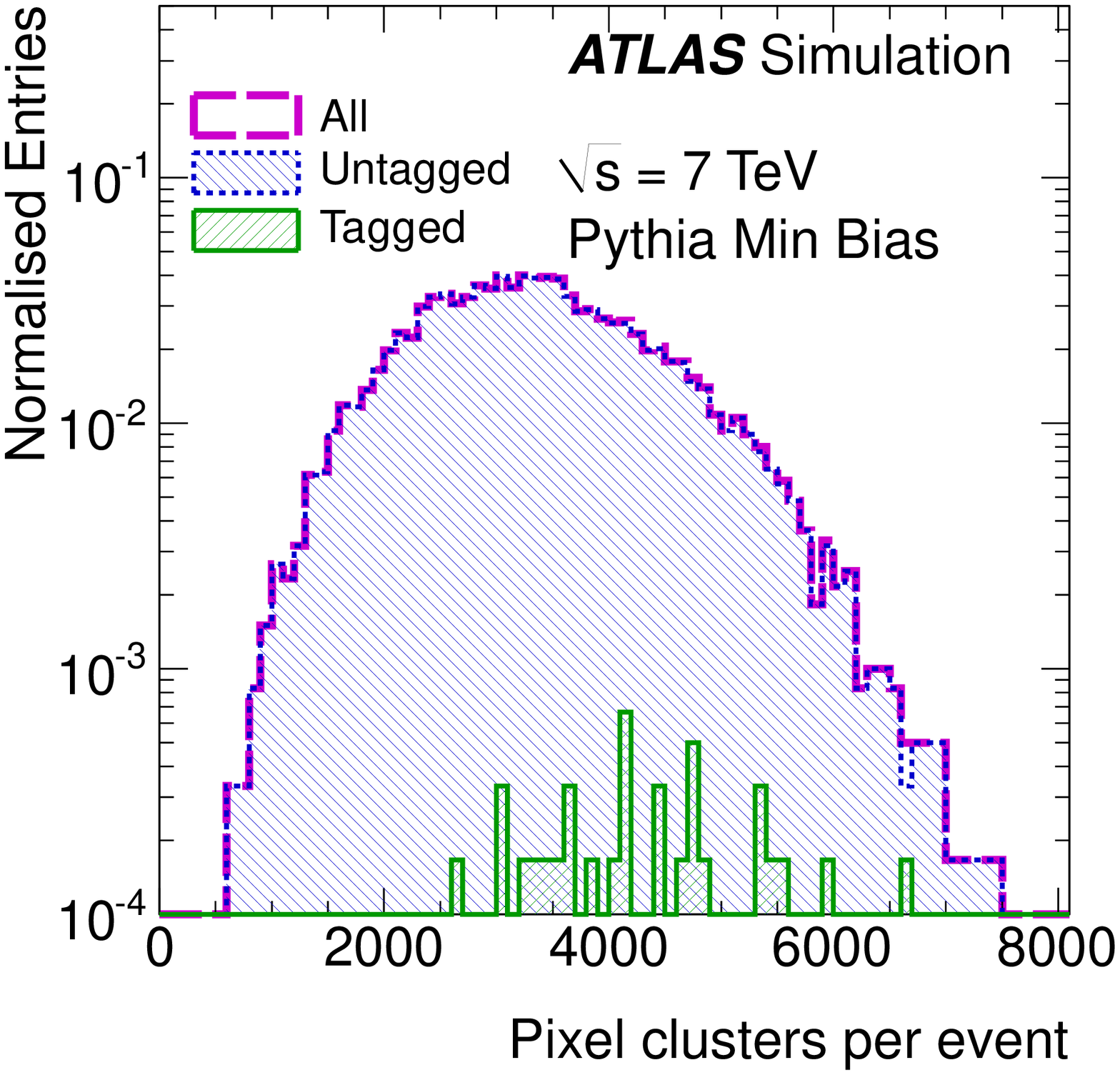}
    \label{fig_pix_pixclusnum_tag_mcminbias}
  }
}
\mbox{
  \subfigure[]{
    \includegraphics*[width=0.49\textwidth]{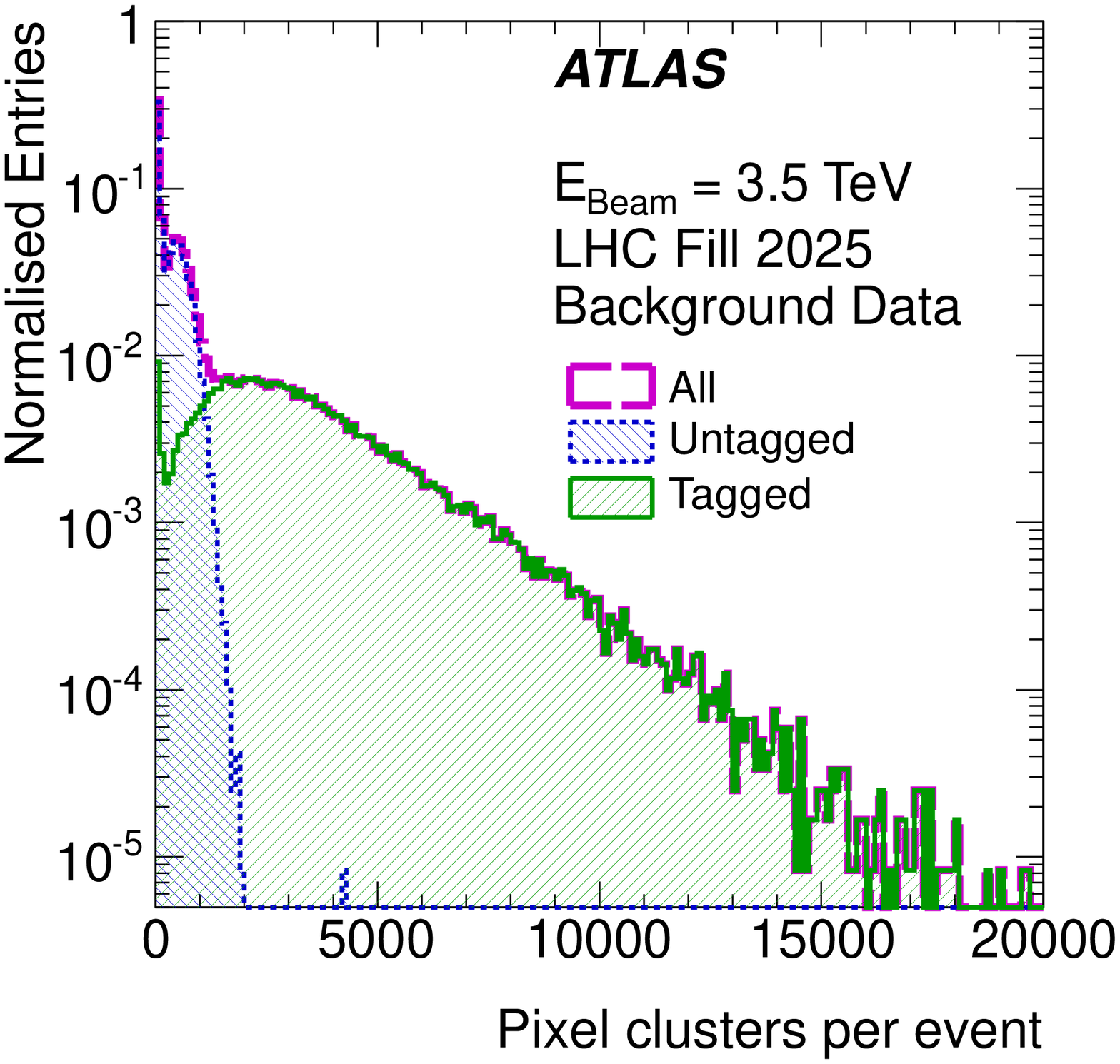}
    \label{fig_pix_pixclusnum_tag_data}
  }
  \subfigure[]{
    \includegraphics*[width=0.49\textwidth]{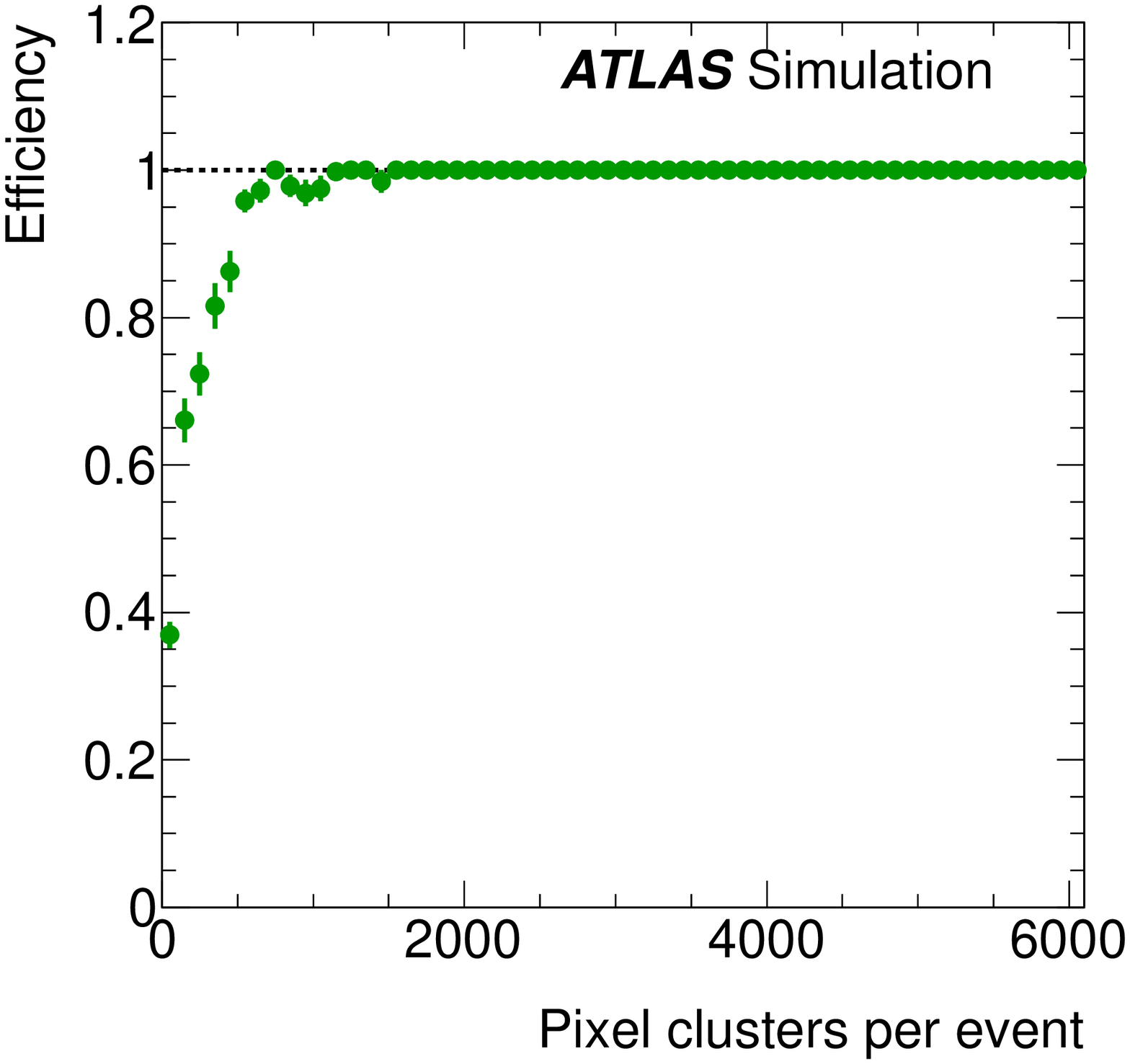}
    \label{fig_pix_pixclusnum_tag_mcefficiency}
  }
}
\caption{The pixel BIB tagging algorithm applied to: Monte Carlo samples of (a) beam-gas events, (b) minimum-bias collisions with pile-up of 21 events per bunch crossing, and (c) 2011 background data. The tagging efficiency as a function of the pixel cluster multiplicity is shown in (d).
The efficiency is evaluated from the beam-gas simulation (a).}
\label{fig_pix_clus_tag_efficiency}
\end{figure}

Figures\,\ref{fig_pix_pixclusnum_tag_mcbeamgas} and \ref{fig_pix_pixclusnum_tag_mcminbias} show the tagging efficiency in simulated beam-gas
events and the mis-tagging rate in simulated collision events, respectively. It should be noted that Fig.\,\ref{fig_pix_pixclusnum_tag_mcminbias} 
is based on an average pile-up of 21 interactions per bunch crossing, which implies that the peak near 3000 clusters corresponds to an average of about 150 
clusters in a single event.
Figure\,\ref{fig_pix_pixclusnum_tag_data} shows the tagged and untagged events in recorded background data, which contain mostly BIB and
sometimes single ghost collisions. The latter are seen as the peak around 200 clusters per event and remain correctly untagged.
The tail extending to a large number of clusters is consistent with the beam-gas simulation and is efficiently tagged as BIB.
Finally, Fig.~\ref{fig_pix_pixclusnum_tag_mcefficiency} shows that the BIB tagging  efficiency 
is above 95\% if there are $\gtrsim 500$ BIB pixel clusters in the event.

\subsection{BIB characteristics seen in 2011 data}

\begin{figure}[htb!]
\centering
\includegraphics*[width=0.49\textwidth]{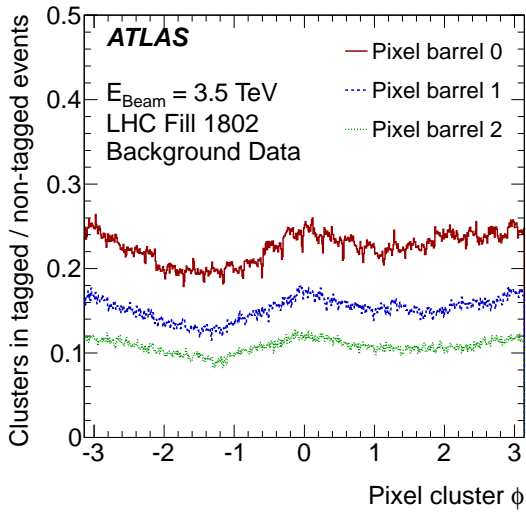}
\caption{Azimuthal distribution of background tagged Pixel clusters, normalised by the cluster distribution in collision events.}
\label{fig_pix_clusphi_data}
\end{figure}

The pixel BIB tagging algorithm, described above, is applied to 2011 data to investigate the distribution of the 
BIB clusters in the Pixel detector and to assess the rate of BIB events 
as a function of the vacuum pressure upstream of the ATLAS detector.

The cluster $\phi$ distribution for each barrel layer is plotted in Fig.~\ref{fig_pix_clusphi_data} for events 
which are selected by the algorithm as containing BIB. The distribution is normalised 
by the number of clusters in collision events, which are not selected by the algorithm, to reduce the geometrical effects of module overlaps 
and of the few pixel modules that were inoperable during this data-taking period. A small excess is observed at 
$\phi=0$ and $\phi=\pi$, corresponding to a horizontal spread of the BIB, most likely 
due to bending in the recombination dipoles. 
An up-down asymmetry is also apparent, which might be an artifact of the vertical crossing angle 
of the beams. Additional simulation studies are required to verify this hypothesis or to identify some
other cause for the effect.\footnote{Since the Pixel detector is very close to the beam-line, the
tunnel floor causing a similar effect in Fig.\,\protect\ref{phidist4} cannot be the cause here.}


\section{BIB muon rejection tools}

The BIB muon rejection tools described in this section are based on timing
and angular information from the endcap muon detectors and the barrel calorimeters,
and are primarily designed to identify fake jets due to BIB.
The events to which the rejection tool is applied are typically selected by
jet or \MET{} triggers.

\subsection{General characteristics}
\label{sec:tertiaryGeneral}

At radial distances larger than those covered by the acceptance of the tracking
detectors, BIB can be studied with the calorimeters and the muon system. 
The LAr barrel has a radial coverage from $1.5$ to $2.0\,\textrm{m}$ and is therefore entirely
covered by the radial range of the Cathode-Strip Chambers (CSC).
The TileCal covers the radial range of $2.2\,\textrm{m} <r<4.3\,\textrm{m}$
which fully overlaps with the acceptance of the inner endcaps of the Monitored Drift Tube (MDT) system.

\begin{figure}
\begin{center}
\includegraphics[width=0.45\columnwidth]{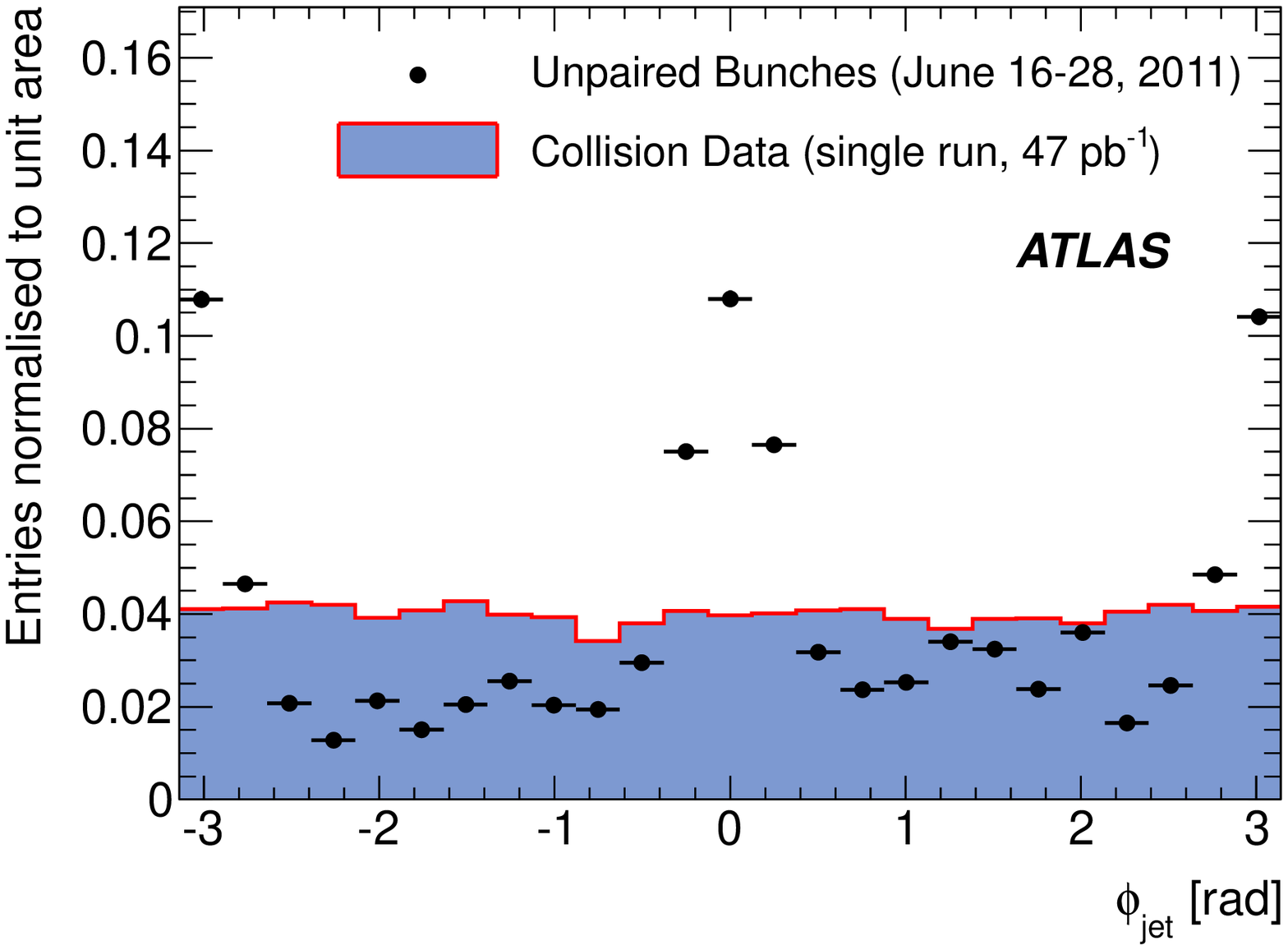}
\includegraphics[width=0.45\columnwidth]{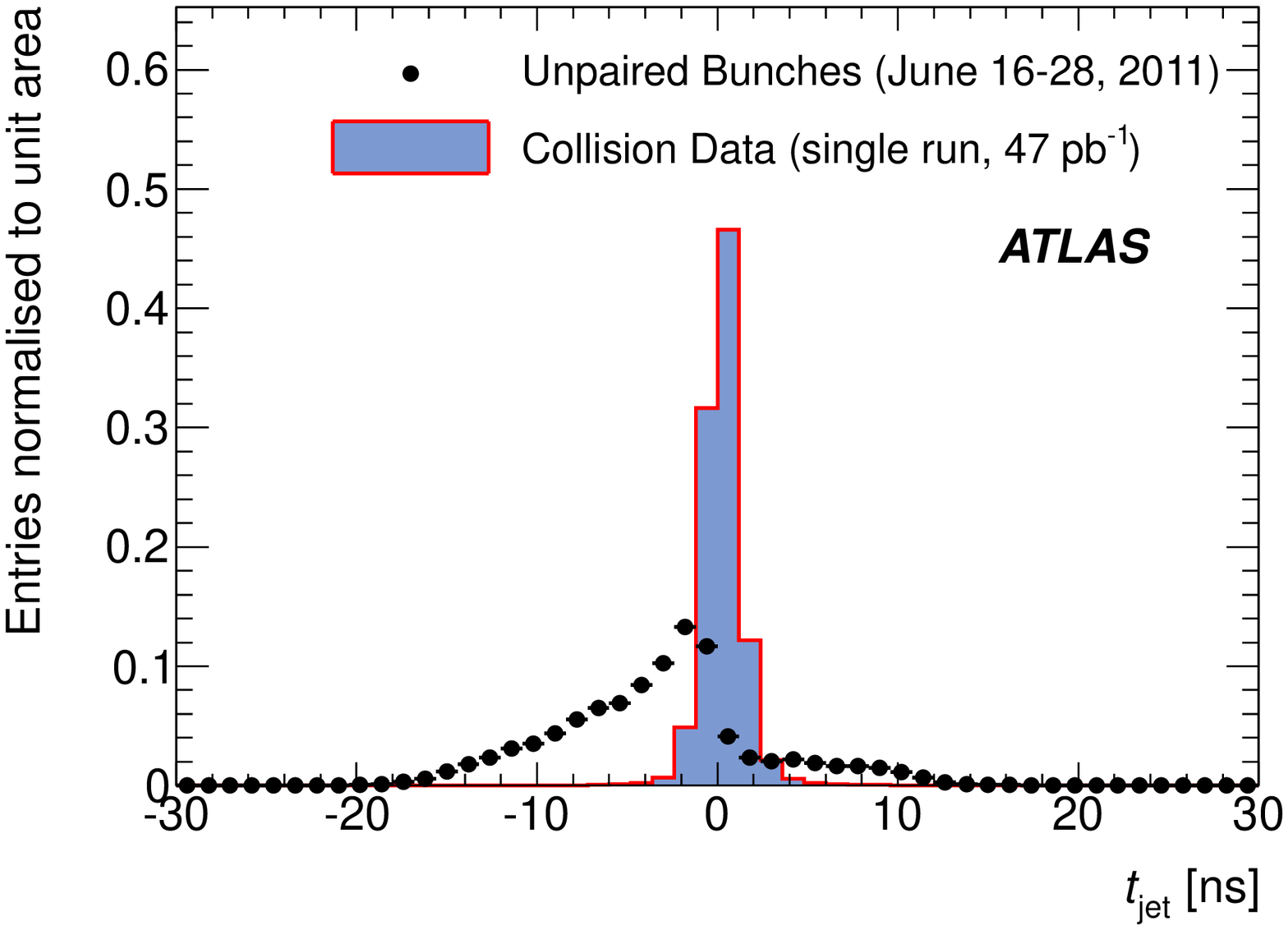}
\end{center}
\caption{Leading jet $\phi$ (left) and time (right) in unpaired bunches and collision data.}
\label{fig:jet1}
\end{figure}

The left plot in Fig.~\ref{fig:jet1} compares the $\phi$ distribution of the leading jets
in data from unpaired bunches and from collisions.
Both samples have general data quality requirements applied.
Furthermore, the unpaired bunches are cleaned from ghost collisions by removing events with
a reconstructed primary vertex. 
A striking difference is observed between
the azimuthal distribution of leading jets from collisions and BIB. Whereas for collisions 
there is no preferred $\phi$ direction of jets, the azimuthal distribution for fake jets from BIB
has two peaks, at $\phi=0$ and $\phi=\pi$. 
The region between the two peaks is somewhat more populated for $\phi>0$  than for 
$\phi<0$. These features are also seen in Fig.\,\ref{phidist4} and are
explained by the arrangement of the dipole magnets and the shielding effect of the tunnel floor, respectively. 
The right plot in Fig.\,\ref{fig:jet1} shows that the reconstructed time of the fake jets from BIB 
is typically earlier than for jets from collisions.
Physics  objects from collisions have time $t\sim\,0\,\textrm{ns}$ since all the time measurements
are corrected for the time-of-flight from the interaction point
\begin{equation}
\label{eq:tof}
t_\textrm{\scriptsize ToF}=\sqrt{r^2+z^2}/c
\end{equation}
where ($r$, $z$) is the position of the physics object and $c$ is the speed of light.
Since the high-energy components of BIB arrive simultaneously with the proton bunch,
the BIB objects have time $t\sim\pm|z|/c$ with respect to the interaction time,
where the sign depends on the direction of the BIB particle.
As the reconstructed times are corrected for the time-of-flight, the reconstructed
time of the BIB objects can be calculated as
\begin{eqnarray}
t_{\textrm{\scriptsize BIB}} &=& -z/c-t_\textrm{\scriptsize ToF} \qquad\textrm{for the A$\rightarrow$C direction},
\label{eq:timeBibAC} \\
t_{\textrm{\scriptsize BIB}} &=& +z/c-t_\textrm{\scriptsize ToF} \qquad\textrm{for the C$\rightarrow$A direction}.\footnotemark
\label{eq:timeBibCA}
\end{eqnarray}
\footnotetext{The $z$-axis in the ATLAS coordinate system points from C to A.}
These equations explain the observed time distribution in Fig.\,\ref{fig:jet1}
as $t_{\textrm{\scriptsize BIB}}$ is negative for the $z$-position where the BIB particle enters
the detector and increases towards $0\,\textrm{ns}$ on its way out of the detector on the other side. 
The entries at $t_\textrm{\scriptsize jet}>0\,\textrm{ns}$ in the unpaired-bunch data are
due to pile-up from the neighbouring interleaved bunches that are separated by only a $25\,\textrm{ns}$ bunch spacing.

\begin{figure}
\begin{center}
\subfigure[][]{
\includegraphics[width=0.35\columnwidth, trim=0 30 0 0, clip]{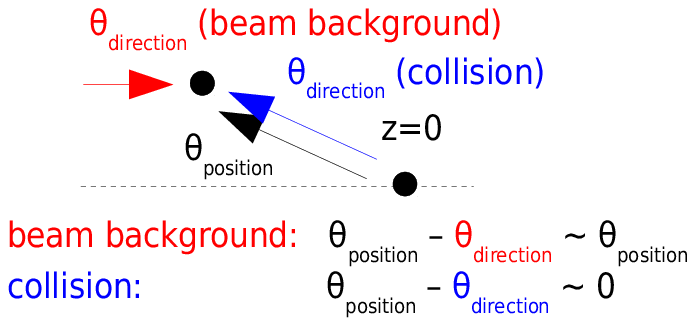}
\label{fig:sketchTheta}}
\subfigure[][]{
\includegraphics[width=0.52\columnwidth]{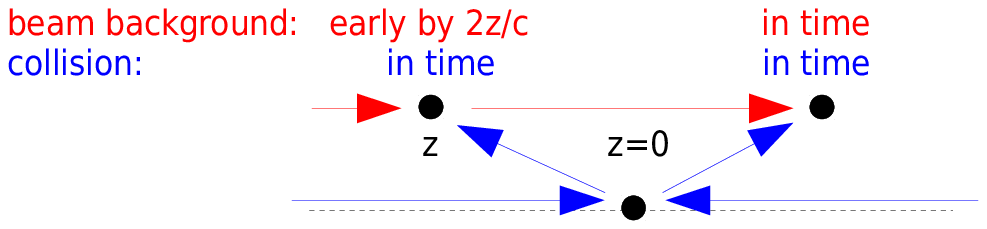}
\label{fig:sketchTime}}
\end{center}
\caption{(a) Polar position and direction and (b) reconstructed time 
of the BIB objects compared to the collision objects.}
\label{fig:sketch}
\end{figure}

The response of the muon chambers to energetic BIB muons differs from that to muons from 
collisions, primarily due to their trajectories but 
also due to the early arrival time of the BIB muons with respect to the 
collision products. Figure~\ref{fig:sketch} shows sketches of both of these characteristic
features of BIB compared to the collision particles.
The BIB particles have direction nearly parallel to the beam-pipe, therefore
$\theta_{\textrm{\scriptsize pos}}-\theta_{\textrm{\scriptsize dir}} \sim \theta_{\textrm{\scriptsize pos}}$,
where $\theta_{\textrm{\scriptsize pos}}$, $\theta_{\textrm{\scriptsize dir}}$ denote the reconstructed polar position and direction, respectively.
The collision products point to the interaction point and hence have $\theta_{\textrm{\scriptsize pos}}-\theta_{\textrm{\scriptsize dir}} \sim 0$.
The reconstructed time of the BIB particles follows from Eqs.\,\eqref{eq:timeBibAC} and \eqref{eq:timeBibCA}.
For the endcap chambers, the BIB particles can arrive either in time or early and
the expected time can be formulated as
\begin{eqnarray}
t_{\textrm{\scriptsize in-time}} &=& +|z|/c-t_\textrm{\scriptsize ToF},
\label{eq:timeBibInTime} \\
t_{\textrm{\scriptsize early}} &=& -|z|/c-t_\textrm{\scriptsize ToF}.
\label{eq:timeBibEarly}
\end{eqnarray}
For $z\gg r$, the time-of-flight correction in Eq.\,\eqref{eq:tof} simplifies to $t_\textrm{\scriptsize ToF}\sim|z|/c$.
As the reconstructed times are corrected for the time-of-flight, the time of the BIB particles is either
$t\sim(+|z|-|z|)/c=0$ or $t\sim(-|z|-|z|)/c=-2|z|/c$,
depending on where along the path of the BIB particle through the detector the object is reconstructed.
This approximation is illustrated in Fig.\,\ref{fig:sketchTime}.

\begin{figure}
\begin{center}
\includegraphics[width=0.45\columnwidth]{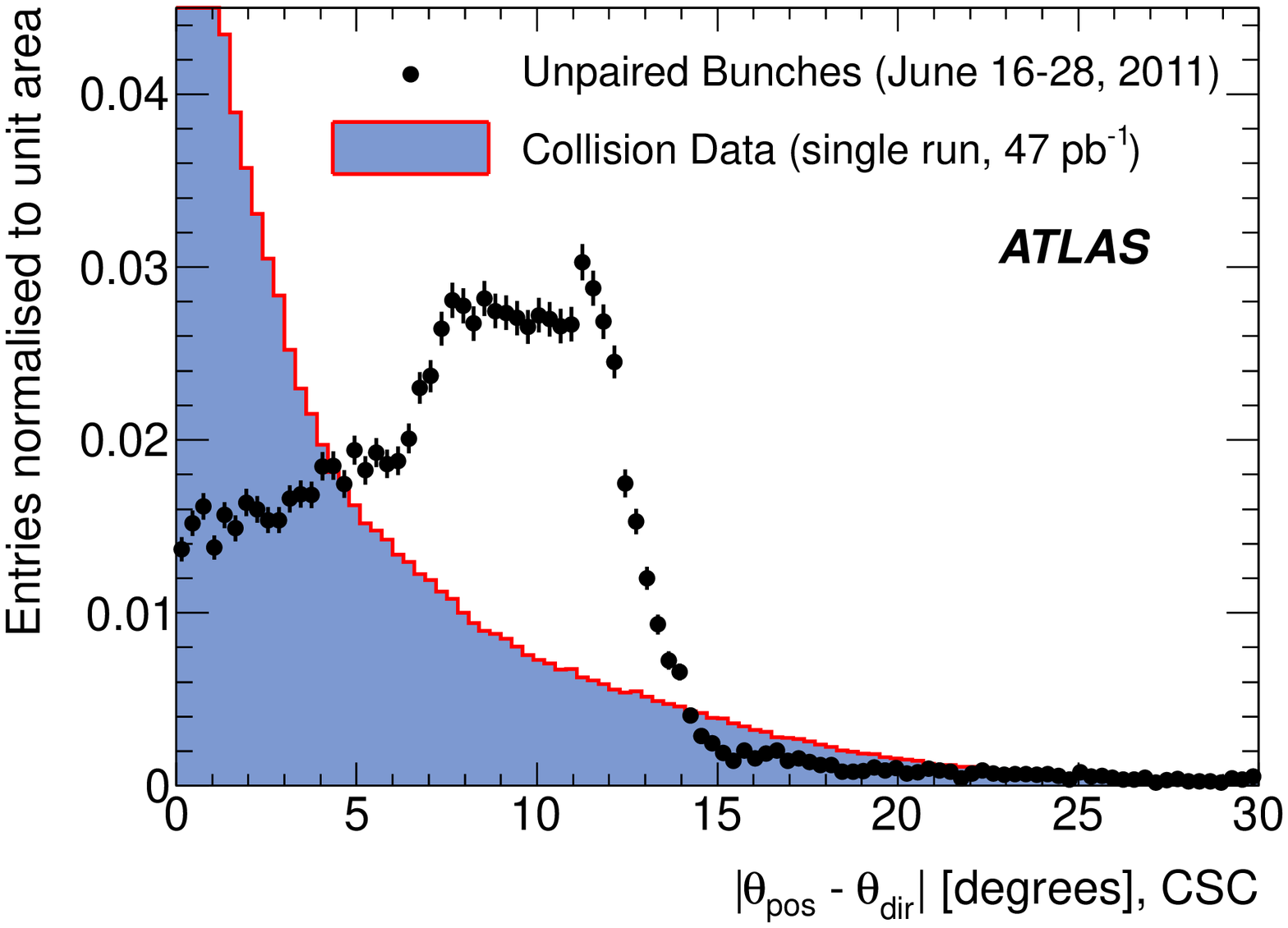}
\includegraphics[width=0.45\columnwidth]{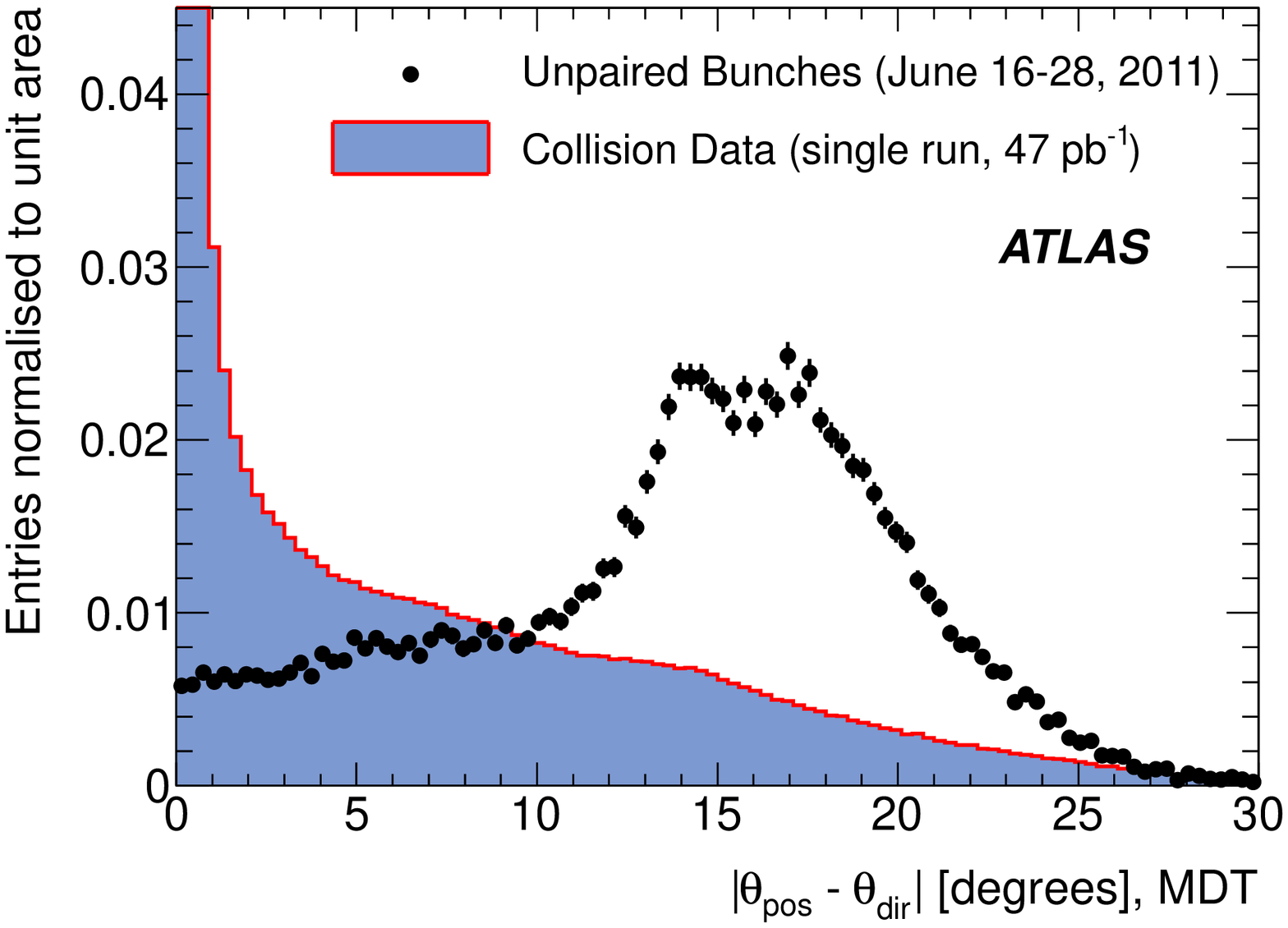}
\end{center}
\caption{Difference between the reconstructed polar position $\theta_{\textrm{\scriptsize pos}}$
and the reconstructed polar direction $\theta_{\textrm{\scriptsize dir}}$ for the muon segments in the CSC (left)
and the inner MDT endcap (right).
Data from cleaned unpaired bunches (points) are compared to collisions (filled histogram).
}
\label{fig:theta}
\end{figure}

Hits in each muon station are grouped into segments which allow the reconstruction of the direction
of the particle causing the hits. 
At least three hits are required in order to form a segment.
Figure~\ref{fig:theta} shows
the difference between the reconstructed polar position $\theta_{\textrm{\scriptsize pos}}$ and the
reconstructed polar direction $\theta_{\textrm{\scriptsize dir}}$ of the muon segments in the CSC and the inner MDT endcaps
in cleaned unpaired bunches and collision data which, as can be seen in Fig.~\ref{fig:sketchTheta}, is expected to be $\sim0$
in collisions. This is indeed seen in Fig.~\ref{fig:theta} where the entries for collisions
at non-zero values are due to angular resolution and particles bending in the toroidal magnetic field.
For BIB,
where $\Delta\theta=|\theta_{\textrm{\scriptsize pos}}-\theta_{\textrm{\scriptsize dir}}|~\sim\theta_{\textrm{\scriptsize pos}}$, 
the expected values are $7^{\circ}<\Delta\theta<14^{\circ}$ for the CSC and $\Delta\theta>14^{\circ}$
for the inner MDT endcaps.
The data clearly support the hypothesis that BIB muons are traversing the detector parallel to the 
beam-line at radii beyond $1\,\textrm{m}$.

\begin{figure}
\begin{center}
\includegraphics[width=0.6\columnwidth]{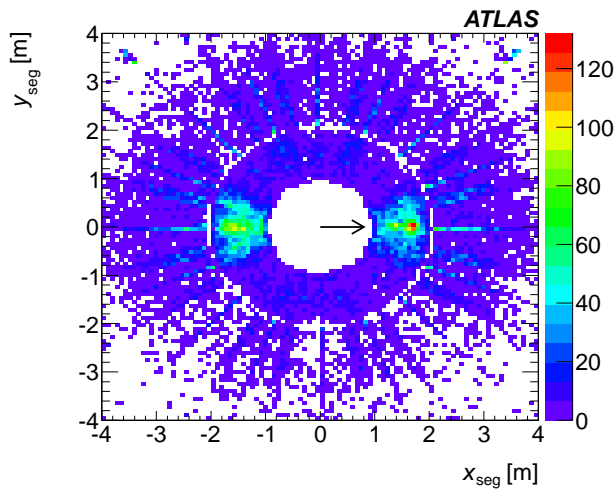}
\end{center}
\caption{Position of the muon segments in the CSC and the inner MDT endcaps
with a direction nearly parallel to the beam-pipe
in the cleaned unpaired bunches.
The arrow indicates the direction towards the centre of the LHC ring (positive $x$-axis).
Units correspond to the number of entries per bin.
}
\label{fig:xySeg}
\end{figure}

Figure~\ref{fig:xySeg} shows the transverse
position of the muon segments that have direction nearly parallel to the beam-pipe in the CSC and the inner MDT endcaps.
This is assured by requiring 
$\Delta\theta>5^{\circ}$ for the CSC and $\Delta\theta>10^{\circ}$ for the inner MDT endcaps.
Only data from unpaired bunches are used
in this plot, and the requirement on the direction of the muon segments
helps to reject contributions coming from ghost collisions and noise.
Such muon segments are referred to as ``BIB muon segments'' in the text below.
It is seen that the charged BIB particles are mostly in the plane of the LHC ring ($y=0$).
Most of the muon segments are located
at $r\sim1.8\,\textrm{m}$ and the distribution is steeply falling further away from the beam-pipe.
The radial dependence and $\phi$-asymmetry are qualitatively consistent with Figs.\,\ref{muon-r-z}
and \ref{phidist4}, respectively. However, for BIB to be seen in data, the events have to
be triggered. This is mostly done by jet triggers, which require calorimeter activity. 
The inner edge of the LAr barrel is at $r=1.5\,\textrm{m}$ which explains why the 
rise of BIB rates towards smaller radii, seen in Fig.\,\ref{muon-r-z},
is not reflected in the data.
The jet triggers predominantly select highly energetic BIB muons that penetrate into the calorimeters
and leave significant energy depositions above the trigger $\pt$ threshold. 
Therefore, the pronounced azimuthal asymmetry of the muon segments observed in Fig.\,\ref{fig:xySeg}
corresponds mainly to high-energy BIB and fully reflects the jet asymmetry seen in Fig.\,\ref{fig:jet1}.

\begin{figure}
\begin{center}
\includegraphics[width=0.45\columnwidth]{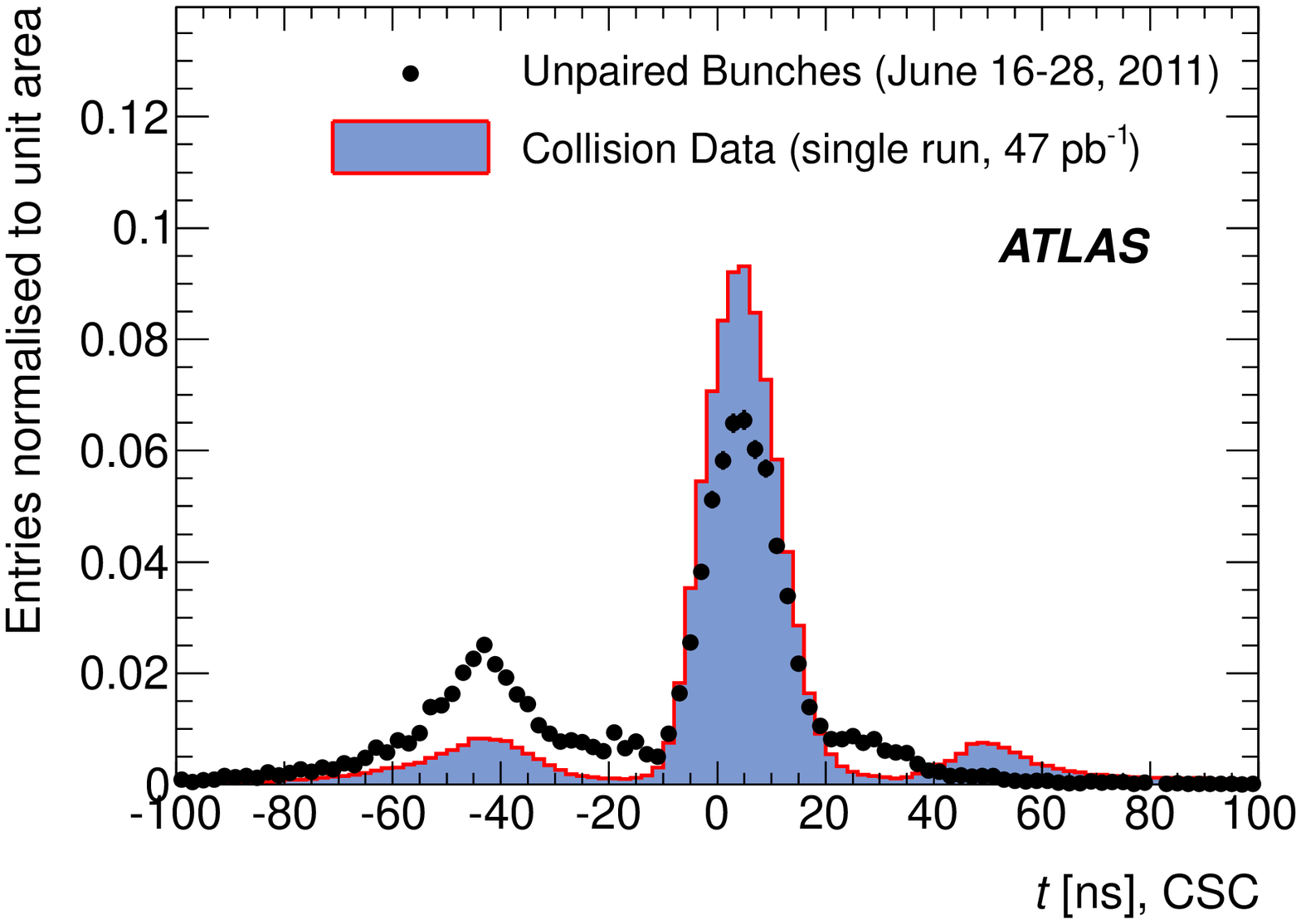}
\includegraphics[width=0.45\columnwidth]{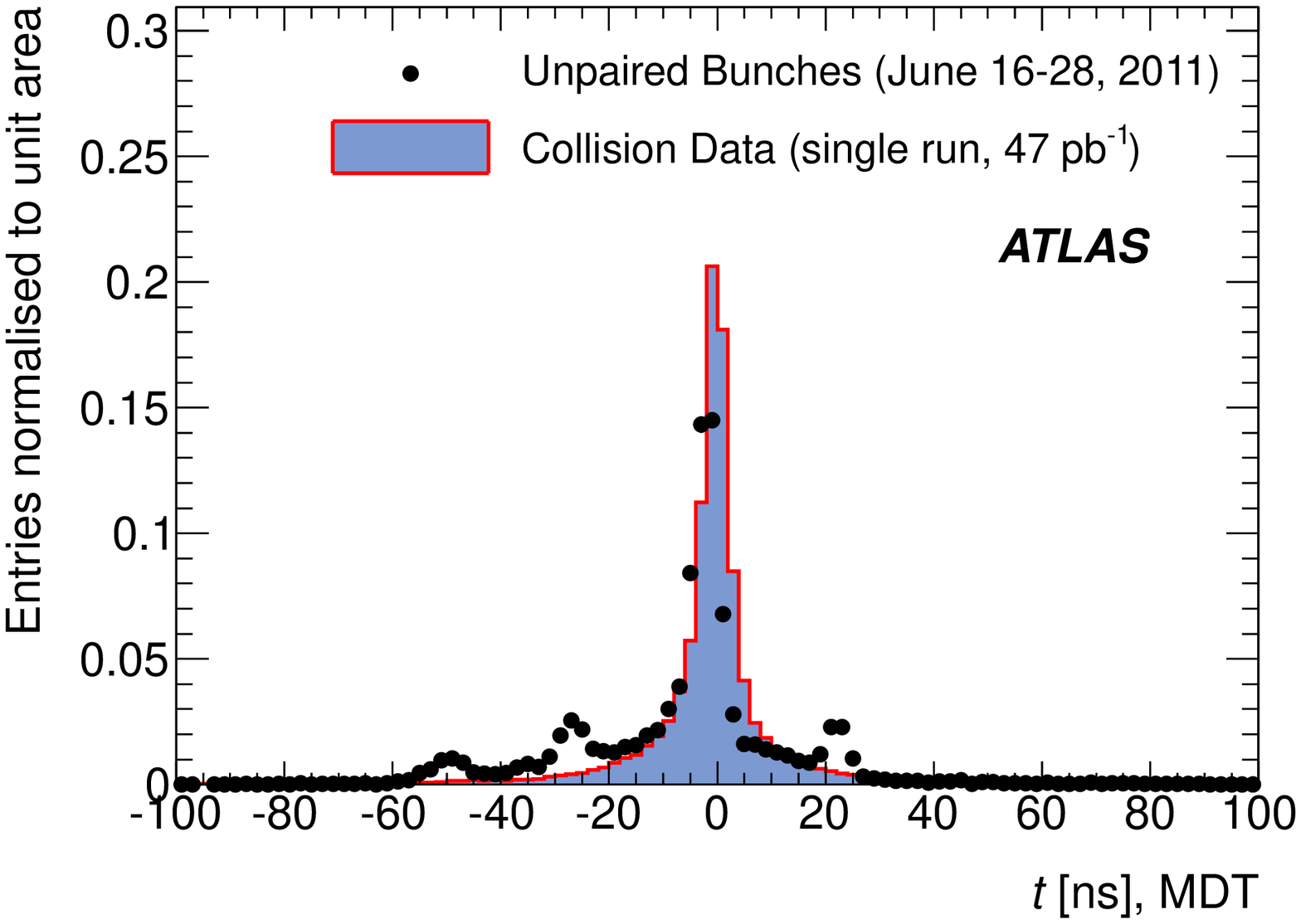}
\end{center}
\caption{Reconstructed time of the CSC muon segments (left) and the inner MDT endcap segments (right)
with a direction nearly parallel to the beam-pipe.
Data from cleaned unpaired bunches (points) are compared to collisions (filled histogram).
}
\label{fig:time}
\end{figure}

Figure~\ref{fig:time} shows the reconstructed time of the BIB muon segments
in cleaned unpaired bunches and collision data.
As stated above, the collision products arrive at $t\sim0\,\textrm{ns}$.
As expected, the time distribution of the muon segments in the inner MDT endcaps from collision data
shows only a peak centred around $0$\,ns.
However, for the CSC muon segments there are two extra peaks in the time distribution
located at $\pm50\,\textrm{ns}$. These peaks 
are related to the out-of-time pile-up due to the $50\,\textrm{ns}$ bunch spacing.
No such peaks are visible for the MDT endcaps since the reconstruction algorithm
for the MDT is written in such a way that the out-of-time objects are suppressed.
Furthermore, it can be seen that the whole time distribution for the CSC is 
shifted by $6.25$\,ns to positive values.\footnote{This is due to the fact that half of the CSC channels
have a $12.5\,\textrm{ns}$ shift that is not corrected.
Therefore, depending on which CSC channels are used for the time reconstruction, the muon segment time is shifted by $0$, $6.25$ or $12.5$\,ns.
The three distinct peaks are not visible in the distribution
due to insufficient time resolution.}
In unpaired bunches, muon segments are expected to be either in time ($t\sim0\,\textrm{ns}$)
or early ($t\sim-50\,\textrm{ns}$) depending on whether the muon segment
is created while exiting or entering the detector (see Fig.~\ref{fig:sketchTime}).
The expected time of $\sim-50\,\textrm{ns}$ corresponds to the time-of-flight between
the muon stations on both sides of the detector that are located at $|z|\sim8\,\textrm{m}$,
and also coincides with the time of the early out-of-time pile-up.

As discussed in Sect.\,\ref{sec:jetrates}, in some of the 2011 LHC bunch patterns, 
interleaved unpaired bunches were created by shifting the bunch trains to overlap with each other.
In these cases bunches in opposite directions were separated by only 25\,ns.
The peaks at $\pm25\,\textrm{ns}$ which are visible in the unpaired bunches
in Fig.~\ref{fig:time} correspond to muon segments reconstructed from the
neighbouring interleaved unpaired bunch.
The amount of data entering these peaks is about $10\%$ of all unpaired-bunch data.

A muon that radiates enough energy to create a fake jet loses a significant fraction of its energy,
which is associated with a non-negligible momentum transfer.
If the deflection, to which the endcap toroid field might also 
contribute in the case of MDT segments, is large enough,
the outgoing muon would not create a muon segment with $\theta_{\textrm{\scriptsize dir}}\sim 0$
on the other side of the detector, or 
it might even miss the CSC or the inner MDT endcap altogether.
Therefore, the number of entries in the early peak is expected to be larger
than in the in-time peak. The fact that fewer early muon segments are seen is due 
to the muon segment reconstruction that is optimised for in-time measurements.
Some of the early CSC segments
are lost due to the fact that the read-out time window is not wide enough
to detect all the early hits.
As for the MDT segments, the out-of-time objects are suppressed by the reconstruction algorithm.

\subsection{BIB identification methods}
\label{sec:bbim}

The characteristic signatures of BIB described above
motivate a set of BIB identification methods.
These either utilise only the basic information (position, direction, time)
of the muon segments, or they try to match the muon segments
to the calorimeter activity.

\subsubsection{Segment method}

The segment method requires the presence of a BIB muon segment,
where $\theta_{\textrm{\scriptsize dir}}\sim 0$,
in the CSC or the inner
MDT endcap. 
This method is very efficient for cleaning the empty bunch-crossings from BIB.
Since the method is completely independent of calorimeter information,
it is suitable for creating background-free empty bunch samples needed to
identify noisy calorimeter cells.

\subsubsection{One-sided method}

The one-sided method requires the BIB muon segments and calorimeter clusters,
with energy larger than $10\,\textrm{GeV}$,
to be matched in relative azimuthal and radial positions. The matching in $\phi$ is motivated
by the fact that BIB muons are not bent azimuthally by the magnetic fields
of the ATLAS detector. 
The matching in $r$ is introduced in order to reduce the mis-identification
probability of this method due to accidental matching.
While the toroidal field does bend the trajectory in $r$, it can be
assumed that the radial deflection remains small
for high-energy incoming muons, or muons at radii below the inner edge of the endcap toroid.
Ignoring the low-energy clusters also helps to suppress accidental matching.
Depending on whether the muon segment is early or in time and on its position,
the direction of the BIB muon may be reconstructed.
The early (in-time) muon segments are selected such that the difference
between the reconstructed time and the expected time
$t_{\textrm{\scriptsize early}}$ ($t_{\textrm{\scriptsize in-time}}$),
defined in Eqs.\,\eqref{eq:timeBibInTime} and \eqref{eq:timeBibEarly},
is less than $25\,\textrm{ns}$, where the value is conservatively chosen as half of the time-of-flight
difference between the muon chambers on side A and side C.
The position of the calorimeter cluster in $z$ and $r$ can be used to estimate
the expected time of the calorimeter energy deposition according to Eqs.\,\eqref{eq:timeBibAC} or \eqref{eq:timeBibCA}.
Since the time resolution of the calorimeter measurements is $\sim1\,\textrm{ns}$
one can precisely compare the reconstructed cluster time with the expected value.
The difference is required to be less than $2.5\,\textrm{ns}$ in order to flag the cluster
as a BIB candidate.

\begin{figure}
\begin{center}
\includegraphics[width=0.45\columnwidth]{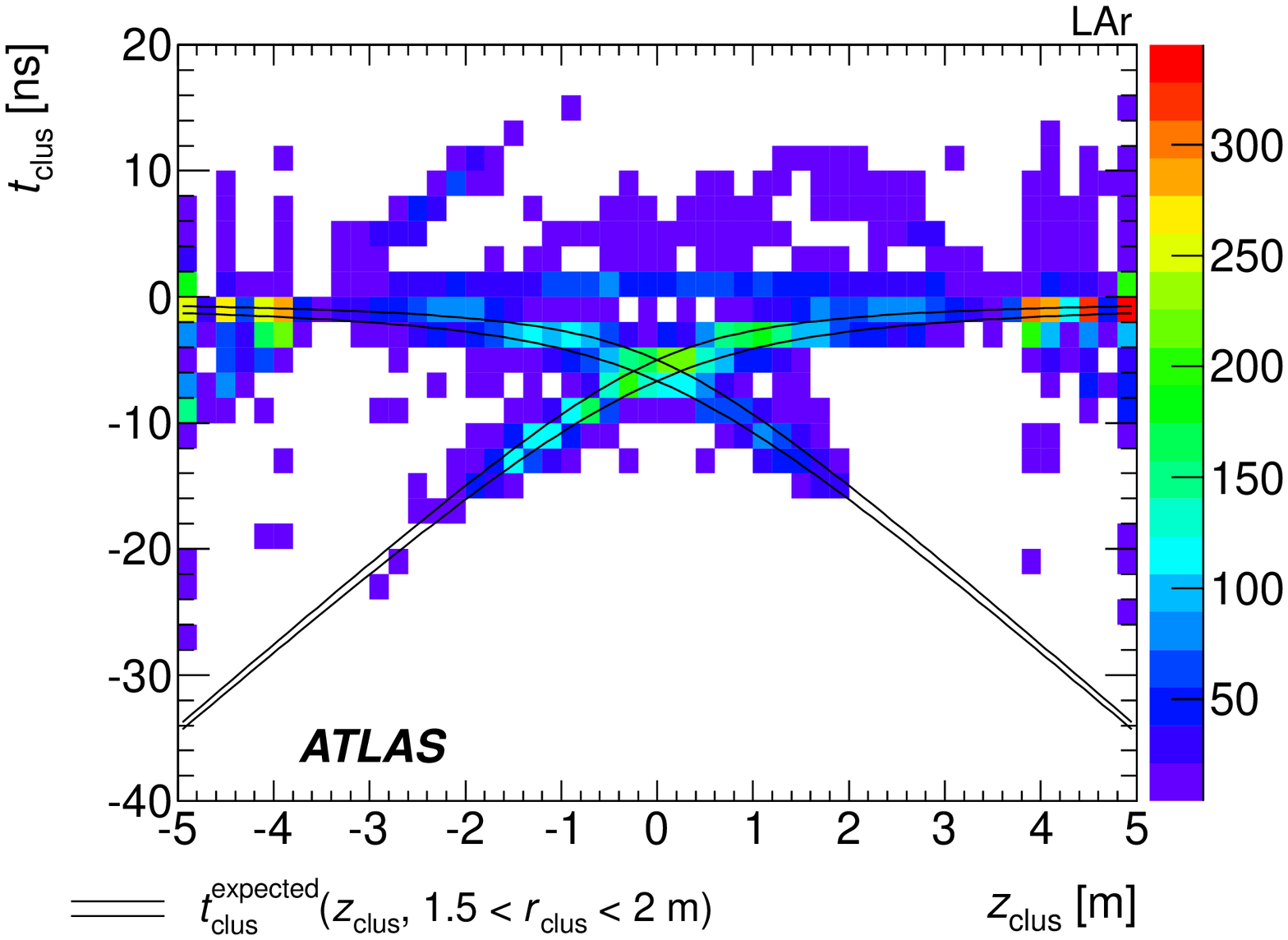}
\includegraphics[width=0.45\columnwidth]{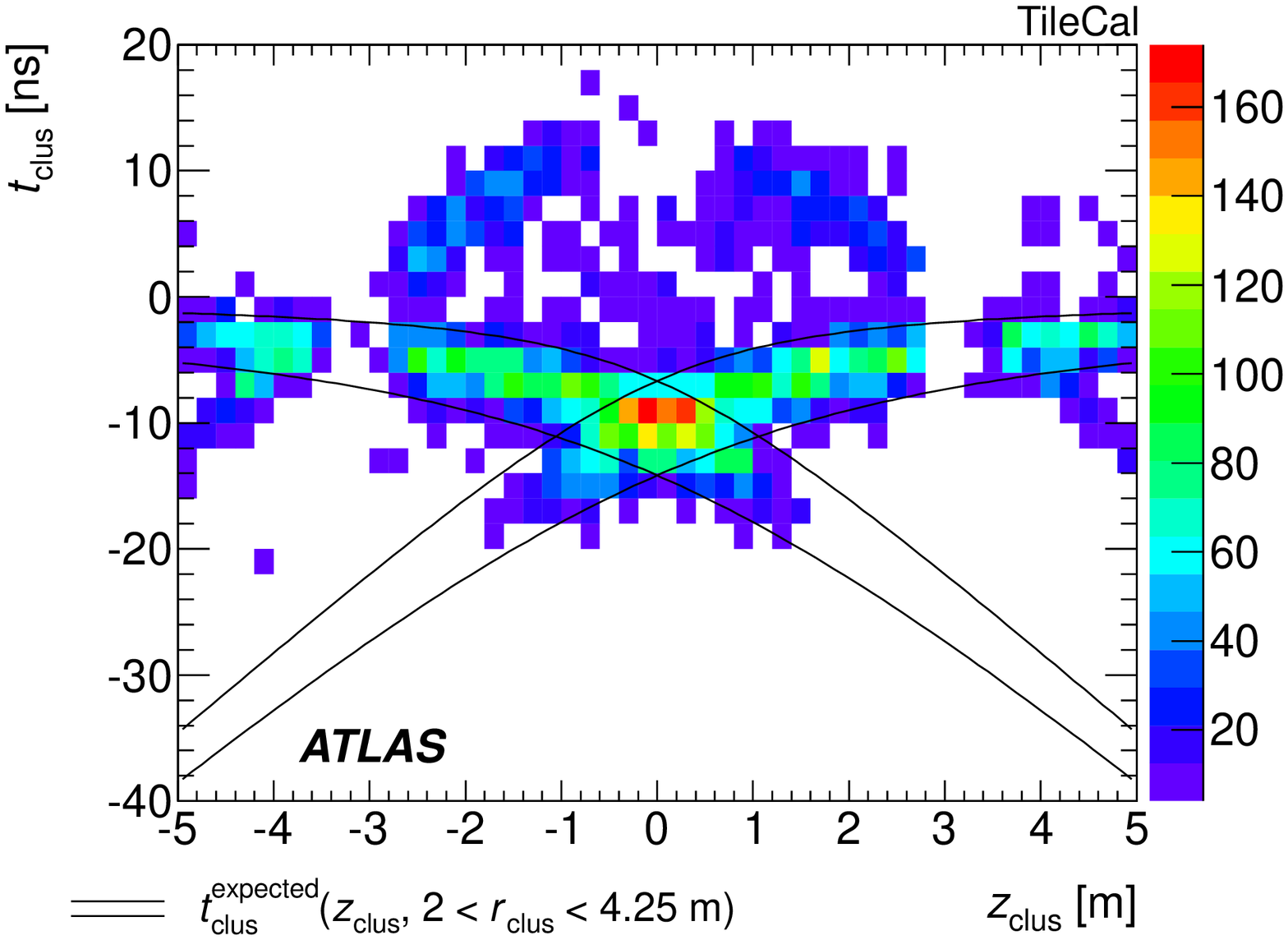}
\end{center}
\caption{Cluster time plotted as a function of its $z$ position in the LAr (left) and TileCal (right)
for the cleaned unpaired bunches.
Only the clusters matching a BIB muon segment are shown.
The two bands, covering the radial extent of the detectors, show the expected time for the BIB 
clusters in $1.5\,\textrm{m} <r<2\,\textrm{m}$ for LAr and $2\,\textrm{m} <r<4.25\,\textrm{m}$ for TileCal
going in the A$\rightarrow$C or C$\rightarrow$A direction.
Units correspond to the number of entries per bin.
}
\label{fig:timeClus}
\end{figure}

Figure~\ref{fig:timeClus} shows the cluster time as a function of the cluster $z$-position
in unpaired bunches separately for the LAr and the TileCal.
Expected cluster times for the radial acceptance of the calorimeters
based on Eqs.\,\eqref{eq:timeBibAC} and \eqref{eq:timeBibCA} are also
indicated for both directions of BIB. The majority of data is seen to fall within the expectation band.
However, there are also other interesting features in the plot:
for the LAr calorimeter, there is a visible set of clusters with $t=0\,\textrm{ns}$ at 
all $z$ positions.
These come from the ghost collisions in the unpaired bunches.
In both plots, one can see a set of clusters
in a pattern similar to the expectation bands
but shifted by $\sim25\,\textrm{ns}$ in time
to positive values. These entries correspond to the clusters 
reconstructed from the neighbouring interleaved bunches, discussed already in Sect.\,\ref{sec:tertiaryGeneral}.

It follows from Eqs.\,\eqref{eq:timeBibAC} and \eqref{eq:timeBibCA} that the expected time for BIB calorimeter 
clusters is close to $0\,\textrm{ns}$ for small $r$ and large $|z|$ on the side where BIB leaves the detector.
Therefore, the one-sided method has large mis-identification probability in the forward region.

\begin{figure}
\begin{center}
\includegraphics[width=0.6\columnwidth]{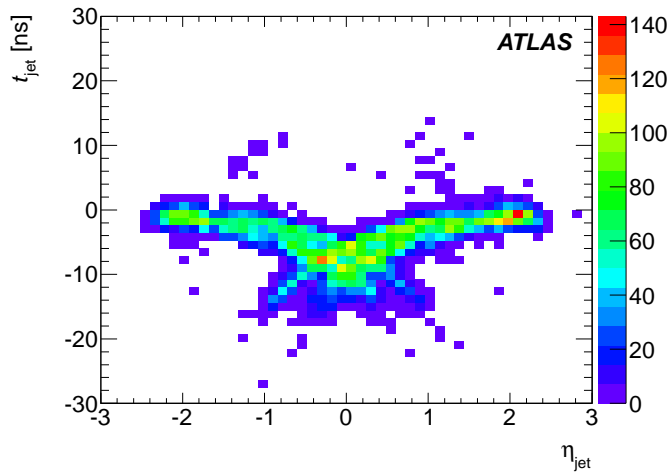}
\end{center}
\caption{Time of the leading jet as a function of its $\eta$ in the cleaned unpaired bunches.
Only the events identified by the one-sided method are shown.
}
\label{fig:jetOneSided}
\end{figure}

Figure~\ref{fig:jetOneSided} shows the leading jet time as a function of its pseudorapidity
in events identified by the one-sided method.
It can be seen that the characteristic timing pattern of the BIB calorimeter clusters
shown in Fig.\,\ref{fig:timeClus} is reflected in the properties of the reconstructed jets due to BIB.

\subsubsection{Two-sided method}

The two-sided method requires a BIB muon segment on both sides to be matched in $\phi$ and $r$
to a single calorimeter cluster of energy above $10\,\textrm{GeV}$. Here, the cluster time is not checked. A corresponding
time difference between the two segments is required instead. 
The expected time difference, due to the relative $z$-position of the muon chambers on both sides of the spectrometer,
is $\Delta t=50\,\textrm{ns}$. 
Since the time resolution of the CSC is about $7\,\textrm{ns}$
(see Fig.~\ref{fig:time}) a conservative cut of $\Delta t>25\,\textrm{ns}$ is applied.

Such an event topology is unlikely to be mimicked by collision products which makes this method
particularly robust against mis-identification.

\subsubsection{Efficiency and mis-identification probability}

The efficiency ($\epsilon$) of the identification methods is evaluated from the whole 2011 unpaired-bunch data.
General data quality assessments are imposed on the sample, and 
ghost collisions are suppressed by vetoing events with one or more reconstructed primary vertices.
Noisy events are further reduced
by requiring a leading jet with a large transverse momentum of $\pt>120\,\textrm{GeV}$.
Jets from the inner part of the calorimeter endcaps, where there is no overlap with any muon chamber,
are suppressed by rejecting events with
the leading jet $|\eta|>2.8$. However, the number of events with the leading jet outside the calorimeter barrel,
$|\eta|>1.5$, is negligible anyway.

The mis-identification probability ($P_\textrm{\scriptsize mis}$) is determined in a back-to-back dijet sample from collision data.
This sample also meets the general data quality requirements and the events with at least two jets
as well as leading jet transverse momentum $\pt>120\,\textrm{GeV}$ and $|\eta|<2$ are selected.
Furthermore, the second leading jet in this sample is required to have a similar transverse momentum to the first one
($\frac{|\pt^{1}-\pt^{2}|}{\pt^{1}+\pt^{2}}<0.2$)
and the two jets are required to be back-to-back in the transverse plane ($\Delta \phi_{\rm j-j} > 2.8$).
An event is mis-tagged as BIB if any of the muon segments or calorimeter clusters satisfy the requirements
of the tagging methods discussed above.

The resulting $\epsilon$ and $P_\textrm{\scriptsize mis}$ are listed in Table~\ref{tab:eff}.
The high efficiency of the segment method ($81.6\%$) makes it useful in preparing background-free
samples or for data quality monitoring.
In physics analyses however, it is important to clean background with a minimum loss of signal events.
Table\,\ref{tab:eff} shows that the two-sided method has high purity, $P_\textrm{\scriptsize mis}=10^{-5}$,
but has an efficiency of only $16.0\%$. The one-sided method has a better efficiency of $54.2\%$, but 
$\sim1.4\%$ of signal events are mis-identified.
However, the numbers given for the mis-identification probabilities also depend on the
final-state topology induced by the signal region cuts in a particular physics analysis.
Therefore, the mis-identification probabilities given here serve only as an illustration
where dijets are chosen as an example. 
The combined efficiency of the one-sided and the two-sided methods yields $56.7\%$ for the \verb+OR+ combination
and $13.5\%$ for the \verb+AND+ combination.

It was shown previously that the interleaved bunches may cause BIB from one BCID to be
reconstructed in a neighbouring BCID with a reconstructed time shifted accordingly
by $25\,\textrm{ns}$.
This introduces a systematic bias to the evaluated efficiencies
of the one-sided and two-sided methods since they select BIB predominantly from the current BCID.
The fraction of BIB, reconstructed from the neighbouring interleaved bunches,
in all unpaired-bunch data is approximately $10\%$ and it is not certain to what extent 
there is double counting of such events in the sample.
Therefore, $10\%$ is also taken as a relative systematic uncertainty.

\begin{table}
\begin{center}
\begin{tabular}{|l|l|l|}
\hline
method & efficiency $\pm\textrm{(stat)}$ & mis-identification probability \\
\hline
\hline
segment    & $0.816 \pm 0.017$ & $0.46$\\
one-sided  & $0.542 \pm 0.013$ & $0.014$\\
two-sided  & $0.160 \pm 0.006$ & $10^{-5}$\\
\hline
\end{tabular}
\end{center}
\caption{Efficiency and mis-identification probability of the BIB identification methods.
The mis-identification probabilities are derived from high-statistics samples,
therefore no statistical uncertainties are given.}
\label{tab:eff}
\end{table}

\subsection{BIB rate in 2011}

\begin{figure}
\begin{center}
\subfigure[BIB rate in filled bunches.]{
\includegraphics[width=\columnwidth]{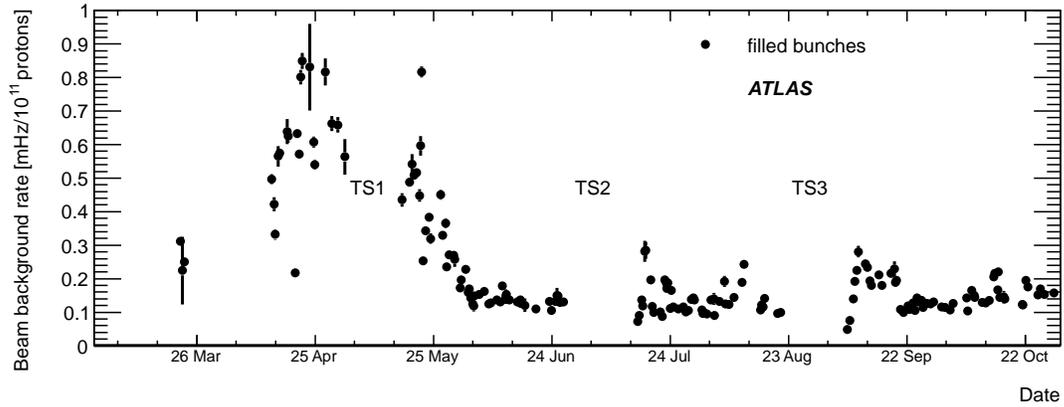}
\label{fig:rateFilled}
}
\subfigure[BIB rate in unpaired isolated and unpaired non-isolated bunches.]{
\includegraphics[width=\columnwidth]{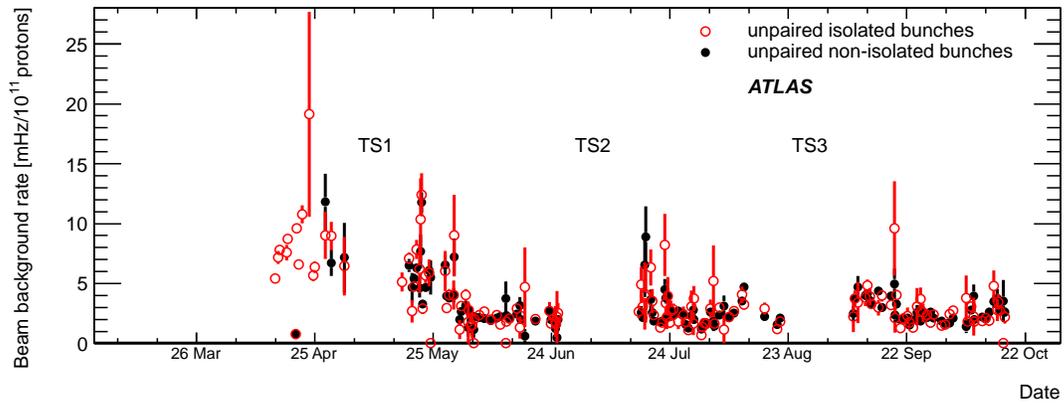}
\label{fig:rateUnpaired}
}
\subfigure[Ratio of the BIB rate in beam-1 and beam-2 in the unpaired bunches.]{
\includegraphics[width=\columnwidth]{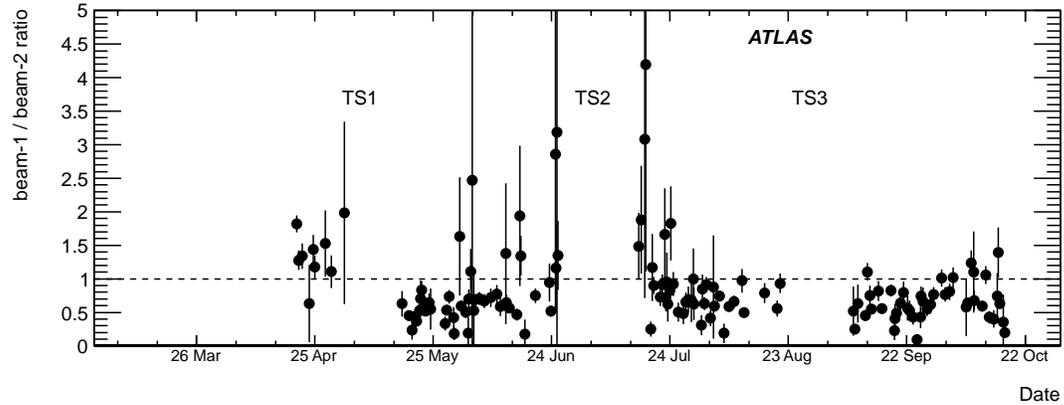}
\label{fig:rate12}
}
\end{center}
\caption{BIB rate in 2011 proton-proton runs.
The rates in filled (a) and unpaired (b) bunches cannot be compared quantitatively because of different trigger requirements.
One entry in the plot corresponds to one LHC fill.
Only the statistical uncertainties are shown.
Technical stops are indicated in the plot.}
\end{figure}

The two-sided method is used to evaluate the rate of BIB
in the whole 2011 collision data set. Figure~\ref{fig:rateFilled} shows the
time evolution of the BIB rate
normalised to the nominal bunch current of $10^{11}$ protons.
The plot shows that the rate was high early in the year 
and then after the first technical stop (TS1) rather rapidly decreased by a 
factor $\sim 3$, staying at a fairly constant level after early June. 
The only exceptions are the first runs after technical stops 2 and 3, where higher 
rates are observed.

In unpaired bunches, the rate is evaluated using the \verb+AND+ combination
of the two-sided and one-sided methods. The former one is chosen in order
to maintain the low mis-identification probability. The latter one helps to remove
the BIB reconstructed from the neighbouring interleaved bunches.\footnote{
Removing the entries from the neighbouring interleaved bunches is important
in particular for evaluating rates for beam-1 and beam-2 separately.
}
Figure~\ref{fig:rateUnpaired} shows the BIB rate in unpaired isolated
and unpaired non-isolated bunches.
As in the filled bunches,
higher rates before May and just after the technical stops are also visible here. 
The rates in filled and unpaired bunches cannot be compared quantitatively
since different triggers were used and no trigger efficiency corrections are applied here.

The data from unpaired isolated and unpaired non-isolated bunches are two
statistically independent samples and the corresponding rates should be in agreement.
The ratio of the measured rates 
for all data after the first technical stop is $0.974\pm0.018$ where only the statistical uncertainties are considered.
Possible explanations for the relative difference are dead time and different
trigger efficiency depending on the relative position of unpaired bunches with respect to
the colliding bunch-trains.

The identification methods also enable the direction of BIB muons to be reconstructed.
This allows the determination of whether the BIB muon originates in beam-1 or beam-2.
Figure\,\ref{fig:rate12} shows the comparison of the BIB rate for beam-1 and beam-2 separately
using the data from unpaired bunches.
Averaged over the entire year the rate in beam-1 is lower than in beam-2 by a factor
of $0.57\pm0.01$, taking only the statistical uncertainty into account.
But it is also evident from Fig.\,\ref{fig:rate12} that the ratio differs from 
fill to fill and the origin of the asymmetry has not yet been identified.
As discussed in the context of LHC collimation, there is no reason to believe that 
the beam halo should be equal for both beams. 
Attempts were
made to correlate the relative rates with beam losses in the cleaning insertions, but
no clear correlations could be found. Most likely other BIB sources, such as variations in
vacuum quality, which can be different for the two beams to some extent, also play a role. 



\section{Removal of non-collision background with jet observables}

The term ``non-collision backgrounds'' refers to the sources of backgrounds that are
not related to the proton-proton collision products. These comprise
BIB, cosmic rays and noise.
This section describes a method to remove non-collision background in physics analyses based on jet observables,
with a special emphasis on BIB. 
A set of jet cleaning cuts, which are commonly used in ATLAS analyses, is introduced first.
It is then shown how non-collision backgrounds can be further reduced and how to estimate their residual levels.
Finally, an example of the monojet signatures search\,\cite{monojet2011} illustrates the
performance of the standard cleaning techniques.

\subsection{Jet cleaning}

The jet selection criteria should effectively reject jets due to background 
processes while keeping high efficiency for jets 
produced in proton-proton collisions.
Since the level and composition of the background depends on the event topology and the jet kinematics, 
several criteria are proposed, corresponding to different levels of fake jet rejection and jet selection efficiency.

\subsubsection{Event samples}
\label{sec:samples}

The selection criteria, based on jet quality, are optimised by
studying event samples enriched in jets from collisions or in fake jets.
Events are classified mainly by the missing transverse momentum significance, defined as
$\Etmiss/\sqrt{\Sigma E_{\rm T}}$, where $\Etmiss$ is the missing transverse momentum\,\cite{Aad:2011re} 
and $\Sigma E_{\rm T}$ is the scalar sum 
of the transverse energies of all energy deposits in the calorimeter. 

\begin{itemize}
\item The collision jet sample requires two jets with
$\ptjet > 20$\,\GeV{} that are back-to-back in the transverse plane ($\Delta \phi_{\rm j-j} > 2.8$)
and have small missing transverse momentum significance $\Etmiss/\sqrt{\Sigma E_{\rm T}} < 2$\,$\GeV^{1/2}$.  
Events are selected by single-jet triggers\,\cite{Aad:2011xs}, where
the threshold is chosen such that the trigger is  
fully efficient ($>$ 99\%) in the considered $\ptjet$-bin.
The selected sample is dominated by dijet events and is called ``sample enriched in collision jets'' in the following.

\item Fake jets are selected from events with only one jet with $\ptjet >$~150\,\GeV, 
large missing transverse momentum $\met>$~150\,GeV and large missing transverse momentum significance $\Etmiss/\sqrt{\Sigma E_{\rm T}} > 3$\,$\GeV^{1/2}$.
The transverse component of the jet momentum is required to be opposite to the missing transverse momentum direction ($\Delta \phi_{\Etmiss -\rm j} > 2.8$).
Events with sub-leading jets with $\ptjet>40$\,\GeV{} or with reconstructed leptons are discarded.
The events are triggered by requiring the presence of a jet and missing transverse momentum. 
The trigger thresholds are chosen to be fully efficient with respect to the selection criteria described above.  
This event sample is dominated by BIB, with a negligible contamination from calorimeter noise 
and physics processes like $Z\rightarrow\nu\nu$+jets and $W\rightarrow \ell \nu$+jets. 
In the following, this event sample is  called ``sample enriched in fake jets''.

\end{itemize}

For both samples, requirements that ensure the quality of beam conditions, detector performance and data processing are imposed. 
After application of these criteria, the total integrated luminosity is about 4.7\,fb$^{-1}$.

\subsubsection{Criteria to remove non-collision background}
\label{sec:JetSelectioncriteria}

\subsubsection*{Beam-induced background and cosmic rays} 

The distribution of energy deposits by the jet, the shower shape and its direction, in particular the pointing to the interaction point, 
can be employed to discriminate collision jets from BIB-induced fake jets. Examples of discriminating variables 
are the electromagnetic energy fraction ($\emf$),
defined as the energy deposited in the electromagnetic calorimeter, divided by the total jet energy,
and ($\fmax$), the maximum energy fraction in any single calorimeter layer.

The vast majority of collision jets contain charged hadrons that are reconstructed by the tracking system. 
In the tracker acceptance, $|\eta|<2.5$, the jet charged particle fraction ($\chf$) is defined
as the ratio of the scalar sum of the \pt{} of the tracks associated with the jet divided by \ptjet{}. This  
is another powerful tool to discriminate collision jets from fake jets, which typically have no associated tracks. 
Finally, BIB and cosmic rays induce jet candidates that are usually not in-time with the collision products.

\subsubsection*{Noise in the calorimeters}
Most of the noise is already identified and rejected by the data quality inspection performed shortly after data-taking, based on standardised quality criteria. 
A small fraction of calorimeter noise remains undetected and needs to be rejected by additional criteria, 
because it can lead to reconstruction of energy deposits not associated with particle interactions in the calorimeter.
As explained in Sect.\,\ref{atlas-sect}, the characteristic pulse shape of real energy deposits in the calorimeter cells can
be used to distinguish a true ionisation signal from noise. This leads to the definition of the quality variables $\hecf$, $\Qmean$, $\LArQuality$ and $\HECQuality$, 
described in Sect.\,\ref{atlas-sect}.

\subsubsection*{Jet quality selections} 
\label{sec:jetquality}
Four sets of jet quality criteria -- ``Looser'', ``Loose'', ``Medium'' and ``Tight'' -- are
defined in order to reject fake jets in 2011 data. These correspond to 
different background rejection factors and jet selection efficiencies.
The selection criteria using jet quality to identify and reject fake jets are listed in 
Table\,\ref{cut_table_looser}.
The Looser criteria are designed to provide a signal efficiency above $99.8 \%$ 
with a fake-jet rejection factor of about $50 \%$, while the Tight criteria
are designed to provide a large fake-jet rejection factor with a signal
inefficiency not larger than a few percent. The two other sets of cuts 
correspond to intermediate rejection factors and selection efficiencies.

\begin{figure}
  \centering
  {\includegraphics[width=0.45\textwidth]{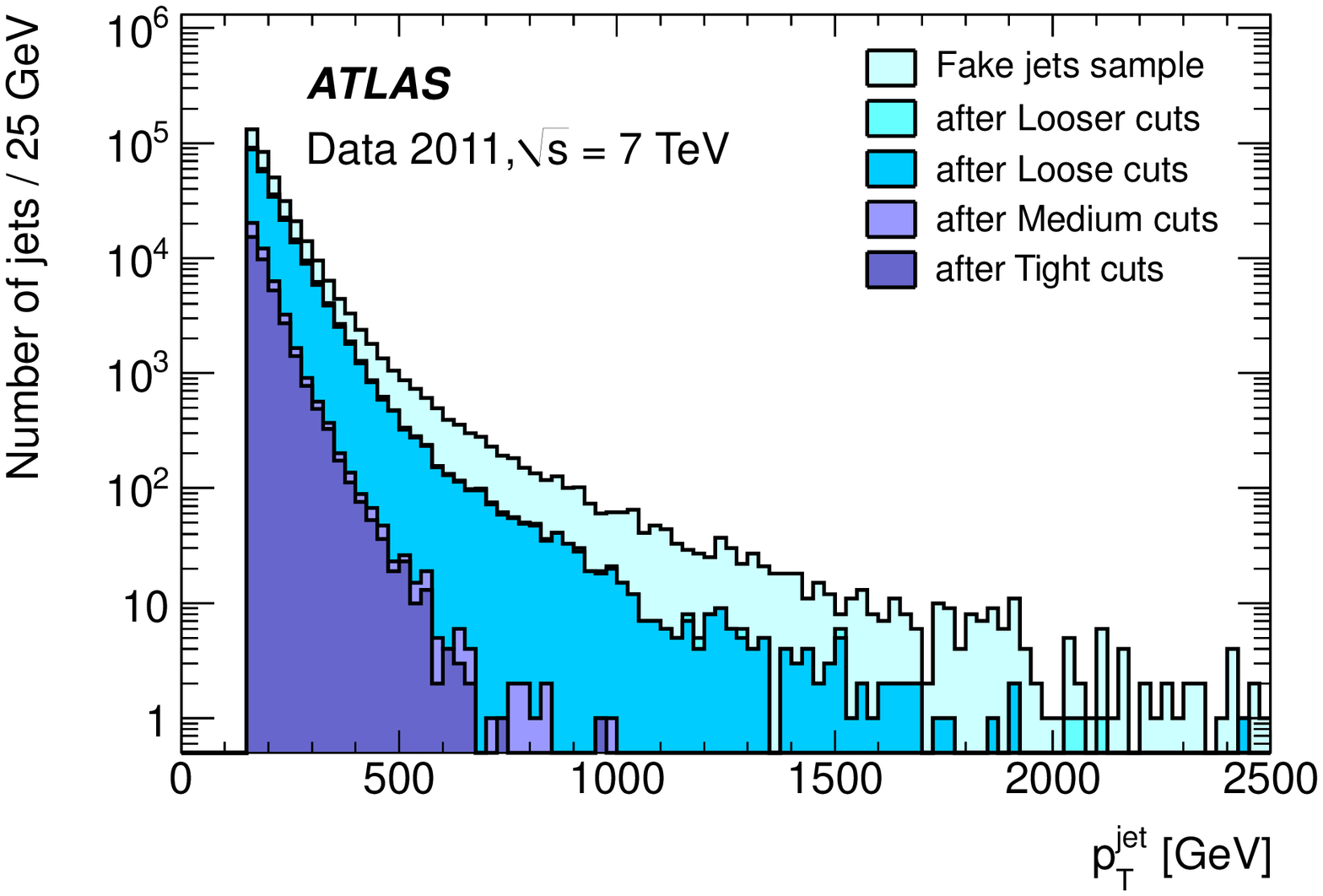}}            
  {\includegraphics[width=0.45\textwidth]{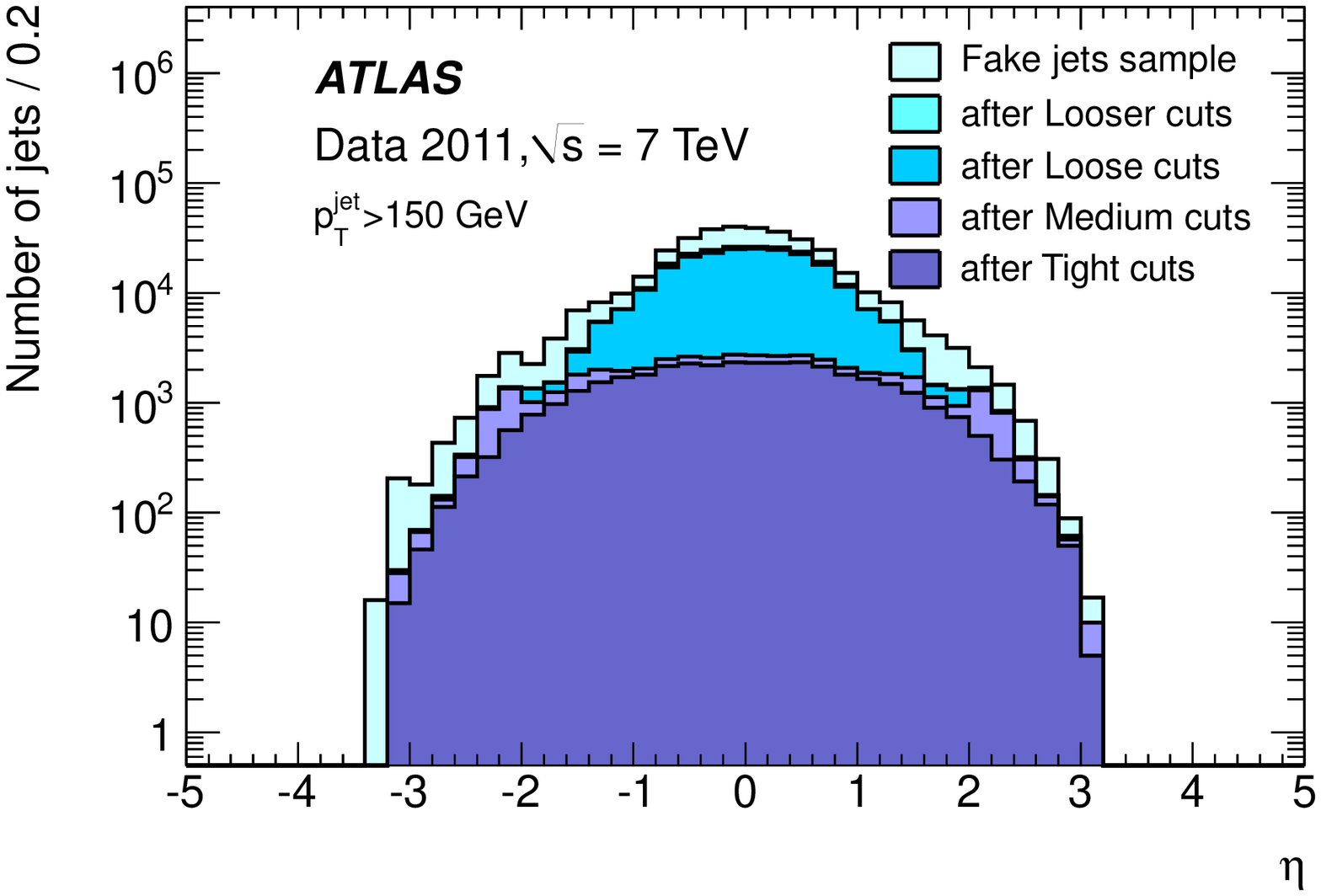}} \\         
  {\includegraphics[width=0.45\textwidth]{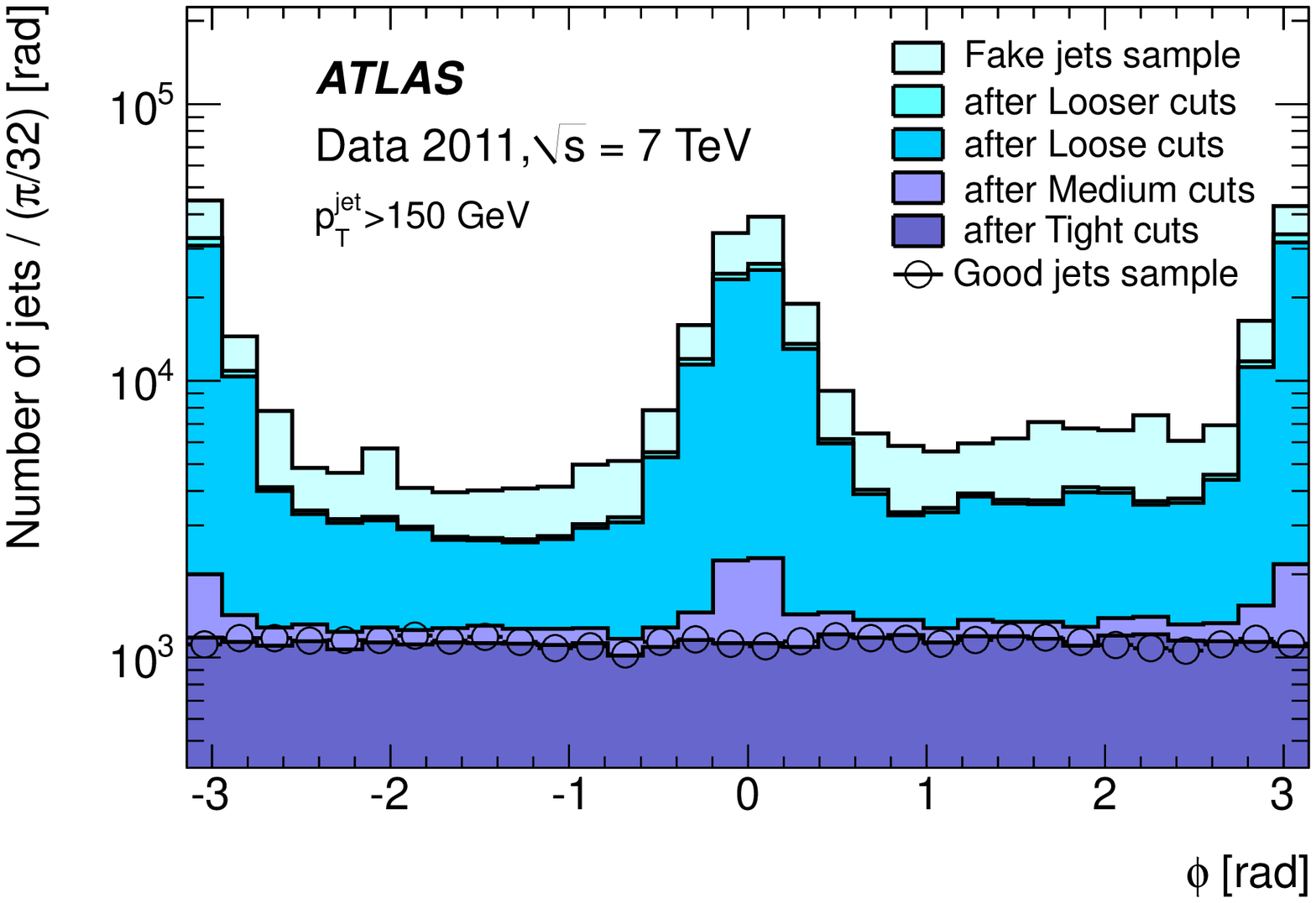}}          
  {\includegraphics[width=0.45\textwidth]{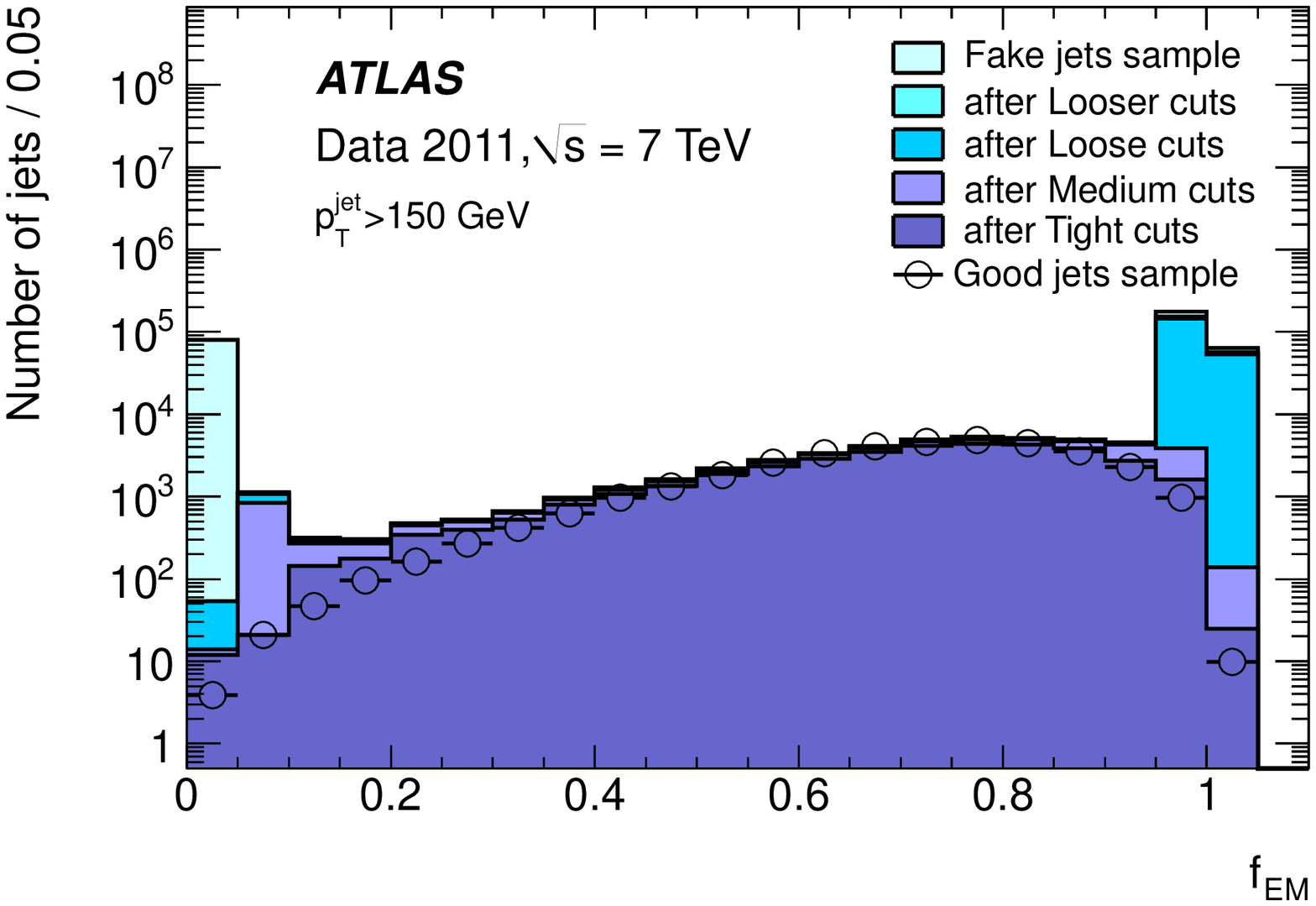}} \\         
  {\includegraphics[width=0.45\textwidth]{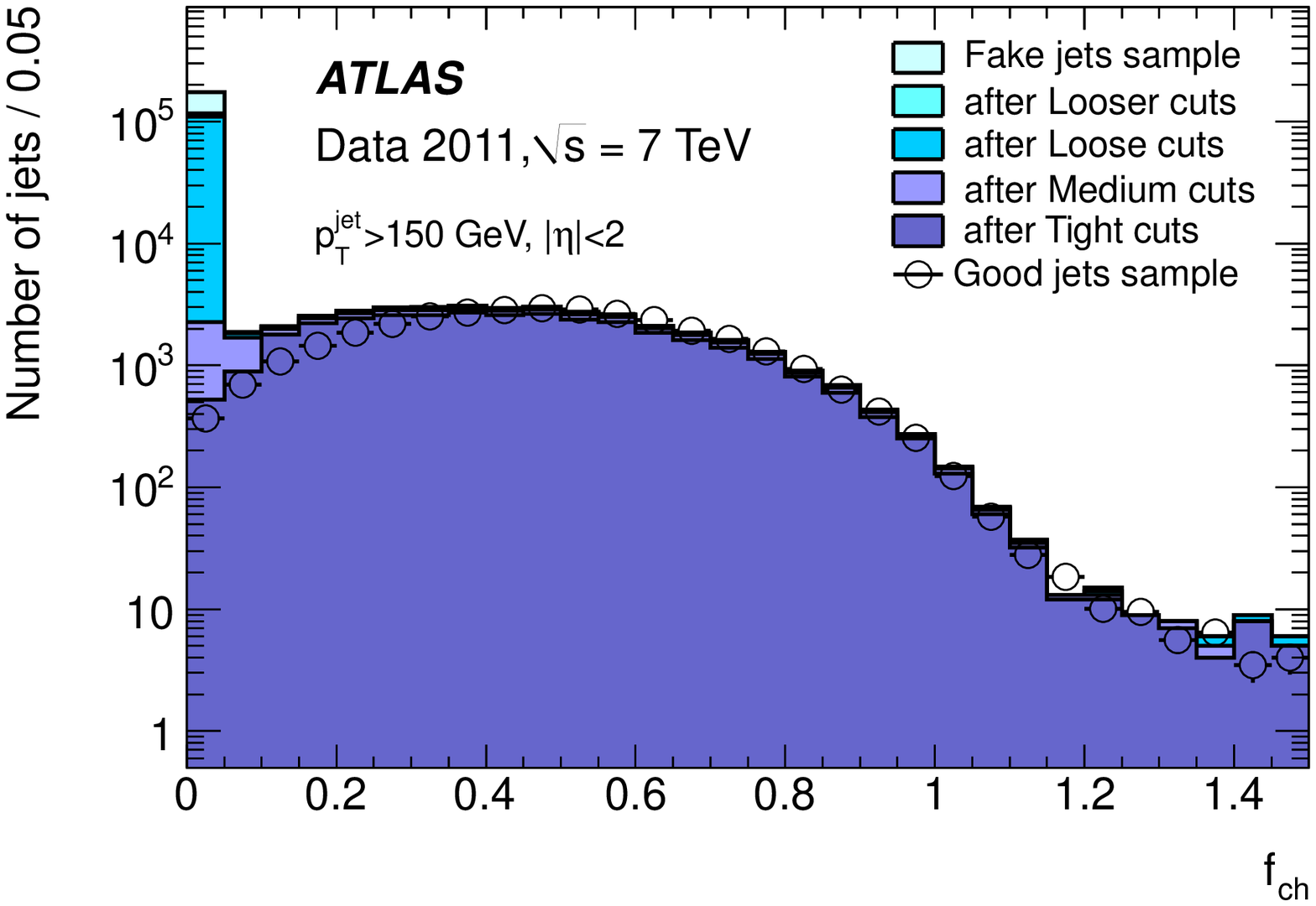}}          
  {\includegraphics[width=0.45\textwidth]{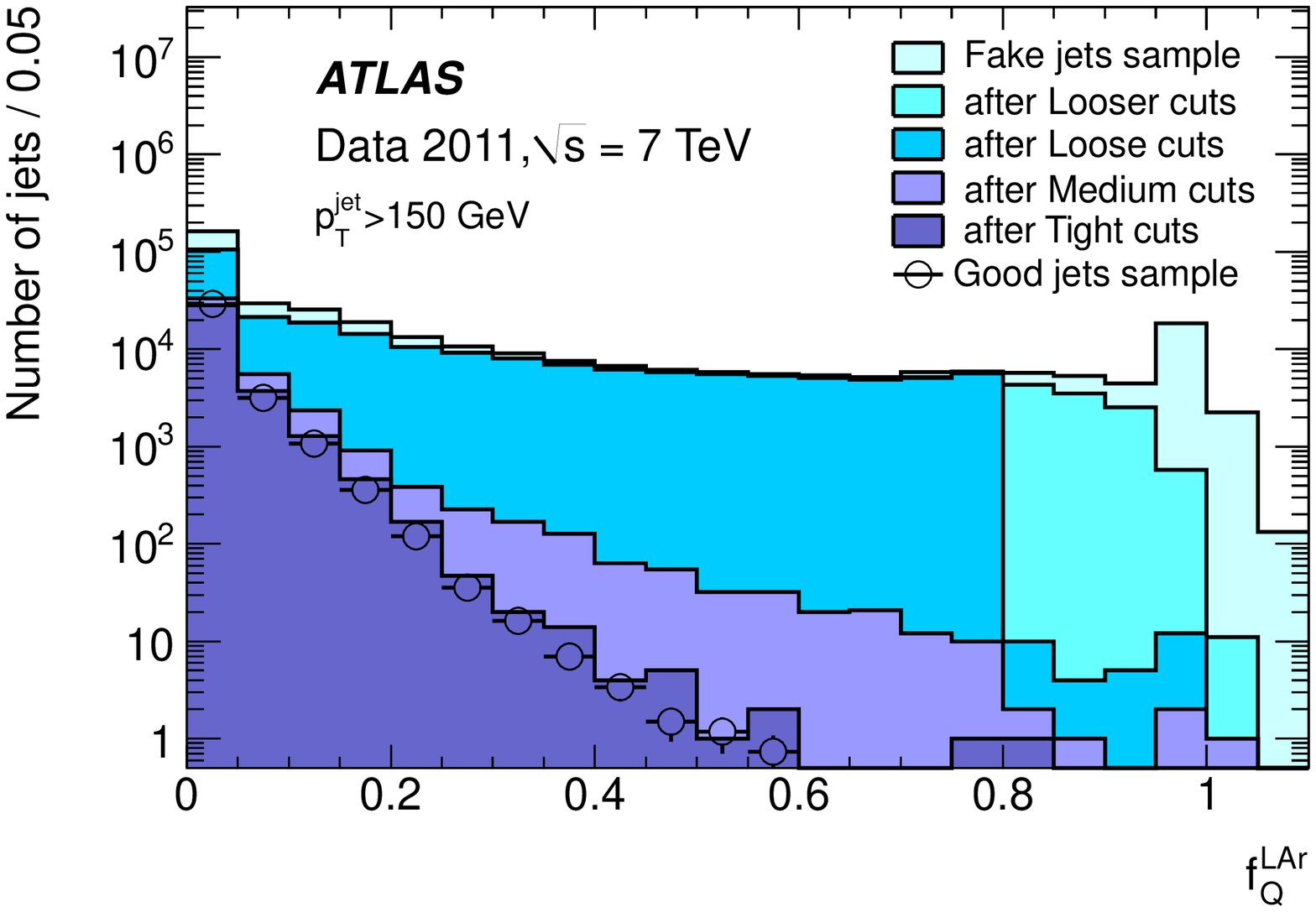}} \\          
  {\includegraphics[width=0.45\textwidth]{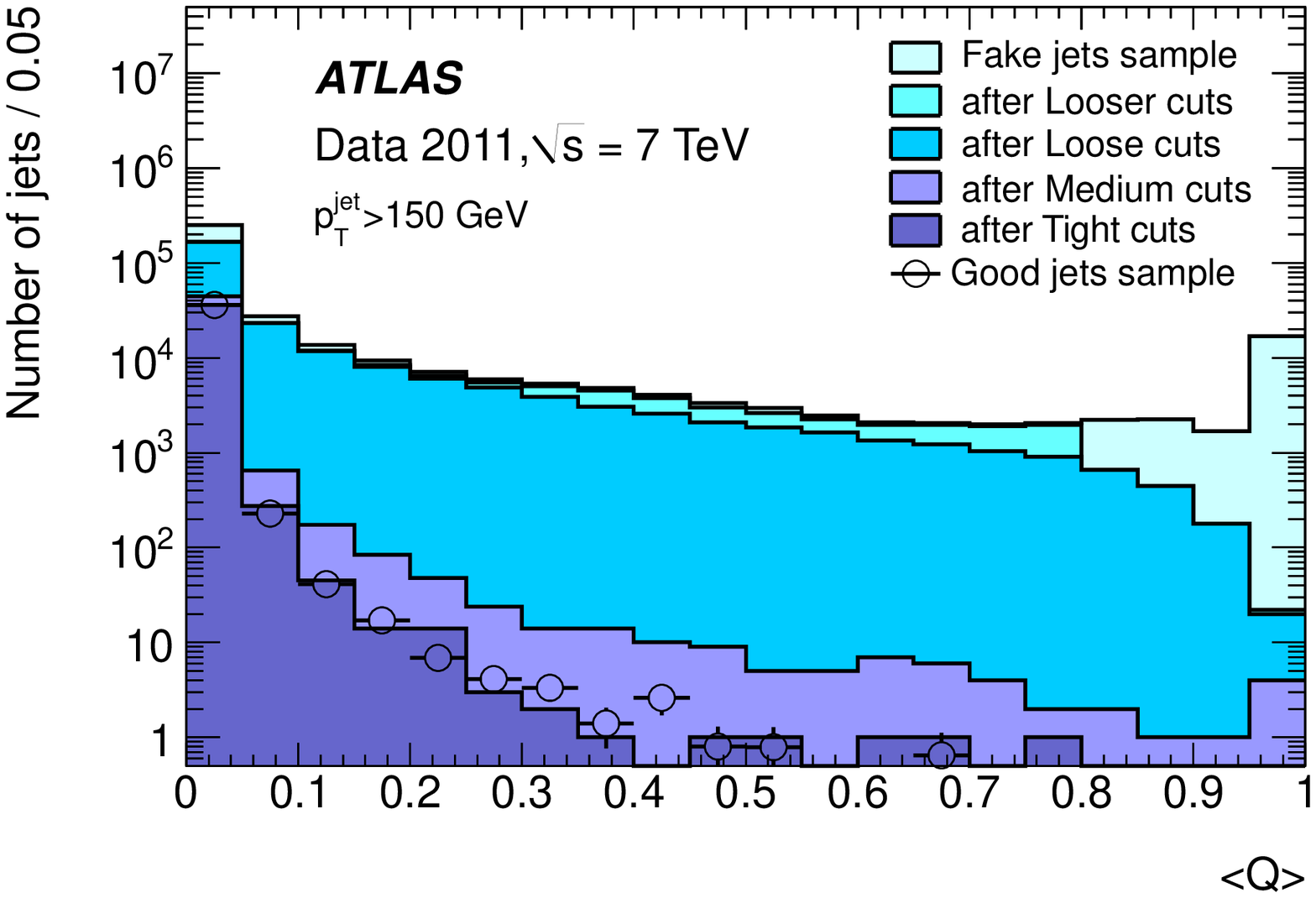}}          
  {\includegraphics[width=0.45\textwidth]{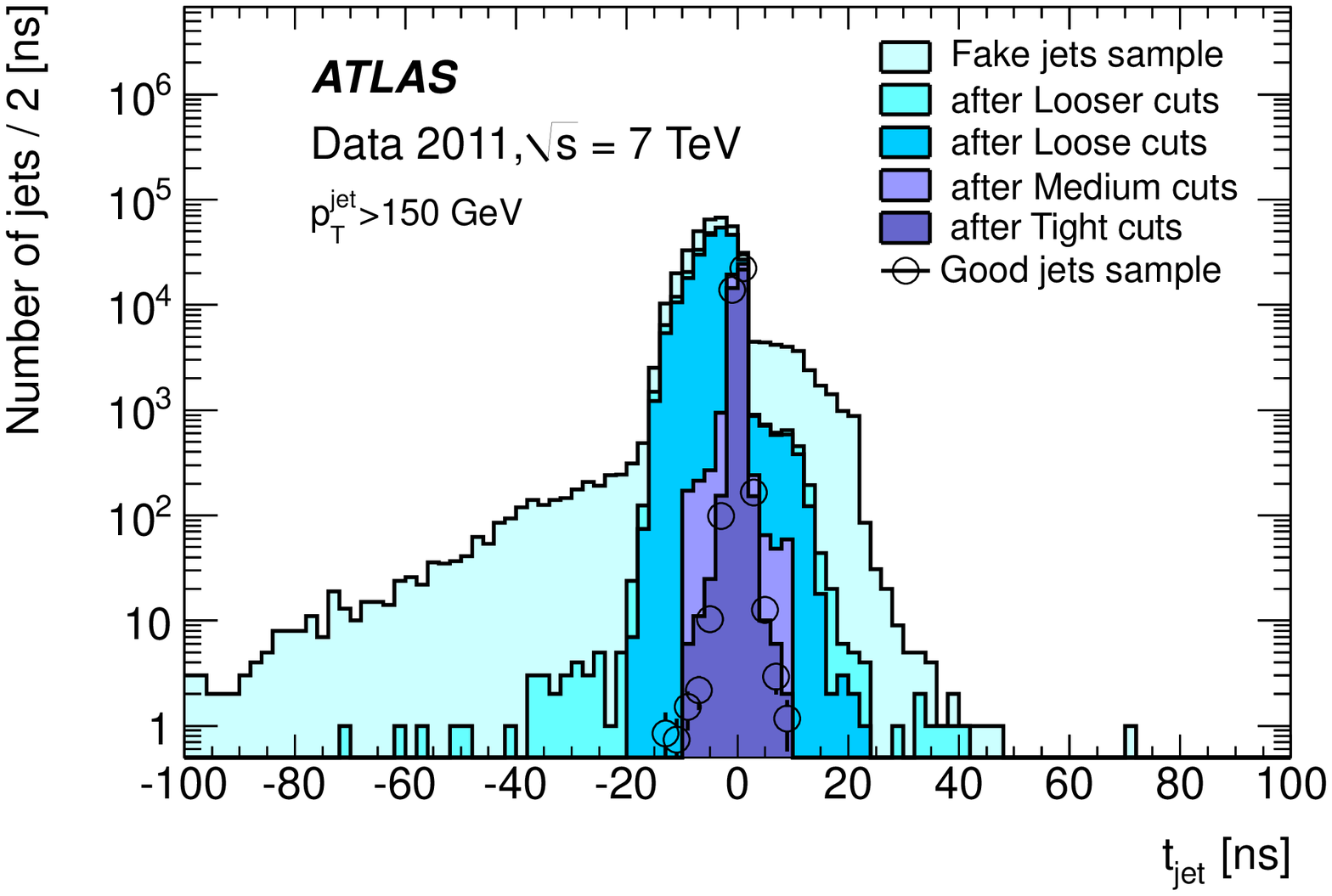}}          

  \caption{Distributions of jet kinematic and discriminating variables for the sample enriched in fake jets before and after applying the jet selection criteria. 
Distributions for the sample enriched in collision jets, labelled as ``good jets sample'' in the figures, are also superimposed where applicable. 
Distributions for jets from collisions are re-weighted in a way to reproduce the two-dimensional jet $\ptjet$ versus jet $\eta$ 
distribution obtained from the sample enriched in fake jets after Tight selection cuts.
}
  \label{fig:jet_var}
\end{figure}

Figure\,\ref{fig:jet_var} shows jet distributions for the sample enriched in fake jets before and after applying the selection criteria listed in 
Table\,\ref{cut_table_looser}.
Distributions from the sample enriched in collision jets are also superimposed where applicable.

As shown before, the two peaks at $\phi=0$ and $\phi=\pi$ are characteristic of BIB and are effectively removed only by the Tight selection criteria. 
The good agreement between the sample enriched in fake jets after the Tight selection criteria and the sample enriched in collision jets shows that 
the fake-jet background contamination is very small once the Tight selection criteria are applied. After this cleaning, the sample enriched in fake jets is dominated 
by physics processes like $Z\rightarrow\nu\nu$+jets and $W\rightarrow \ell \nu$+jets. 

An ``out-of-time'' sub-set of the sample enriched in fake jets is selected by requiring 5$<|\timing|<$10\,ns.
Since this time cut is not used in the fake-jet sample selection,
this sub-sample provides a fake-jet sample that can be used to compute an independent estimate of the fake-jet rejection. 
The timing cut helps to reduce significantly the collision jet contamination in the sample enriched in fake jets (see Fig.\,\ref{fig:jet_var}). 
The Looser criteria reject 37.8\% (68.6\%) of the out-of-time fake jets with \ptjet{}$>$150\,GeV (500\,GeV), while the Tight criteria reject more than 99.9\% of the jets in the out-of-time sub-set of the sample enriched in fake jets. 
The results are summarised in Table\,\ref{tab:rej}.
The valid $\eta$ ranges for all cuts are indicated in Table\,\ref{cut_table_looser}.
Only overall efficiencies, integrated over the whole $\eta$ range, are given here, although variations depending on $\eta$ have to be expected.
The efficiency of the $\chf$ and $\emf$ selection criteria for fake jets is expected 
to be degraded with increasing pile-up compared to the 2011 data studied here,
because the characteristic peaks at $\chf = 0$ and $\emf = 0$ and $1$ become broader.

\begin{table*}[ht!]
\begin{center} 
\begin{tabular}{|p{4.0cm}|p{7.0cm}|}
\hline
              & \textbf{Looser} \\
\hline
BIB and cosmic rays      & ($\fmax > 0.99$ and $\mid \eta \mid < 2 $) \\
   & or ($\emf < 0.05$ and $\chf < 0.05$ and $\mid \eta \mid < 2 $) \\
              & or ($\emf < 0.05$ and $\mid \eta \mid \geq 2 $) \\
\hline
Calorimeter noise      & ($\hecf > 0.5$ and $\mid \HECQuality \mid > 0.5 $ and $\Qmean > 0.8$) \\
             & or $\mid \negE \mid > 60$\,\GeV \\
                  & or ($\emf > 0.95$ and $\LArQuality > 0.8$ and $\Qmean > 0.8$ and $\mid \eta \mid < 2.8 $ ) \\
\hline
\end{tabular}
\\
\begin{tabular}{|p{4.0cm}|p{7.0cm}|}
\hline
              &  \textbf{Loose} \\
\hline
BIB and cosmic rays      & Looser or\\
   & $\mid \timing \mid > 25$ {\rm ns} \\
\hline
Calorimeter noise      & Looser or  \\
             & ($\hecf > 0.5$ and $\mid \HECQuality \mid > 0.5 $)\\
                  & or ($\emf > 0.95$ and $\LArQuality > 0.8$ and $\mid \eta \mid < 2.8 $)  \\
\hline
\end{tabular}
\\
\begin{tabular}{|p{4.0cm}|p{7.0cm}|}
\hline 
           & \textbf{Medium} \\
\hline

BIB and cosmic rays      & Loose or \\
   & $\mid \timing \mid > 10$ {\rm ns} \\
              & or ($\emf < 0.05$ and $\chf < 0.1$ and $\mid \eta \mid < 2 $) \\
              & or ($\emf > 0.95$ and $\chf < 0.05$ and $\mid \eta \mid < 2 $) \\
\hline
Calorimeter noise   & Loose or \\
         & $\hecf > 1- \mid \HECQuality \mid $ \\
              & or ($\emf > 0.9 $ and $\LArQuality>0.8$ and $\mid \eta \mid < 2.8$) \\
\hline
\end{tabular}
\\
\begin{tabular}{|p{4.0cm}|p{7.0cm}|}
\hline 
           & \textbf{Tight} \\
\hline

BIB and cosmic rays      & Medium or\\
   & ($\emf < 0.1$ and $\chf < 0.2$ and $\mid \eta \mid < 2.5 $)  \\
              & or ($\emf > 0.9$ and $\chf < 0.1$ and $\mid \eta \mid < 2.5 $) \\
              & or ($\chf < 0.01$ and $\mid \eta \mid < 2.5 $) \\
              & or ($\emf  < 0.1$ and $\mid \eta \mid \geq 2.5 $)\\
\hline
Calorimeter noise  & Medium or \\
         & $\LArQuality>0.95$ \\
              & or ($\emf > 0.98 $ and  $\LArQuality>0.05$)  \\
\hline
\end{tabular}
\caption{Selection criteria used to identify fake jets. 
They are classified from the loosest to the tightest one: Looser, Loose, Medium and Tight selection criteria.} 
\label{cut_table_looser}
\end{center}
\end{table*}

\begin{table*}[ht!]
\begin{center}
\begin{tabular}{|c|c|c|c|c|c|}
\hline
& Total & Looser & Loose & Medium & Tight\\
\hline 
$\ptjet>$150\,GeV & 124890  &  77675 (37.81\%)&  70226 (43.76\%)  & 663 (99.46\%) & 38 (99.99\%)\\
$\ptjet>$500\,GeV &  2140   &    671 (68.64\%)&  652   (69.53\%)  &  10 (99.53\%) &    0 (100\%)\\
\hline
\end{tabular}
\caption{Number of jet candidates in the out-of-time sub-set of the sample enriched in fake jets before and after applying the jet selection criteria. 
Numbers in parentheses are the fraction of jets identified as fake jets.}
\label{tab:rej}
\end{center}
\end{table*}

\subsubsection{Evaluation of the jet quality selection efficiency}
\label{sec:efficiency}

The efficiency of the jet selection criteria is measured using 
the ``tag-and-probe'' method. 
Collision dijet events are selected as described in Sect.\,\ref{sec:samples}.
The tagging jet ($\pt^{\rm ref}$)
is required to pass the Tight selection criteria, 
and to be back-to-back with the probe jet ($\pt^{\rm probe}$).
The probe-jet sample is used to measure the jet selection efficiency defined as the fraction of probe 
jets selected, as a function of $\eta$ and $\ptjet$ of the probe jets.

The efficiency for the selection of good jets using the Looser criteria is better than
99.8\% over all $\ptjet$ and $\eta$ bins while a slightly lower efficiency is measured for the Loose criteria in particular at low $\ptjet$ and for 2.5$<|\eta|<$3.6. The Medium and Tight selection criteria have lower jet selection efficiency due mainly  to the cuts on the jet charged particle fraction.
For jets with \ptjet{} of about $25$\,\GeV{}, the Medium and Tight criteria have inefficiencies of $4 \%$ and $15 \%$ respectively.
For $\ptjet >$ 50\,\GeV, the Medium and Tight selection criteria have efficiencies better than $99 \%$ and $98 \%$, respectively.

The event selection (using $\Delta \phi_{\rm j-j}$ and $\Etmiss/\sqrt{\Sigma E_{\rm T}}$ cuts) and the Tight selection of the tag jet are varied to study the systematic uncertainties. For the Loose and Looser criteria, the jet selection efficiency is almost unchanged (variations are smaller than 0.05\%) when varying the selection cuts. 
For the Medium (Tight) criteria the size of the variation is at most 0.1\% (0.5\%).

The jet selection efficiency is measured in multijet Monte Carlo samples and compared to the data driven estimates.
Very good agreement is observed for the Looser and Loose criteria. For the Medium (Tight) selection criteria differences not larger than 0.2\% (1\%) are observed for $\ptjet >$ 40\,\GeV. 
Differences at lower $\ptjet$ values are at most 1\% (2\%) for the Medium (Tight) selection criteria. 

\subsection{Monojet analysis}

Events with a single jet balanced by large missing transverse momentum
are often exploited to search for  signatures of new physics. The monojet
analysis searches for new exotic phenomena such as Supersymmetry,
Large Extra Dimensions, an invisibly decaying Higgs boson or Dark Matter candidates.
The analysis is carried out on data from proton-proton collisions at $\sqrt{s}=7\,\textrm{TeV}$
taken in 2011, corresponding to an integrated luminosity of $4.7\,\textrm{fb}^{-1}$.
A detailed description of the analysis can be found in\,\cite{monojet2011}.
Only the BIB rejection methods are discussed here.

The dominant Standard Model physics processes that form an irreducible
background in this analysis are $Z\rightarrow\nu\nu+\textrm{jets}$, where a jet from initial-state radiation
is detected and the two neutrinos create large \MET, and $W\rightarrow \ell\nu+\textrm{jets}$,
where the lepton is out of the acceptance of the detector or badly reconstructed.
Other backgrounds in the analysis, in decreasing order of importance, are
top-quark decays, multijet production, non-collision backgrounds and diboson production ($WW$, $WZ$, $ZZ$).

The events in the monojet analysis signal region are selected by \MET{} triggers
and must have a reconstructed primary vertex.
Furthermore, events with reconstructed leptons are rejected.
A leading jet with $\ptjet>120\,\textrm{GeV}$, accompanied by $\MET>120\,\textrm{GeV}$, is required.
Events with a third jet with $\ptjet$ above $30\,\textrm{GeV}$ are vetoed.
The veto on additional jets is less stringent than in the previous ATLAS monojet search\,\cite{monojet2010}
as it was shown that allowing a second jet in the event reduces systematic uncertainties
from initial- or final-state radiation and increases signal selection efficiencies.
If a second jet exists, the difference between the azimuthal angle of the second jet and \MET{} is 
required to be larger than $0.5\,\textrm{rad}$.
This cut suppresses back-to-back QCD dijet events where one of the jets is mis-measured
resulting in $\MET$ pointing in its direction.
This set of cuts is referred to as the ``monojet selection''.

The monojet selection and the selection for the sample enriched in fake jets,
defined in Sect.\,\ref{sec:samples}, are remarkably similar.
Indeed, it is shown below that the monojet selection tends to select predominantly  non-collision 
background events and therefore the analysis requires efficient cleaning of BIB and cosmic rays.

While most ATLAS physics analyses require only the Looser jet selection criteria
introduced in Table\,\ref{cut_table_looser}, the monojet analysis requires the Medium criteria.
This applies to all jets with $\ptjet>20\,$GeV in an event. 
An additional cut on the leading jet charged particle fraction $\chf>0.02$
and electromagnetic energy fraction $\emf>0.1$ is imposed
in order to obtain even higher rejection power.

\begin{figure}
\begin{center}
\includegraphics[width=0.45\columnwidth]{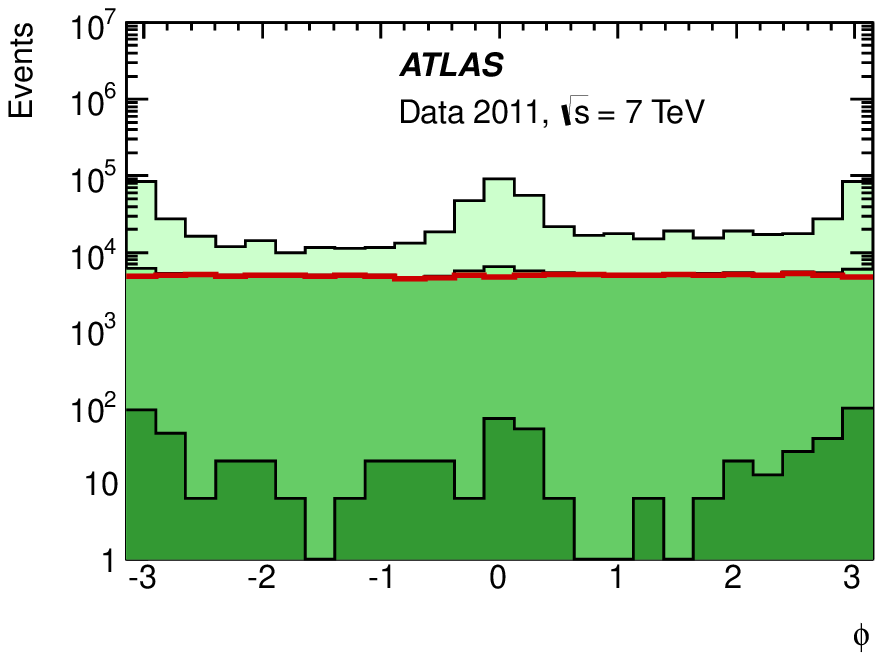}
\includegraphics[width=0.45\columnwidth]{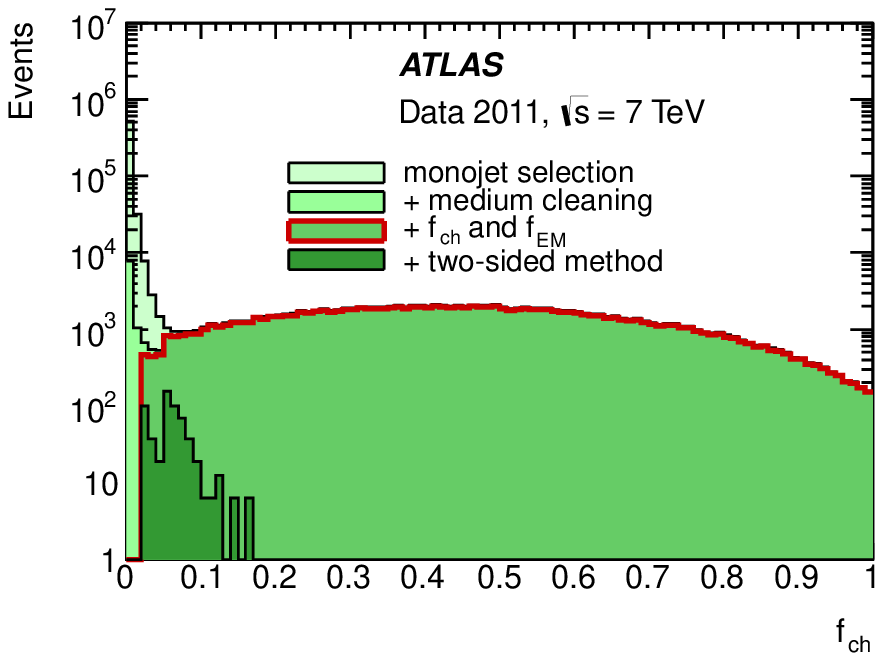}
\end{center}
\caption{Azimuthal distribution (left) and the charged particle fraction (right)
of the leading jet in the monojet analysis signal region
before and after the cleaning cuts.
The monojet analysis signal region events are emphasised by the red line.
The residual level of BIB as estimated by the two-sided method
is also shown.}
\label{fig:monojet}
\end{figure}

Figure\,\ref{fig:monojet} illustrates the importance of the cleaning cuts in the monojet analysis.
The leading jet $\phi$ distribution, before applying any cleaning cuts, clearly shows
the typical azimuthal signature of BIB, as described in Sects.\,\ref{sect-simulation} and \ref{sec:tertiaryGeneral}.
Here, the total number of selected events is $\sim694000$.
The Medium jet cleaning reduces the amount of BIB significantly by removing $\sim560000$ events from the sample, which is $\sim80\%$ of the original sample size.
However, as discussed in Sect.\,\ref{sec:jetquality}, it
identifies $99.5\%$ of the fake jets, which means that a certain residual
amount of BIB after the cleaning is still expected.
Indeed, the contamination from BIB in the remaining $\sim134000$ events
after this cleaning is visible as a slight excess at $\phi=0$ and $\phi=\pi$.
Therefore, even stronger cleaning is needed and the additional cuts on the leading jet charged particle fraction and
electromagnetic energy fraction are applied.
The resulting $\phi$ distribution looks flat which demonstrates
the rejection power of these cleaning cuts.
The flat $\phi$ distribution suggests that the sample is dominated by
physics processes as indicated in Fig.\,\ref{fig:jet1}.
The number of events in the monojet analysis signal region, i.e. after the monojet selection
with all the cleaning cuts, is $124704$
which corresponds to $\sim18\%$ of the size of the original sample without any cleaning applied.
These selected events  correspond mainly to physics processes 
but there may still be a small fraction of BIB events left.

\begin{table}
\begin{center}
\begin{tabular}{|l|l|l|}
\hline
selection & total number of events & non-collision background \\
\hline
\hline
monojet selection & $\sim694000$ & $\sim557000$ \\
after Medium jet cleaning & $\sim134000$ & $\sim7000$ \\
\hline
after $\chf$ and $\emf$ cuts & $124704$ & $575\pm60\textrm{(stat)}\pm57\textrm{(sys)}$ \\
\hline
\end{tabular}
\end{center}
\caption{Number of events in the monojet signal region before and after the cleaning cuts.
Non-collision background levels are also indicated.
The last row corresponds to the monojet analysis signal region where the non-collision background
is the BIB contamination determined using the two-sided method.
In the other two cases, the estimate is based on the jet selection inefficiency evaluated in Monte Carlo simulations.}
\label{tab:monojet}
\end{table}

The dominant Standard Model backgrounds ($Z$ and $W$ boson plus jet production)
are estimated in a data-driven way in dedicated control regions.
Multi-jet backgrounds are also estimated from data,
while the diboson and top-quark backgrounds are obtained from Monte Carlo simulations.
Since the monojet analysis searches for rare events (beyond Standard Model physics),
even the smallest backgrounds need to be  estimated accurately in order to quantify
how many of the events may be due to new phenomena.
The two-sided method, described in Sect.\,\ref{sec:bbim},
is completely independent of the jet cleaning criteria applied in the monojet signal region selection,
and is used to quantify the residual number of BIB
events present in the $124704$ monojet signal region events.
As shown in Table\,\ref{tab:monojet},
the method estimates the BIB level to be $575\pm60\textrm{(stat)}\pm57\textrm{(sys)}$ events.
This residual background is also indicated in Fig.\,\ref{fig:monojet}.
As expected, the distribution of the leading jet charged particle fraction shows that a majority of the
events tagged as BIB  have leading jets without tracks pointing to them.

A Monte Carlo study reveals that the Medium cleaning selection criteria applied to all jets with $\ptjet>20$\,GeV
removes $\sim7\%$ of physics events passing  the monojet selection.
Introducing the additional cuts on the leading jet $\chf$ and $\emf$ reduces the number of physics events in the Monte Carlo samples
by an additional $\sim2\%$.
These estimates for the jet selection inefficiency can be used to determine the number of non-collision background
events in the sample after the monojet selection with and without the Medium cleaning selection criteria applied (see Table\,\ref{tab:monojet}).

The total number of non-collision background events in the data sample just after the monojet selection is $\sim557000$,
which corresponds to $80\%$ of the sample size. 
After applying all the cleaning cuts the number of BIB
events in the sample is reduced to 575, corresponding to a rejection power of $\sim10^3$ for this analysis.

\begin{figure}
\begin{center}
\includegraphics[width=0.45\columnwidth]{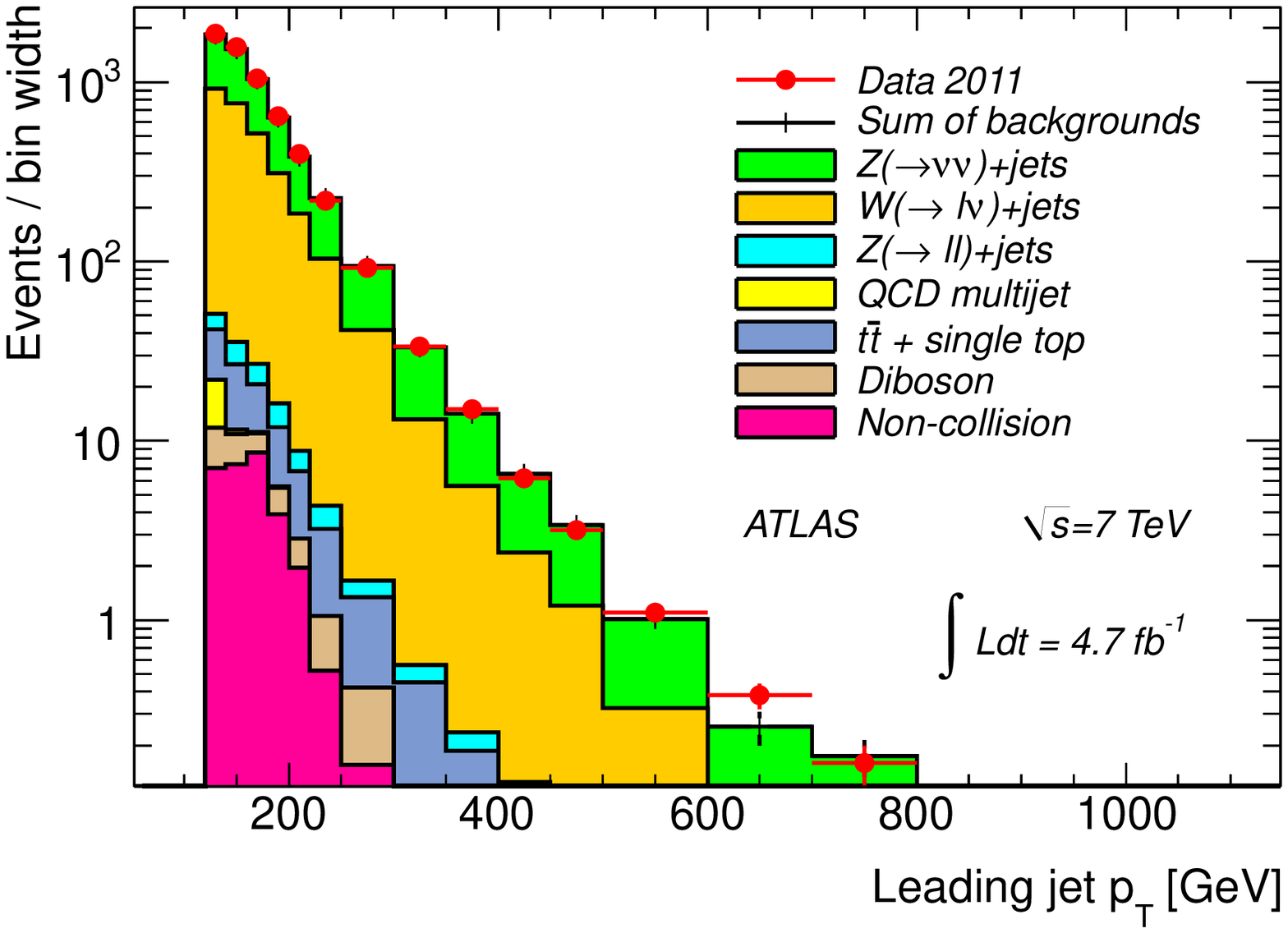}
\includegraphics[width=0.45\columnwidth]{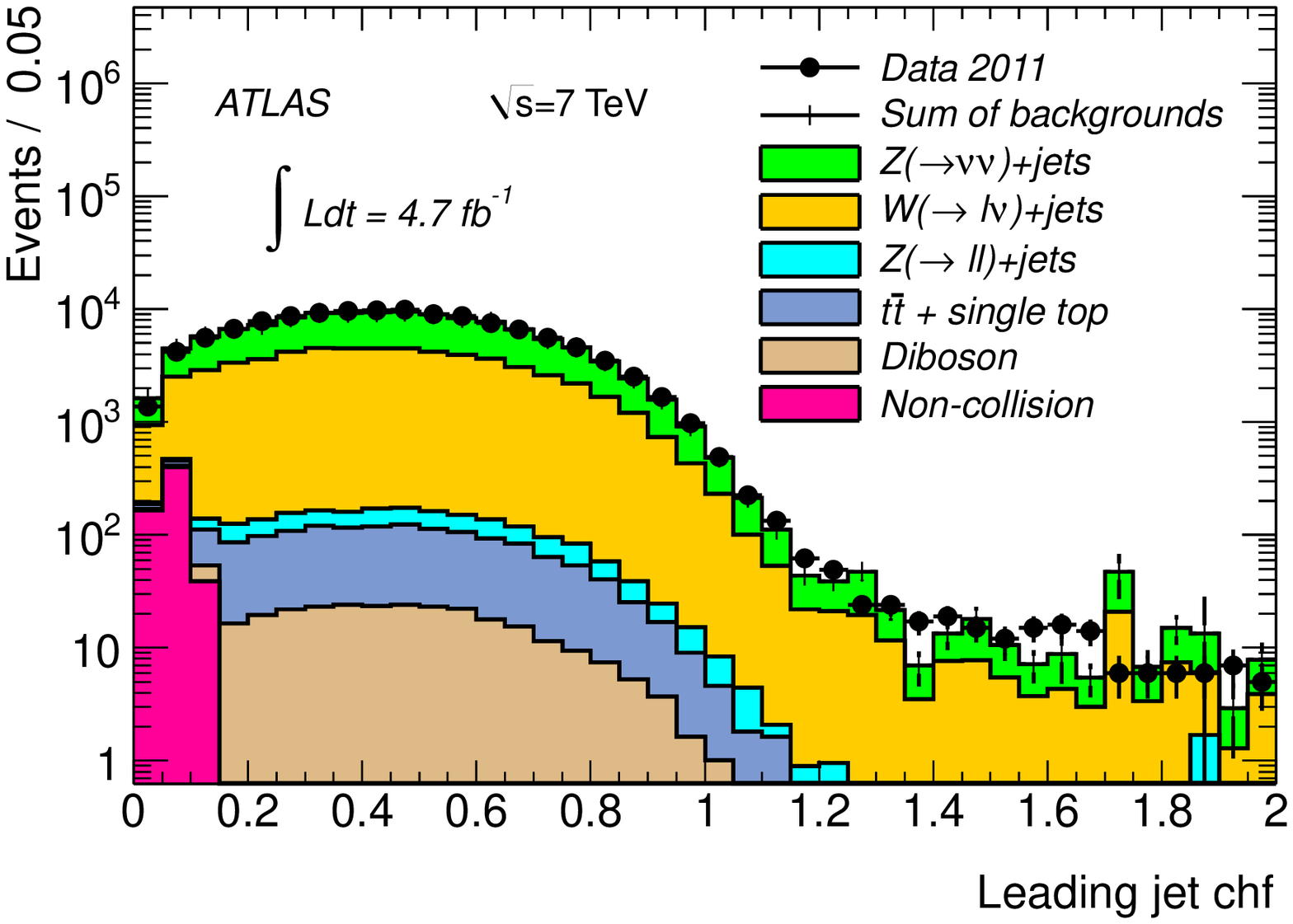}
\end{center}
\caption{Transverse momentum (left) and the charged particle fraction (right)
of the leading jet in the monojet analysis signal region.
The non-collision background is the BIB evaluated by the two-sided method.}
\label{fig:sr}
\end{figure}

\begin{figure}
\begin{center}
\includegraphics[width=\columnwidth]{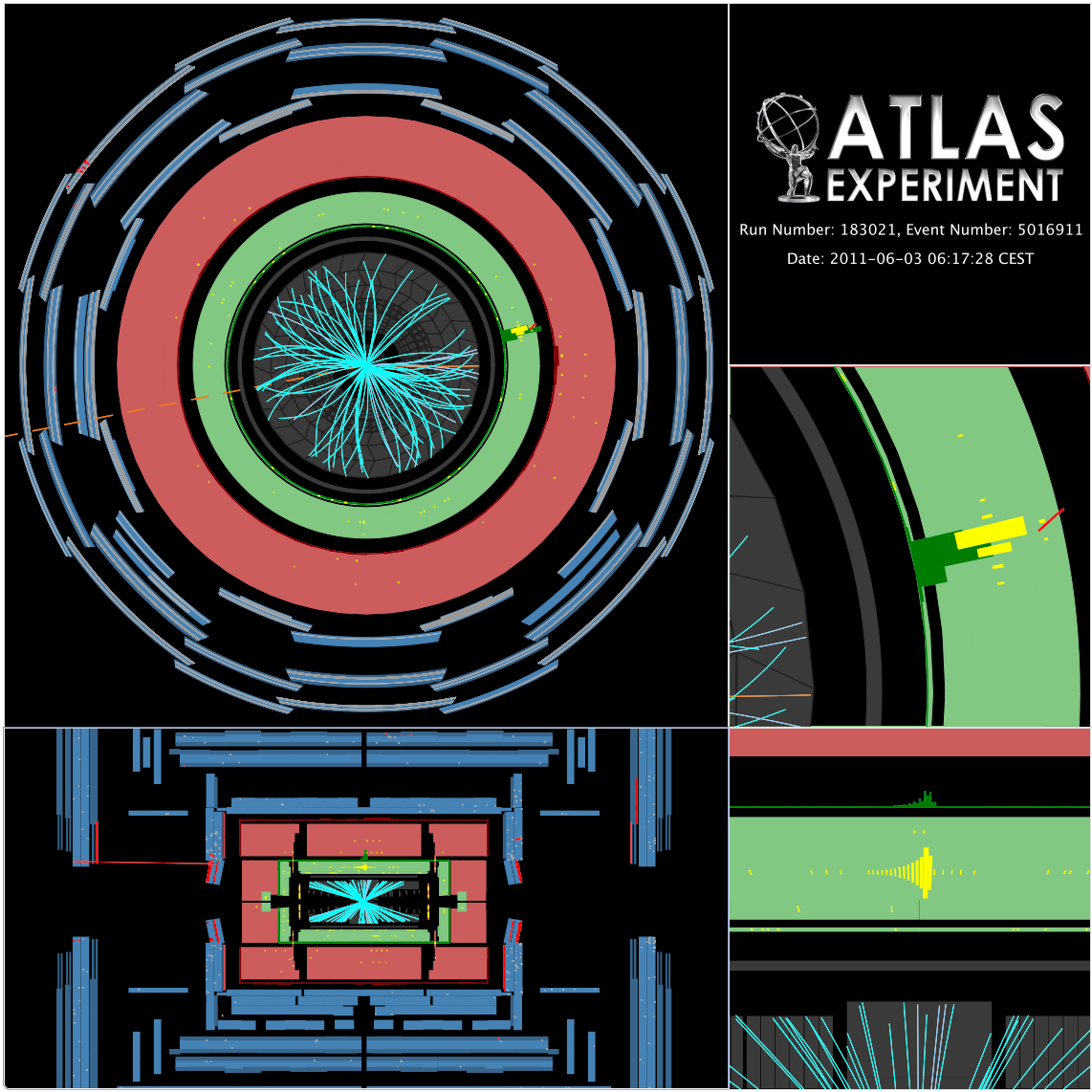}
\end{center}
\caption{Example of an event in the monojet analysis signal region
with a BIB muon entering from the right and causing a fake jet.
In the longitudinal projection (bottom left), CSC chambers with hits (highlighted in red) are seen on both sides.
LAr calorimeter cells (yellow) in-between contain large energy (green towers) that forms a fake jet.
A muon track (red line) parallel to the $z$-axis is reconstructed on side C.
The transverse projection (top left) shows \MET (dashed line) opposite to the fake jet.
The reconstructed tracks (blue) in the inner tracking detector do not point towards the fake jet.
A detailed view (middle right) shows that the calorimeter cells and the muon track are aligned in $\phi$.
Focusing on the LAr energy depositions in the longitudinal projection (bottom right) reveals
that the fake jet consists of a cluster elongated in the $z$ direction.
}
\label{fig:evdncb}
\end{figure}

Finally, Fig.\,\ref{fig:sr} shows the leading jet \pt{} distribution and the leading jet charged particle fraction distribution
for the monojet signal events together with various sources of Standard Model backgrounds.
The residual BIB, which amounts to only $0.5\%$ of the signal region events,
is also illustrated in the figure.
The other events in the monojet signal region sample of $124704$ events
are in agreement with the background estimates for Standard Model processes.
No evidence for physics beyond the Standard Model is found in the 2011 data.
All the fake jets tagged by the two-sided method in this analysis have $\chf<0.2$ and
have a \ptjet{} lower than $300\,\textrm{GeV}$.
These events are typically BIB muons overlaid on top of a minimum bias process.
An example of such a BIB event in the monojet analysis signal region
is shown in Fig.\,\ref{fig:evdncb} where a BIB muon travels in the A$\rightarrow$C
direction leaving hits in the CSC detectors on both sides of ATLAS.
A LAr calorimeter cluster stretched along the $z$-axis is seen in-between, leading to
a fake jet with $\ptjet\sim270\,\textrm{GeV}$ 
with the corresponding missing transverse momentum in the opposite direction.
No collision tracks point towards this jet.
The energy of BIB muons can be up to the TeV level, 
and a few cases have been seen in data where the energy deposition of such a BIB muon 
has been reconstructed as a jet with $\ptjet>1\,\textrm{TeV}$.

A tighter cut on the leading jet charged particle fraction could clearly remove non-collision
background events even further. The two-sided method allows studies of
the efficiency and the mis-identification probability of different cleaning cuts.
Such studies, using Monte Carlo simulation samples, reveal that tighter cleaning cuts
also significantly reduce the signal acceptance, which is not desired in searches.
The set of cleaning cuts used in the monojet analysis is a balance between large background rejection and small physics signal suppression.

Since the efficiency of the jet charged particle fraction cut is expected to decrease with increasing
pile-up, the independent methods of BIB removal described here are expected to become
more important in LHC runs after 2011.

\subsection{Summary of jet cleaning techniques}
The selection method  based on jet observables to remove non-collision backgrounds
is particularly powerful and widely used in ATLAS physics analysis. 
The Looser criteria already provide good background rejection,
while having a negligible loss of efficiency for jets originating from  proton-proton collisions.
The collision-jet selection efficiency is better than 99.8\% for $\ptjet>$~20\,GeV and
its performance is well reproduced by the Monte Carlo simulation.
When larger rejection factors of non-collision backgrounds are needed,
further selection criteria based on the electromagnetic fraction and the charged particle fraction of the jets can be applied. 
Such tighter cleaning cuts have been successfully applied in new physics searches,
for instance the monojet signatures search.
There, the topology of the signal region events is similar to the signatures of jets due to non-collision backgrounds,
and it has been shown that $\sim80\%$ of the selected data come from non-collision backgrounds if no cleaning cuts are applied.
Dedicated cleaning reduces the non-collision background contamination to $0.5\%$,
where the estimate of the residual BIB level is carried out using the methods
described in Sect.\,\ref{sec:bbim}, which are independent of the cleaning cuts.


\section{Conclusions}

During the 2011 proton run the LHC delivered more than 5\,fb$^{-1}$ of luminosity, of
which about 4.7\,fb$^{-1}$ is usable for physics analyses. The number of colliding
bunches increased during the year from a few hundred to 1331. Each physics fill of the LHC also contained
on the order of 50 unpaired, i.e. non-colliding, 
bunches to monitor the beam induced background (BIB).
The events in those unpaired bunches were triggered by 
dedicated algorithms and stored in a special background stream at a rate of a few Hz.

Due to the large event rate, the Level-1 trigger rates before prescaling allowed detailed monitoring of
backgrounds, while the recorded events formed the basis for developing dedicated background
tagging tools to be used in physics analyses.

The main detector used for beam-gas monitoring in ATLAS is the Beam Conditions Monitor (BCM), located very close to
the beam-line. A special background-like trigger was implemented for BCM hits, which selected
only events with an early hit on one side and an in-time hit on the other side of the 
ATLAS Interaction Point (IP). The rates of this trigger are shown to correlate very well
with residual gas pressure close to the experiment (the pressure measured at $|z|=22$\,m), but
 have much less sensitivity to beam losses further away, i.e. pressure at $z=58$\,m.

The BCM also provides a collision-like trigger, i.e. an in-time coincidence on both 
sides of the IP. The rates from this trigger are used to study the ghost charge distribution
by looking for collisions of protons in unpaired bunches with protons in nominally empty bunches. 
These studies reveal that non-negligible ghost charge can extend as far as 150\,ns from the 
filled bunches.
This result is supported by similar findings for the Level-1 J10 (jet with $\pt>10$\,GeV)
trigger rates.

In order to gain a deeper understanding of BIB sources and formation, dedicated
simulations have been performed\,\cite{roderick-in-prep}. The main results of these
simulations are presented in this paper and characteristic features of
the BIB, such as radial and azimuthal distributions, are shown. Some of these characteristic features have been
observed in 2011 data as well.

The various ATLAS sub-detectors allow accurate studies of the BIB to be performed. 
A particularly well-suited detector for studying BIB at small radii is the ATLAS Pixel detector. 
Since the Pixel barrel is coaxial with the beam-line and BIB tracks are predominantly
parallel to the beam, a characteristic feature of BIB events in the Pixel detector is
the presence of elongated clusters with large total charge deposition. This feature
has been used to develop an algorithm for tagging BIB events. Comparison of
data and simulations indicates very good agreement for both collisions and BIB.
The tagging tool has been used to produce background samples that show that
the BIB rate in the Pixel detector correlates very well with the residual pressure
at $|z|=22$\,m. This shows that the Pixel detector, like the BCM, is
sensitive mainly to beam-gas events close to the detector. However, the background data also show
a slight $\phi$-asymmetry. The BIB simulations suggest that such an asymmetry is created
by bending in the magnets of the inner triplet and beyond. Thus a fraction of the 
background seen by the Pixel detector seems to originate from a larger distance.

The main impact of BIB on physics analyses is the production of fake jets due to radiative energy losses
of high-energy muons passing through the calorimeters. This affects mainly the analyses relying on large \MET{} signatures.  
The simulations indicate that such
muons have to originate far from the detector ($>100$\,m) in order to reach the calorimeter radii. In 
addition they are predicted to show a very pronounced $\phi$-asymmetry with a strong preference
for muons to be in the horizontal plane.
Such an asymmetry is clearly seen in the distribution of fake-jet candidates. A special tool,
based on identifying the incoming/outgoing muon in the CSC and the inner MDT endcap muon chambers, has been
developed to remove such events from physics analysis. This tool comprises several algorithms,
yielding different efficiency and mis-identification probabilities. In addition to using
the position and direction information from the muon detectors (both polar and azimuthal) it also uses
timing information of both the muon detectors and the calorimeter.

The standard jet  cleaning algorithms used in ATLAS to reject non-collision background events have been summarised and
their application in the monojet signatures search has been
presented. It has been shown that after the jet cleaning criteria are applied, the event sample still contains 
BIB events, which are identified by a special analysis tool and 
taken into account in the background estimates in this analysis.
Without this dedicated cleaning, BIB events would represent a serious background for some searches for new phenomena.

\clearpage

\appendix
\section{Alternative methods for BIB identification in the calorimeters}

This appendix outlines two alternative methods 
for BIB identification in the calorimeters in addition to those described in Sect.\,\ref{sec:bbim}.
The first method uses the time signature of energy depositions due to BIB in the TileCal.
The other checks the shape and orientation of a calorimeter cluster in order to differentiate between
BIB and collision products.
Both methods are presently under study.

\subsection{Beam background signatures in the Tile calorimeter}

\begin{figure}
\begin{center}
\includegraphics[width=0.3\columnwidth]{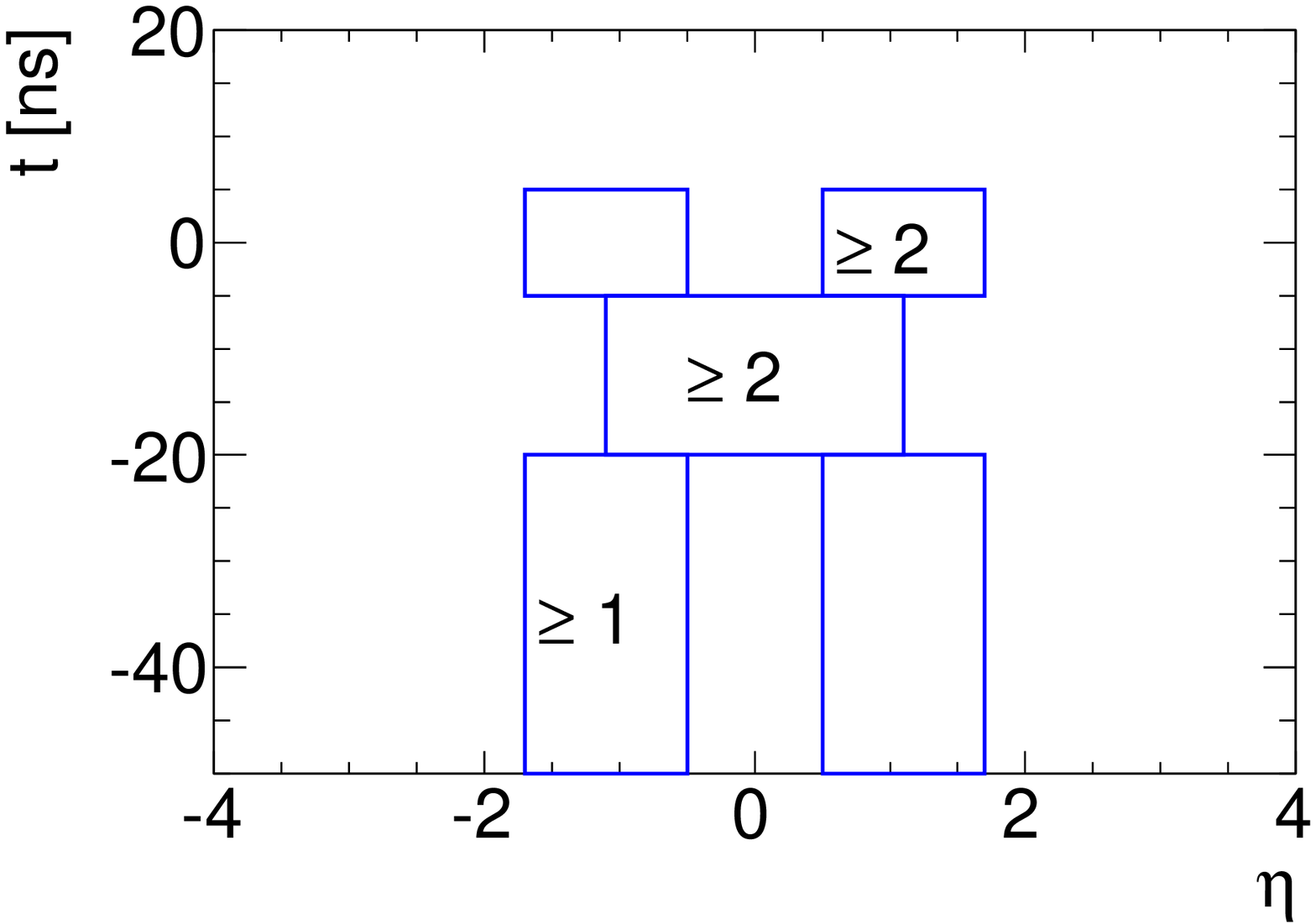}
\includegraphics[width=0.3\columnwidth]{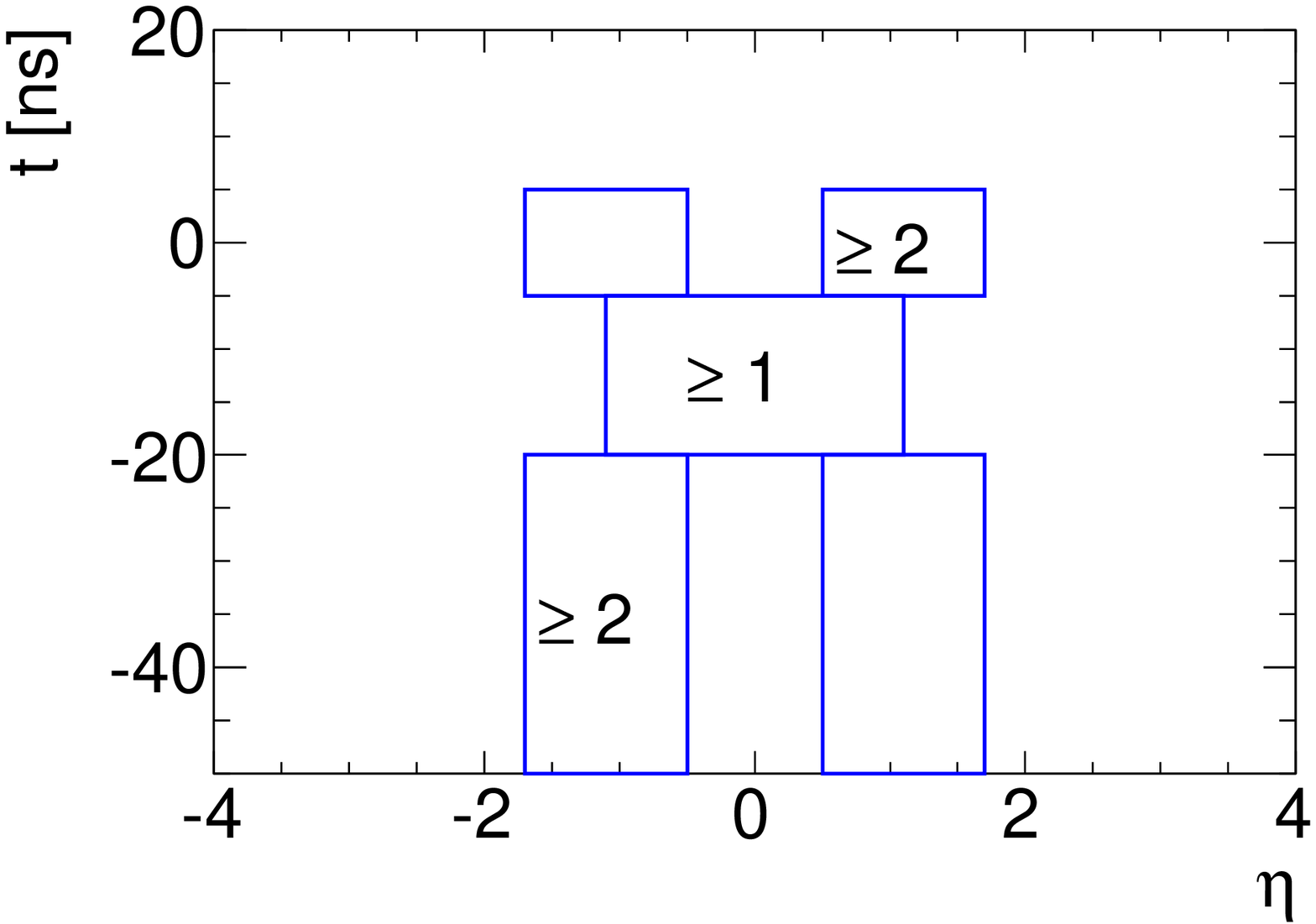}
\includegraphics[width=0.3\columnwidth]{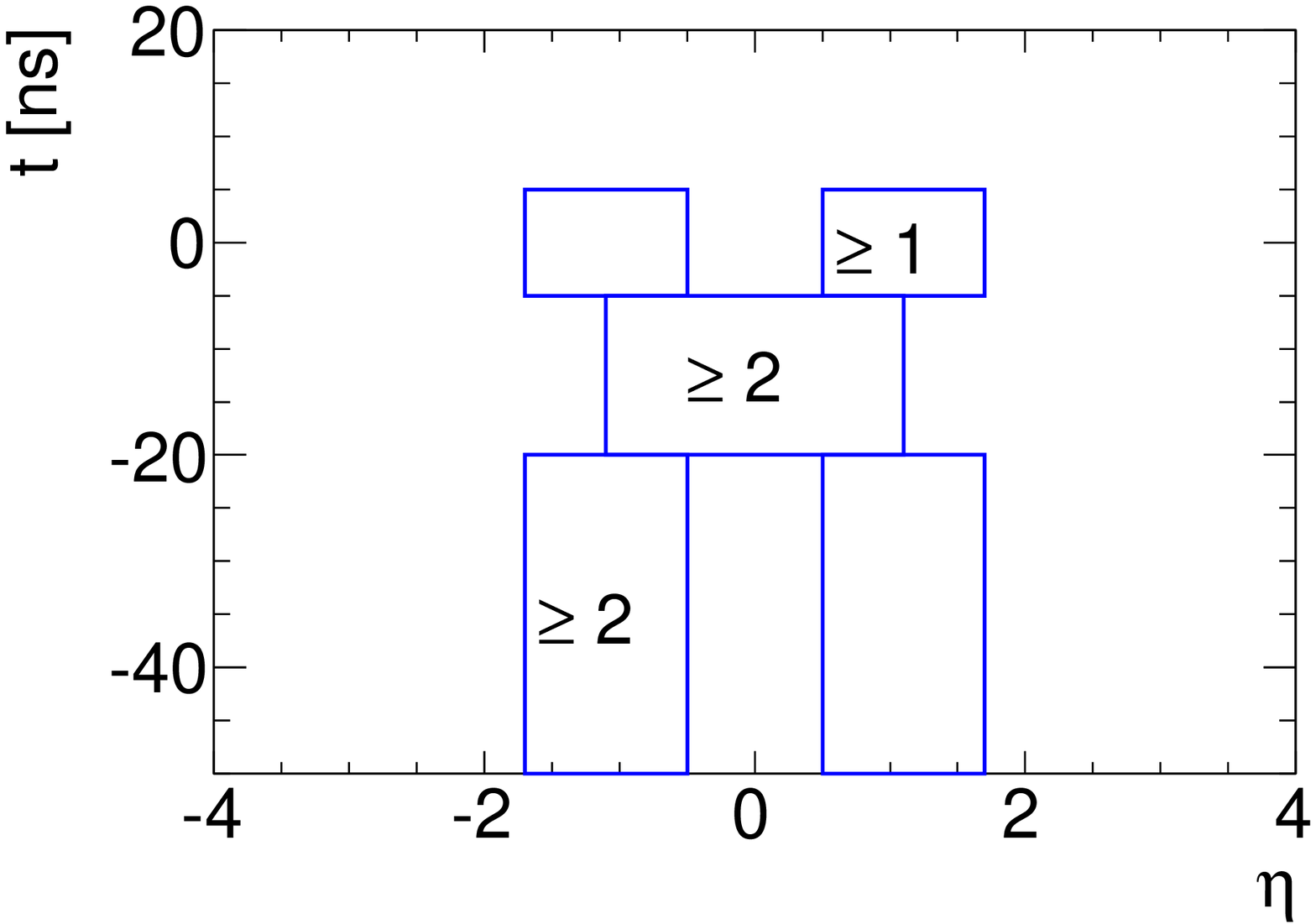}
\end{center}
\caption{Selection criteria for the TileCal muon filter.
  The blue rectangles correspond to the $\eta$--$t$ regions used by the TileCal muon filter to select the events.
The numbers correspond to the minimum number of selected clusters required in each region.
}
\label{fig:regions}
\end{figure}

\begin{figure}
  \center
  \includegraphics[width=0.4\textwidth]{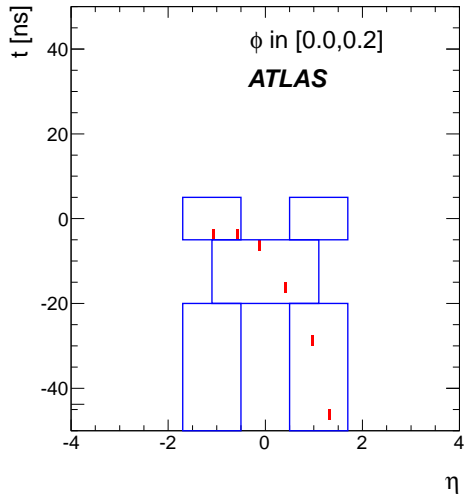}
  \caption{Example of an event selected by the TileCal muon filter in unpaired bunches.
The clusters  are shown in red marks and belongs to the same $\phi$ slice.}
  \label{fig:filterRegionsNonCol}
\end{figure}

The  timing measurements with 1\,ns resolution, and spatial information for the measured energy depositions of the TileCal provide an analysis tool capable of reconstructing muons
which may originate from BIB.
A possible signature of such muons is a series of aligned energy depositions parallel to the beam direction,
starting on one side of the TileCal and propagating to the other.
The time measurement of the energy deposits has to be consistent with the hypothesis of
a particle travelling parallel to the beam direction at the speed of light,
while having roughly the same azimuthal angle ($\phi$) in the detector.
The  criteria used to identify such distinct patterns are the following:
\begin{itemize}
\item Select calorimeter clusters with a fraction of energy in the TileCal of at least 90\%.
\item The TileCal is divided in 64 overlapping slices in $\phi$, such that the $n^{\rm th}$ slice covers
$\phi \in [n\frac{\pi}{32} , n\frac{\pi}{32}+\frac{\pi}{16}]$.
The width of each $\phi$ slice, $\frac{\pi}{16}$, corresponds to two consecutive TileCal modules,
which define the angular resolution of the TileCal in $\phi$.
In each $\phi$ slice, clusters are selected if their pseudorapidity and time measurement are compatible with the hypothesis 
of a particle travelling parallel to the beam axis. 

\item  A minimum number of selected clusters is required to tag a BIB muon candidate
and it is required that they have a specific pattern in $\eta$--$t$.
Figure\,\ref{fig:regions} illustrates the $\eta$--$t$ regions that are defined to tag BIB muons
and shows the minimum cluster multiplicity requirement in each of the regions along the muon path.
No additional selection criterion is applied, i.e. these $\eta$--$t$ regions 
have no segmentation in the radial direction.
Three sets of cuts on the number of clusters in each region are defined in the figure.
The illustrated selection criteria apply to muon background travelling in the C$\rightarrow$A direction.
For the A$\rightarrow$C direction, 
the diagonal of the $\eta$--$t$ regions is reversed.
\end{itemize}
An example of a BIB event tagged in an unpaired bunch is shown in Fig.\,\ref{fig:filterRegionsNonCol}.  

The efficiency of the selection criteria is evaluated in the data from unpaired bunches,
requiring exactly one jet with $\ptjet>30$\,GeV, $|\eta|<1.5$. A timing cut of $t < -5$~ns is applied
in order to reduce the contamination from ghost collisions in the unpaired bunch data, and
90\% of the jet energy should belong to TileCal channels in order to ensure that the BIB particle deposits its energy in the TileCal.
Since collision data samples are always contaminated by BIB events, the mis-identification rate of the TileCal muon filter is estimated with a multijet Monte Carlo sample that reproduces the pile-up conditions of the data. 
The BIB-enriched and Monte Carlo samples described above are composed of 2101 and   $1.4\cdot 10^{6}$ events, respectively.
The efficiency for the selection criteria depicted in Fig.\,\ref{fig:regions} is about 12\%, and the mis-identification rate
is about $2\cdot10^{-4}$.
By requiring only one hit in each of the relevant $\eta$--$t$ regions, the efficiency is higher by about a factor three and the mis-identification rate increases by more than two orders of magnitude.
If the minimum number of required hits in the regions is increased to two,
the mis-identification rate drops to about $10^{-6}$ and the selection efficiency decreases to only about 1\%.
The relatively low efficiency can be explained by the fact that the selection criteria require the muon to cross the calorimeter completely from side to side. 
Muons that enter on one side and are stopped inside the calorimeter are not tagged, but they contribute to the inefficiency of the method.

\subsection{Cluster shape}

\begin{figure}
\begin{center}
\includegraphics[width=0.6\columnwidth]{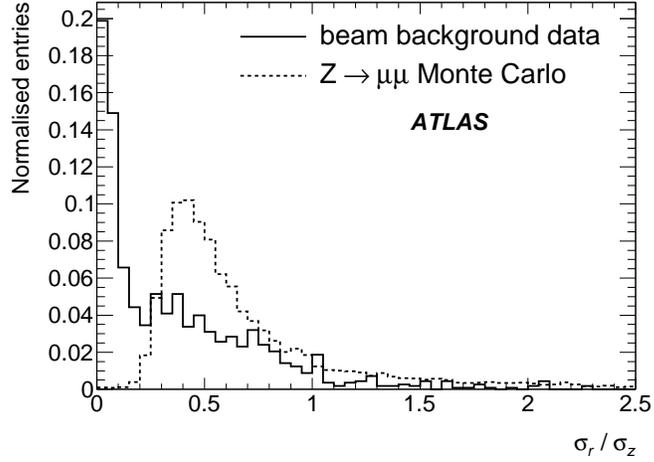}
\end{center}
\caption{Ratio of the standard deviation of $r$ and $z$ position of the cells contained within a cluster
in unpaired bunches (solid) and simulated collision events (dashed).
Taken from\,\cite{conf-note-2010}.}
\label{fig:cellrz}
\end{figure}

Because BIB muons travel parallel to the beam-pipe, their cluster shapes in the calorimeter
are different from those generated by collisions.
The particle shower develops mainly along the $z$ direction for BIB,
whereas for collisions it develops in the direction from the interaction point.
In order to distinguish between BIB and collision products based on the cluster shape,
one can compare the standard deviations of the $z$ and $r$ positions
of the cells contained within a cluster.
The ratio
\begin{equation}
\frac{\sigma_r}{\sigma_z} = \frac
{\sum (r_\textrm{\scriptsize cell}-r_\textrm{\scriptsize clus})^2}
{\sum (z_\textrm{\scriptsize cell}-z_\textrm{\scriptsize clus})^2}
\end{equation}
is defined,
where $z_\textrm{\scriptsize cell}$, $r_\textrm{\scriptsize cell}$ and $z_\textrm{\scriptsize clus}$, $r_\textrm{\scriptsize clus}$ are the positions
of cells and clusters, respectively.
Only the cells with a well-measured time and an energy deposition
above $100\,\textrm{MeV}$ are considered in the sum in order to suppress noise.
Figure\,\ref{fig:cellrz} compares the ratio of standard deviations in $z$ and $r$, $\sigma_r/\sigma_z$,
for BIB muon data from unpaired bunches with simulated collision data.
The distributions motivate a cut on $\sigma_r/\sigma_z<0.15$ to select BIB muons.
It can be seen that some of the clusters from collisions also satisfy the cut on $\sigma_r/\sigma_z<0.15$.
Given the large number of clusters per event, this leads to non-negligible mis-tagging rates.
In order to reduce mis-identification of collisions due to this fact,
selection criteria based on other quantities need to be applied as well.

\clearpage


\section*{Acknowledgements}

We thank CERN for the very successful operation of the LHC, as well as the
support staff from our institutions without whom ATLAS could not be
operated efficiently.

We thank, especially, G. Bregliozzi and G. Lanza from the LHC Vacuum group
for providing us with pressure maps and explanations. We are also grateful
for the help of the FLUKA team at CERN, who contributed to the geometry
models of the LHC interaction region simulations.

We acknowledge the support of ANPCyT, Argentina; YerPhI, Armenia; ARC,
Australia; BMWF and FWF, Austria; ANAS, Azerbaijan; SSTC, Belarus; CNPq and FAPESP,
Brazil; NSERC, NRC and CFI, Canada; CERN; CONICYT, Chile; CAS, MOST and NSFC,
China; COLCIENCIAS, Colombia; MSMT CR, MPO CR and VSC CR, Czech Republic;
DNRF, DNSRC and Lundbeck Foundation, Denmark; EPLANET, ERC and NSRF, European Union;
IN2P3-CNRS, CEA-DSM/IRFU, France; GNSF, Georgia; BMBF, DFG, HGF, MPG and AvH
Foundation, Germany; GSRT and NSRF, Greece; ISF, MINERVA, GIF, DIP and Benoziyo Center,
Israel; INFN, Italy; MEXT and JSPS, Japan; CNRST, Morocco; FOM and NWO,
Netherlands; BRF and RCN, Norway; MNiSW, Poland; GRICES and FCT, Portugal; MERYS
(MECTS), Romania; MES of Russia and ROSATOM, Russian Federation; JINR; MSTD,
Serbia; MSSR, Slovakia; ARRS and MIZ\v{S}, Slovenia; DST/NRF, South Africa;
MICINN, Spain; SRC and Wallenberg Foundation, Sweden; SER, SNSF and Cantons of
Bern and Geneva, Switzerland; NSC, Taiwan; TAEK, Turkey; STFC, the Royal
Society and Leverhulme Trust, United Kingdom; DOE and NSF, United States of
America.

The crucial computing support from all WLCG partners is acknowledged
gratefully, in particular from CERN and the ATLAS Tier-1 facilities at
TRIUMF (Canada), NDGF (Denmark, Norway, Sweden), CC-IN2P3 (France),
KIT/GridKA (Germany), INFN-CNAF (Italy), NL-T1 (Netherlands), PIC (Spain),
ASGC (Taiwan), RAL (UK) and BNL (USA) and in the Tier-2 facilities
worldwide.

\bibliographystyle{epjc}

\onecolumn

\clearpage

\begin{flushleft}
{\Large The ATLAS Collaboration}

\bigskip

G.~Aad$^{\rm 48}$,
T.~Abajyan$^{\rm 21}$,
B.~Abbott$^{\rm 111}$,
J.~Abdallah$^{\rm 12}$,
S.~Abdel~Khalek$^{\rm 115}$,
A.A.~Abdelalim$^{\rm 49}$,
O.~Abdinov$^{\rm 11}$,
R.~Aben$^{\rm 105}$,
B.~Abi$^{\rm 112}$,
M.~Abolins$^{\rm 88}$
% ...
\bigskip
\\
% ...
$^{11}$ Institute of Physics, Azerbaijan Academy of Sciences, Baku, Azerbaijan\\
$^{12}$ Institut de F{\'\i}sica d'Altes Energies and Departament de F{\'\i}sica de la Universitat Aut{\`o}noma de Barcelona and ICREA, Barcelona, Spain\\
% ...
$^{21}$ Physikalisches Institut, University of Bonn, Bonn, Germany\\
% ...
$^{48}$ Fakult{\"a}t f{\"u}r Mathematik und Physik, Albert-Ludwigs-Universit{\"a}t, Freiburg, Germany\\
$^{49}$ Section de Physique, Universit{\'e} de Gen{\`e}ve, Geneva, Switzerland\\
% ...
$^{88}$ Department of Physics and Astronomy, Michigan State University, East Lansing MI, United States of America\\
% ...
$^{105}$ Nikhef National Institute for Subatomic Physics and University of Amsterdam, Amsterdam, Netherlands\\
% ...
$^{111}$ Homer L. Dodge Department of Physics and Astronomy, University of Oklahoma, Norman OK, United States of America\\
$^{112}$ Department of Physics, Oklahoma State University, Stillwater OK, United States of America\\
% ...
$^{115}$ LAL, Universit{\'e} Paris-Sud and CNRS/IN2P3, Orsay, France\\
% ...
\end{flushleft}

\end{document}